\documentclass[10pt,letterpaper]{article}
\usepackage[top=0.85in,left=2.75in,footskip=0.75in]{geometry}

\usepackage{amsmath,amssymb}

\usepackage{changepage}

\usepackage[utf8x]{inputenc}
\usepackage{hyperref}
\usepackage{textcomp,marvosym}

\usepackage{cite}

\usepackage{nameref}

\usepackage[right]{lineno}

\usepackage{microtype}
\DisableLigatures[f]{encoding = *, family = * }

\usepackage[table]{xcolor}

\usepackage{array}
\usepackage{booktabs}
\usepackage{multirow}

\newcolumntype{+}{!{\vrule width 2pt}}

\raggedright
\usepackage[aboveskip=1pt,labelfont=bf,labelsep=period,justification=raggedright,singlelinecheck=off]{caption}

\usepackage{graphicx}
\usepackage{epstopdf}

\usepackage{subcaption}

\begin{document}
\vspace*{0.2in}
\begin{flushleft}
{\Large
\textbf\newline{Long-Range Dependence in Financial Markets:\\ a Moving Average Cluster Entropy approach} 
}
\newline
\\
Pietro Murialdo\textsuperscript{1}
Linda Ponta\textsuperscript{2},
Anna Carbone\textsuperscript{1},
\\
\bigskip
\textbf{1} Politecnico di Torino, corso Duca degli Abruzzi 24, 10129 Torino, Italy
\\
\textbf{2}  LIUC-Universit\`a Cattaneo, corso Giacomo Matteotti 22, 21053 Castellanza, Italy
\\
\bigskip
\end{flushleft}
\section*{Abstract}
A perspective is taken on the intangible complexity of economic and social systems by investigating  the underlying dynamical processes  that produce, store and transmit information in financial time series in terms of the \textit{moving average cluster entropy}. An extensive analysis has evidenced market and horizon dependence of the \textit{moving average cluster entropy} in real world financial assets.  The origin of the behavior is scrutinized by applying   the \textit{moving average cluster entropy} approach to long-range correlated stochastic processes as the  Autoregressive Fractionally Integrated Moving Average (ARFIMA) and Fractional Brownian motion (FBM). To that end, an extensive set of  series is generated with a broad range of values of the Hurst exponent  $H$ and of the autoregressive, differencing and moving average parameters $p,d,q$. A systematic relation between \textit{moving average cluster entropy}, \textit{Market Dynamic Index}  and long-range correlation parameters $H$, $d$ is observed. This study  shows that the characteristic behaviour exhibited by the horizon dependence of the cluster entropy  is related to long-range positive correlation in  financial markets. Specifically, long range positively correlated ARFIMA processes with differencing parameter $ d\simeq 0.05$, $d\simeq 0.15$ and $ d\simeq 0.25$  are consistent with \textit{moving average cluster entropy} results obtained in time series of DJIA,  S\&P500 and  NASDAQ.




\section{Introduction}
 In recent years, much effort has been spent on studying complex interactions in financial markets by means of information theoretical measures from different standpoints.   The information flow  can be probed by observing a relevant quantity  over a certain temporal range (e.g. price and volatility series of financial assets).
 Socio-economic complex systems exhibit remarkable features related to patterns  emerging  from the
seemingly random structure in the observed time series, due to the interplay of long- and short-range correlated decay processes.
The correlation degree is intrinsically linked to the information embedded in the patterns, whose extraction and quantification allow one to add clues to the underlying complex phenomena \cite{grassberger1983characterization,crutchfield1989inferring,crutchfield2012between,ormos2014entropy,Yang2017Information,ghosh2017what,backus2014sources,zhou2013applications,shalizi2004quantifying,carbone2004analysis,carbone2007scaling,carbone2013information,zhao2018multiscale,humeau2015multiscale,niu2015quantifying}.
\par
 An information measure $S(x)$  was proposed by Claude Shannon to the aim of quantifying the degree of uncertainty of strings of elementary random events in terms of their probabilities \cite{shannon1948mathematical}. The elementary stochastic events are related to a relevant variable $x$ whose values are determined by the probability $\left\{p_{i}\right\}$. For example, the  size $\ell$ of a string (block),  corresponding to a particular realization within the sequence, can be associated to the probability $p_{i}\left(\ell \right)$ that, for stationary processes, does not depend on the actual position of the string (block) in the sequence.  The Shannon measure is then given by the expectation value of every possible event $S(\ell)=\sum_i p_{i}(\ell) \log p_{i}(\ell)$.
 The Shannon entropy is calculated over all possible strings and the entropy density  $s_{\ell}=\lim _{\ell \rightarrow \infty} {S(\ell)}/{\ell}$ quantifies the rate at which the process produces unexpected information as a function of the size $\ell$.
 \par
 A  complexity measure $K(x)$ was proposed by Kolmogorov to  quantify the amount of information contained in the string $x$ \cite{kolmogorov1965three}. The relation between Kolmogorov complexity and Shannon entropy has been  extensively investigated in \cite{li2008introduction}. The entropy density $s_{\ell}$ for a stationary process is equal to the Kolmogorov entropy rate.
\par
 The first step required for the practical implementation of  entropy and complexity measures is a suitable partition of the sequence which is  critical to unbundle random and deterministic blocks of given length (decryption). The method  usually adopted for  partitioning a sequence and estimating its entropy is based on a uniform division in blocks with same length \cite{marcon2014generalization,rubido2018entropy,darbellay1999estimation,steuer2001entropy}.

\par
The \textit{cluster entropy method} \cite{carbone2004analysis,carbone2007scaling,carbone2013information} implements the partition via a moving average process. The \textit{clusters} correspond to blocks of different sizes, defined as the portion between  consecutive intersections of a given time series and moving average.
The \textit{cluster entropy method} has been applied to financial markets in \cite{ponta2018information,ponta2019quantifying}. Cumulative information measures (indexes) have been worked out with the ability to provide deep insights  on heterogeneity and dynamics.  In particular:
\begin{itemize}
    \item \textbf{Heterogeneity}.  Volatility series have been analysed by using the cluster entropy approach over constant temporal horizon (six years of tick-by-tick data sampled every minute). An information measure of  heterogeneity, the \textit{Market Heterogeneity Index} $I(T,n)$, where $T$ and $n$ are respectively the volatility and moving average windows, has been developed  by integrating the cluster entropy curves of the volatility series over the cluster length $\tau$. It has been also shown that the \textit{Market Heterogeneity Index} can be used to yield the weights of an efficient portfolio as a complement to Markowitz and Sharpe traditional approaches for markets not consistent with  Gaussian conditions  \cite{ponta2018information}.
       \item  \textbf{Dynamics}. Prices series have been investigated by using the cluster entropy approach  over several temporal horizons (ranging from one to twelve months of tick-by-tick data with sampling interval between 1 up to 20 seconds depending on the specific market). The study has revealed a systematic dependence of the cluster entropy over time horizons in the investigated markets. The \textit{Market Dynamic Index} $I(M,n)$, where $M$ is the temporal horizon and $n$ is the moving average window, defined as the integral of the cluster entropy over $\tau$, demonstrates its ability to quantify the dynamics of  assets' prices over consecutive time periods in a single figure \cite{ponta2019quantifying}.
\end{itemize}
\par
The present  study is motivated by the results obtained in \cite{ponta2019quantifying} showing that cluster entropy of real-world financial markets (NASDAQ, DJIA and S\&P500) exhibits significant \textit{market and horizon dependence}.
According to classical financial theories, subsequent price deviations are identically and independently distributed (\textit{iid}). All the information would be immediately reflected into markets, thus hampering  past observations to predict future outcomes.  If that were true, correlation would be negligible and prices would be simply modelled in terms of \textit{fully uncorrelated Brownian motion}.
However,  several studies have shown that real world markets only partially behave according to the standard theory of perfectly informed and rational agents.
\par
Here, we add further clues to the microscopic origin of the horizon dependence of the cluster entropy in financial markets. To this purpose, the cluster entropy approach is applied to an extensive set of artificially generated series with the aim of shedding light on the characteristic behaviour of real world assets \cite{ponta2019quantifying}.
We report results of the cluster entropy in  \textit {Geometric Brownian Motion} (GBM), \textit {Fractional Brownian Motion} (FBM) and \textit{Autoregressive Fractionally Integrated Moving Average} (ARFIMA) processes.   Those are well-known processes characterized either by hyperbolically decaying  or exponentially decaying correlation functions, features reflected in long-range or short-range dependent dynamics of the elementary random events.  The performance of the \textit{Autoregressive Fractionally Integrated Moving Average} (ARFIMA) process and its variants are receiving a lot of attention and are under intense investigation in the financial research community  \cite{vera2020long,graves2017systematic,bhattacharyya2020the,Bhardwaj2006An,BAILLIE2012Prediction}.
 This work clearly demonstrates  the relationship between the endogenous dynamics of the time series and their  long-range dependence.
 \par
It is shown that deviations of the moving average cluster entropy behaviour in comparison to simple Brownian motion is unequivocally related to the long-range dependence of real-world market series.
In particular,  moving average cluster entropy results obtained on Fractional Brownian Motion with Hurst exponent $H$ in the range $0\leq H \leq 0.5$ (negatively correlated series) show no time horizon dependence. Conversely cluster entropy curves  with Hurst exponent $H$ in the range $0.5\leq H \leq 1$ (positively correlated series) exhibit some dispersion in the horizon dependence in analogy with the real-world financial markets. Results obtained on ARFIMA series confirm and extend the findings reported for FBMs. Horizon dependence of the cluster entropy is observed for a differencing parameter $0\leq d \leq 0.5$. Fine tuning of the horizon dependence is obtained by varying the autoregressive $p$ and moving average $q$ components in the ARFIMA series.
\par
The organisation of the work is as follows. The cluster entropy method used for the analysis and the investigated market and artificial data are described in Section \ref{Sec:Methods and Data}.  Results  on  cluster entropy and market dynamic index estimated  over \textit {Geometric Brownian Motion} (GBM), \textit {Fractional Brownian Motion} (FBM) and \textit{Autoregressive Fractionally Integrated Moving Average} (ARFIMA) series, are reported in Section \ref{Sec:Results}. Finally, results are discussed, conclusions are drawn and a path for future work is suggested in Section \ref{Sec:Discussion and Conclusions}.

\section{Methods and Data}
\label{Sec:Methods and Data}
In  this section the  cluster entropy approach developed in \cite{carbone2007scaling,carbone2004analysis} is  briefly recalled. The second part of this section is devoted to the description of financial market data used in \cite{ponta2019quantifying}. For the sake of completeness, we also recall the main definitions related to the Fractional Brownian Motion and Autoregressive Fractionally Integrated  Moving Average processes.
\subsection{Cluster Entropy Method}
It is well-known that the general idea behind  Shannon entropy is to measure the amount of information embedded in a message to identify  the shortest subsequence actually carrying the relevant information and the degree of redundancy which is not necessary to reproduce the initial message.
The Shannon functional is written as:
\begin{equation} \label{eq:shannonentropy}
S(\tau,n) = \sum P(\tau,n) \log P(\tau,n),
\end{equation}
where $P(\tau,n)$ is a probability distribution associated with the time series $y(t)$. To the aim of estimating the probability distribution $P(\tau,n)$,  it is necessary to partition the continuous phase space into disjoints sets. The  traditionally adopted methods divide the sequence into segments of equal lengths (blocks). Here, we follow another approach.
\par
In \cite{carbone2007scaling,carbone2004analysis} the time sequence $y(t)$,  is partitioned in \textit{clusters} by the intersection with its moving average $\tilde{y}_n(t)$, with $n$ the size of the moving average. The simplest type of moving average is defined at each $t$ as the average of the $n$ past observation from $t$ to $t-n+1$,
\begin{equation} \label{eq:simplema}
\tilde{y}_n(t) = \frac{1}{n} \sum_{k = 0}^{n - 1} y(t-k).
\end{equation}
Note that while the original series is defined from $1$ to $N$, the moving average series is defined from $1$ to $N-n+1$ because $n$ samples are necessary to initialize the series. The original series and the moving average series are indicated as $\{y(t)\}_{t=1}^N$ and $\{\tilde{y}_n(t)\}_{t=1}^{N-n+1}$ respectively. Consecutive intersections of the time series and of the moving average series yield a partition of the phase space into a series of \textit{clusters}. Each cluster is defined as the portion of the time series $y(t)$ between two consecutive intersection of $y(t)$ itself and its moving average $\tilde{y}_n(t)$ and has length (or duration) equal to:
\begin{equation} \label{eq:duration}
\tau_j \equiv || t_j -t_{j-1} ||,
\end{equation}
where $t_{j-1}$ and $t_j$ refers to two subsequent intersections of $y(t)$ and $\tilde{y}_n(t)$.
For each moving average window $n$ the probability distribution function $P(\tau, n)$ which associates the length of a cluster $\tau$ with its frequency can be obtained by counting the number of clusters $\mathcal{N}_j(\tau_j,n)$ with length $\tau_j$, $j \in \{1,N-n-1\}$. The probability distribution function $P(\tau, n)$ results:
\begin{equation} \label{eq:probdistr}
P(\tau,n) \sim \tau^{-D} \mathcal{F}(\tau,n) \quad,
\end{equation}
where $D = 2-H$  indicates  the fractal dimension with  $H$ the Hurst exponent of the sequence. In this framework long-range correlation implies  that the clusters are organized in a similar way along the time series \textit(self-organized), even for clusters far away in time from each other.
The term $\mathcal{F}(\tau,n)$ in Equation (\ref{eq:probdistr}) takes the form:
\begin{equation} \label{eq:expdecay}
\mathcal{F}(\tau,n) \equiv e^{-\tau/n} \quad,
\end{equation}
to account for the drop-off of the power-law behavior for $\tau < n$ and the onset of the exponential decay when $\tau \geq n$ due to the finiteness of $n$. When $n \rightarrow 1$ the lengths $\tau$ of clusters tend to be centered around a single value. When $n \rightarrow N$, that is when $n$ tends to the length of the whole sequence, only one cluster with $\tau = N$ is generated. For middle values of $n$ however a broader range of lengths is obtained and therefore the probability distribution spreads all values.
When the probability distribution in Eq. (\ref{eq:probdistr}) is fed into the Shannon functional in Eq. (\ref{eq:shannonentropy}) the result is the following:
\begin{equation}
S(\tau,n) = S_0 + \log \tau^D - \log \mathcal{F}(\tau,n),
\end{equation}
which, after substituting Eq. (\ref{eq:expdecay}), becomes:
\begin{equation} \label{eq:carboneentropy}
S(\tau,n) = S_0 + \log \tau^D + \frac{\tau}{n},
\end{equation}
where $S_0$ is a constant, $\log \tau^D$ accounts for power-law correlated clusters related to $\tau^{-D}$ and $\tau/n$ accounts for exponentially correlated clusters related to the term $\mathcal{F}(\tau,n)$.
The term $S_0$ can be evaluated in the limit $\tau \sim n \rightarrow 1$, which results in $S_0 \rightarrow -1$ and $S(\tau, n) \rightarrow 0$, that corresponds to the fully deterministic case, where each cluster has size equal to 1. On the other hand, when $\tau \sim n \rightarrow N$, the maximum value for the entropy is obtained with $S(\tau,n) = \log N^D$, which corresponds to the case of maximum randomness, where there is one cluster coinciding with the whole series.
Equation (\ref{eq:carboneentropy}) shows that power-law correlated clusters, characterized by having length $\tau < n$, are described by a logarithmic term as $\log \tau^D$, and their entropy do not depend on the moving average window $n$. However, for values of $\tau \geq n$, which represent exponentially correlated clusters, the term $\tau/n$ becomes predominant. Cluster entropy increases linearly as $\tau/n$, with slope decreasing as $1/n$. Hence, due to the finite size effects introduced by the partitioning method, in $\tau = n$ the behavior of entropy changes and its values exceeds the curve $\log \tau^D$. In other words, clusters that are power-law correlated does not depend on $n$, are said to be \textit{ordered} and represent deterministic information. Clusters that are exponentially correlated does depend on $n$, are said to be \textit{disordered} and represent random clusters.

The meaning of entropy in information theory can be compared with the meaning of entropy in thermodynamics. In an \textit{isolated system}, the entropy increase $dS$ refers to the irreversible processes occurring spontaneously within the system. In an \textit{open system} however a further increase in entropy $dS_{ext}$ occurs due to the irreversible processes spontaneously occurring with the external environment.
\par
The term $\log \tau^D$ should be interpreted as the entropy of the isolated system. It is independent on $n$, that is it is independent on the partitioning method. It takes the form of the Boltzmann entropy, that can be written as $S = \log \Omega$, with $\Omega$ the volume of the system. Therefore the quantity $\tau^D$ corresponds to the volume occupied by the fractional random walker.
\par
The term $\tau/n$ represents the excess entropy caused by the external process of partitioning the sequence.  The excess entropy depends on the moving average window $n$. If same size boxes were chosen, the excess entropy term $\tau/n$ would vanish and entropy would reduce to the logarithmic term. When a moving average partition is used, the term $\tau/n$ emerges to account for the additional heterogeneity introduced by the randomness of the process. Thence, for exponentially correlated clusters entropy exceeds the logarithmic asymptotic.

One important step is to quantify the  property of the entropy result. In order to improve the accuracy of the method, one can consider the integral of the entropy function over the clusters length $\tau$, a cumulative measure able to embed all information in a single figure:
\begin{equation}
I(n) = \int S(\tau, n)d\tau \quad ,
\label{index}
\end{equation}
which for discrete sets reduces to
$I(n) = \sum_\tau S(\tau, n)$.
Eq. (\ref{index}) can be written as:
\begin{equation}
I(n) = \int_1^{\tau(n)} S(\tau, n)d\tau + \int_{\tau(n)}^{\infty} S(\tau, n)d\tau \quad.
\label{eq:realmdi}
\end{equation}
The first integration is referred to the power law regime of the cluster entropy, the second integration is referred to the linear regime of the cluster entropy (i.e. the excess entropy term).

\subsection{Data}
\subsubsection{Financial Data}
The objective of this work is to investigate and shed light on the characteristic features exhibited by  cluster entropy of financial markets. In particular here our focus is on the systematic dependence of the cluster entropy of the price series over time horizon $M$.
\par
In  \cite{ponta2019quantifying}  the cluster entropy is applied to a large set of tick-by-tick data of the USA's indexes (S\&P500, NASDAQ and DJIA). NASDAQ is an index resulting from all the public firms quoted on the market, DJIA and S\&P500  indexes are representative of a selected number of public firms. For each index, investigated data include tick-by-tick prices from January 2018 to December 2018. More information about the markets can be found at the Bloomberg terminal. \par
To study the dynamics of financial series different time horizons need to be compared.   As explained in the Introduction,  entropy is  sample-size dependent by definition, thus in order to rule out spurious results the length of the investigated sequences must be the same. Therefore, cluster entropy analysis requires the comparison to be implemented on series with same length.  Raw data have been downloaded from the Bloomberg terminal in the form of  tick-by-tick data. The lengths of the raw series vary due to  different number of trading days and  transactions per time unit. It is therefore necessary, as first computational step, to implement a sampling of the raw data to make the length of the series exactly the same.  The first raw series ranges from the first transaction of January 2018 to the last one of January 2018; the second ranges from the first transaction in January 2018 to the last of February 2018, $\ldots$, the twelfth ranges from the first transaction in January 2018 to the last of December 2018, a period equivalent to the whole year. Because each raw series ranges from the first tick of 2018 to the last tick of the relative month, the twelve series have very different lengths.  The series are sampled to obtain twelve \textit{series} with same length as described in the following.
\par
Twelve sampling time intervals and corresponding frequencies must be defined, i.e. twelve integers indicating for each series the interval of skipped data. Sampling intervals are obtained by dividing the length of each raw series by the length of the shortest raw one and then rounding to the inferior  integer. Thence, each raw series is sampled with the relative sampling interval to yield a \textit{sampled series}: for each sample in the sampled series, a number of samples equal to the sampling frequency has been discarded in the raw series.
The sampled series obtained are \textit{approximately} of equal lengths. To obtain twelve series of \textit{exactly} equal length, a few observations are cut off, when exceeding the length of the shortest series.
The result consists in twelve sampled series that are equal in length and refer to time horizons varying from one month ($M=1$) to twelve months  ($M=12$). In more details, the length of the series corresponding to the $M$  different horizons is $N_M$ (where $M$ ranges from 1 to 12 for one year of data). Among the monthly series, the shortest month is used to evaluate the minimum value of $N_{min}$ and, correspondingly, of the sampling frequency. Then, the sampling intervals  for the multiple periods is derived by dividing the multiple period lengths (i.e. the sum of multiple consecutive $N_M$)  by the value $N_{min}$. In Table  \ref{tab:sampling} a few examples of sampling intervals and lengths $N_M$ are shown for clarify the procedure.
  It is worth noting that the length of sampled series should be at least $10^5$ to ensure enough accuracy of the results.
\subsection{Artificial Data}
Artificial series have been generated by using FBM and ARFIMA processes with  same  temporal structure corresponding to the different horizons of the financial market data reported in  \cite{ponta2019quantifying}. Artificial series were generated  with length equal to those of the twelfth cumulative series analysed in \cite{ponta2019quantifying}. Thence, we divided the series in the respective cumulative series according to the lengths obtained for NASDAQ cumulative series reported in \cite{ponta2019quantifying}. Then the sampling method proceeds analogously from the calculation of the sampling frequency. Such sampling method was applied to series generated by artificial financial models to make sure that the information content would be comparable to that of real-world financial series.
In the remainder of this section, we recall the main definitions for Fractional Brownian Motion and Autoregressive Fractionally Integrated Moving Average processes.
\subsubsection{Geometric Brownian Motion}
The Geometric Brownian Motion is the basis of the \textit{Black-Scholes-Merton} model used to price options and is defined by the following difference equation:
\begin{equation} \label{eq:gbm}
dX_t = \mu(t) X_t dt + D(t,X_t) \sigma (t) dB_t,
\end{equation}
where $\mu(t)$ indicates the level of return, $\sigma (t)$ the volatility and $dB_t$ is a simple Brownian motion. Volatility is deterministic and constant and there are no jumps. Increments are independent on previous states.
\subsubsection{Fractional Brownian Motion}
The \textit{Fractional Brownian Motion} is a long memory process introduced in \cite{mandelbrot_fractional_1968}:
\begin{equation} \label{eq:fbm}
    \begin{split}
        B_H(t) = B_H(0) &+ \frac{1}{\Gamma(H+1/2)} \bigg(\int_{-\infty}^{0} \big( (t-s)^{H-1/2} \\&- (-s)^{H-1/2} \big) dB(s) +\int_0^t (t-s)^{H-1/2} dB(s) \bigg).
    \end{split}
\end{equation}
It is also referred to as a \textit{self-similar} process. A stochastic process $X_t$, with $t \in \mathbb{R}$, is said to be self-similar if there exist $H > 0$ such that for any \textit{scaling factor} $c > 0$,
\begin{equation}
X_{ct} \stackrel{\mathcal{L}}{=} c^HX_t,
\end{equation}
with $H$ the Hurst exponent and ($\stackrel{\mathcal{L}}{=}$) equivalence in distribution. Self-similar processes are stochastic models where a scaling in time is equivalent, \textit{in term of distribution}, to an appropriate scaling in space. Moreover, if, for any $k$, the distribution of $(X_{t_1 + c} - X_{t_1 + c -1}, ..., X_{t_k + c} - X_{t_k + c -1})$ does not depend on $c$, $X_t$ is said to be self-similar with \textit{stationary increments}. 
So, a Gaussian process $B_H(t)$ is called a \textit{Fractional Brownian Motion}, if it satisfies:
1. $B_H(t)$ is self-similar with $0 < H < 1$;
2. $B_H(t)$ has stationary increments.
When $H = 0.5$  a simple Brownian Motion with independent increments is recovered. When $0 < H < 0.5$ the \textit{Fractional Brownian Motion} is said to be anti-persistent, which means that increments tend to be opposite signed. Conversely, when $0.5 < H < 1$ it is said to be persistent, which means that increments tend to be equally signed.
\subsubsection{Autoregressive Fractionally Integrated Moving Average}
The model of an \textit{Autoregressive Fractionally Integrated Moving Average} process (\textit{ARFIMA}) of a time series of order $(p,d,q)$ with mean $\mu$, may be written, using the lag operator $L$, as:
\begin{equation}
    \Phi(L)(1-L)^d(y_t - \mu) = \Theta(L)\epsilon_t,
    \label{arfimamod}
\end{equation}
with $\epsilon_t ~ i.i.d.$ and  $\sim (0, \sigma_\epsilon^2) $. The autoregressive component of the process is represented by the factor:
\begin{equation}
    \Phi(L) = 1 - \phi_1L - ... - \phi_pL^p,
\end{equation}
where the lag operator of order $p$ shifts the value of $y_t$ back to $p$ observations, so that one obtains:
\begin{equation}
    \Phi(L)y_t = (1 - \phi_1L - ... - \phi_pL^p)y_t = y_t - \phi_1y_{t-1} - ... -\phi_py_{t-p}.
\end{equation}
The moving average component of the process is represented by the factor:
\begin{equation}
    \Theta(L)\epsilon_t = (1 + \theta_1L + ... + \theta_qL^q)\epsilon_t = \epsilon_t + \theta_1\epsilon_{t-1} + ... + \theta_{q}\epsilon_{t-q} \quad .
\end{equation}
The fractionally differencing operator $(1 - L)^d$  is defined as:
\begin{equation}
    (1 - L)^d = \sum_{n=0}^{\infty} \frac{\Gamma(k - d)L^k}{\Gamma(-d)\Gamma(k + 1)}.
\end{equation}
Note that the process is stationary only for $-0.5<d<0.5$ .
For $d < |0.5|$ the ARFIMA process is said to exhibit long memory.
\par
The power spectral representation $f(\lambda)$ of Fractional Brownian Motions and Autoregressive Fractionally Integrated Moving Average Processes provides further details  regarding  their power law behavior and the relation between the characteristic exponents.
It is :
\begin{equation}\begin{aligned}
f(\lambda) & \sim |\lambda|^{-2d}  \hspace{15pt} \quad \text{(ARFIMA)}\\
f(\lambda) & \sim |\lambda|^{1-2H} \hspace{15pt} \quad \text{(FBM)}\end{aligned}
\end{equation}
yielding:
\begin{equation}
H=d+1/2
\end{equation}
Among financial models, the \textit{autoregressive fractionally integrated moving average} is one of the most common processes used to model prices of long-range correlated assets.
\section{Results}
\label{Sec:Results}
In this section, the results of the application of the cluster entropy method to several FBM and ARFIMA  series are presented.
The moving average cluster entropy can be implemented via the MATLAB codes available  at the repository  \cite{DropboxPolito}.
\par
\begin{figure}
    \centering
    \includegraphics[width=4cm]{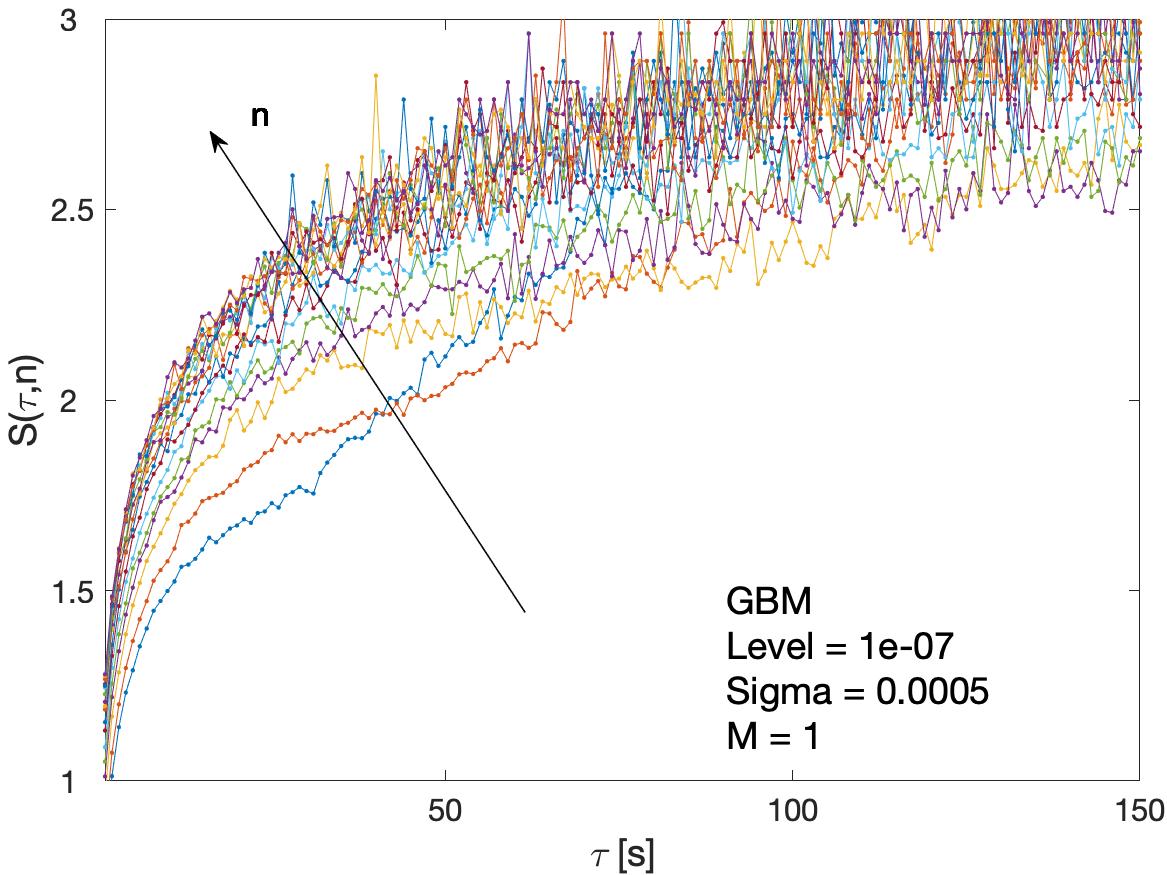}
    \includegraphics[width=4cm]{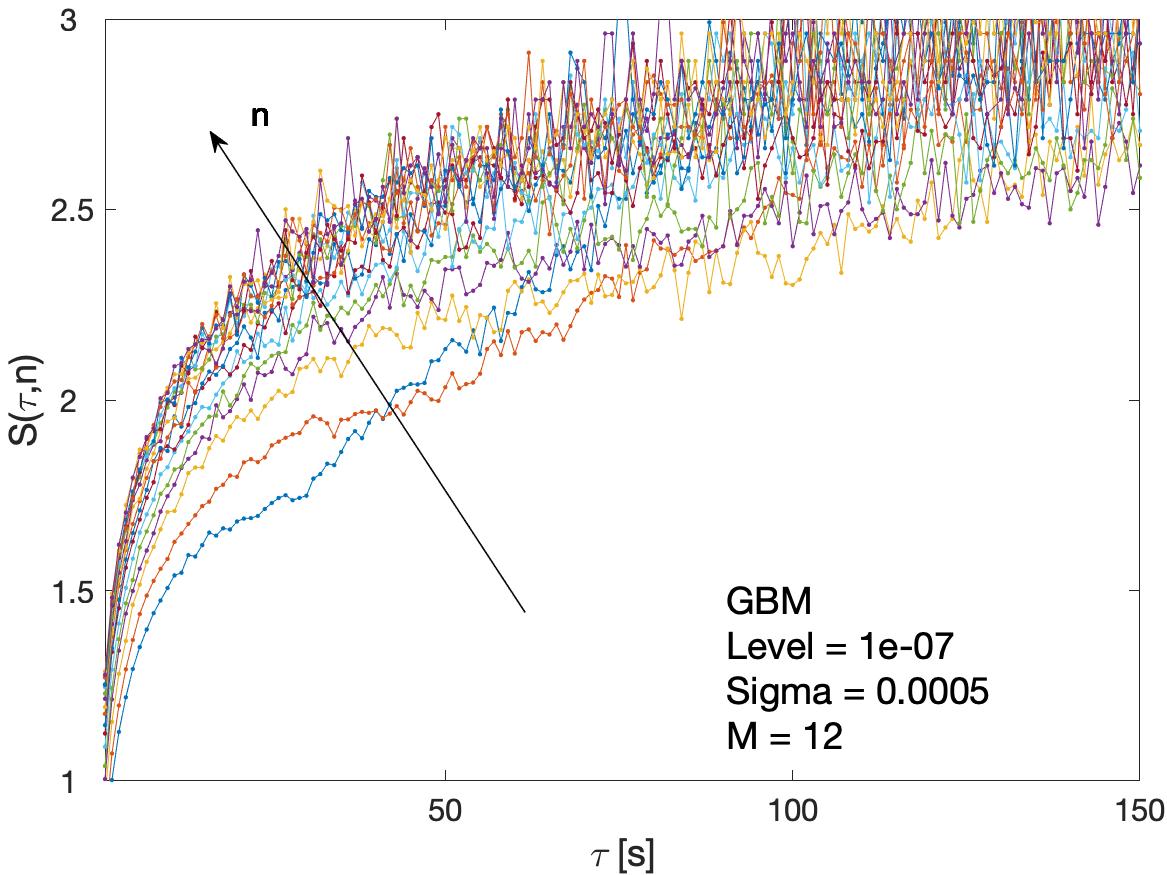}
    \includegraphics[width=4cm]{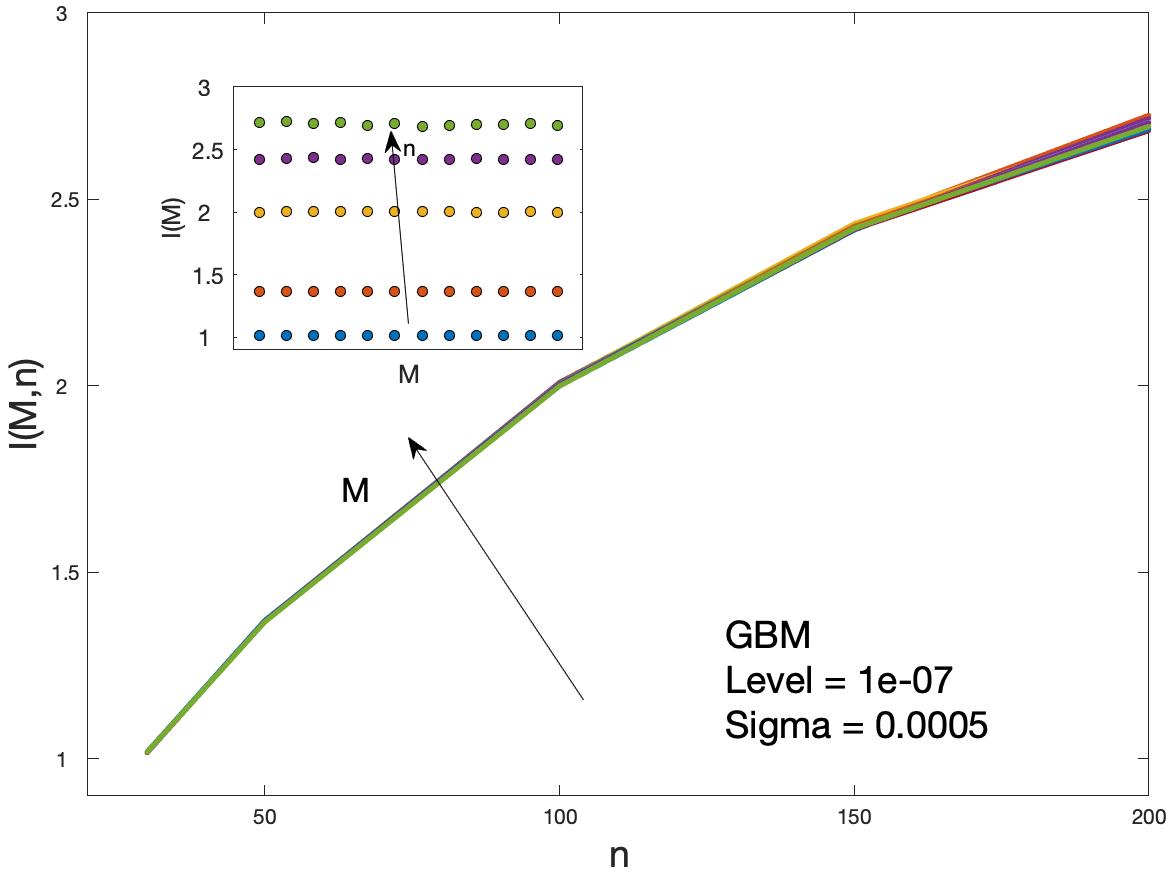}
    \caption{Cluster entropy results for  Geometric Brownian Motion series,  generated with following parameters: $\mu = 1 \cdot 10^{-7}$ and $\sigma = 5 \cdot 10^{-4}$ (left and middle). The market dynamic index $I(M,n)$ is also shown (right). One can note that $I(M,n)$ is practically independent on the temporal horizon $M$. }
    \label{fig:GBM}
\end{figure}

To the purpose to have a set of generic benchmark values for the cluster entropy, first Geometric Brownian Motion series are analysed.
Geometric Brownian Motion series are generated by means of the MATLAB tool available at \cite{GBM}. Several  Geometric Brownian Motion processes are analysed with parameters varying in the range $ 0 \leq \mu \leq 1 \cdot 10^{-7}$ and $5 \cdot 10^{-4} \leq \sigma \leq 5 \cdot 10^{-6}$. Figures \ref{fig:GBM} reports cluster entropy and market dynamic index results on GBM series with parameters: $\mu = 1 \cdot 10^{-7}$ and $\sigma = 5 \cdot 10^{-4}$.

\begin{figure}[t]
    \centering
    \includegraphics[width=4cm]{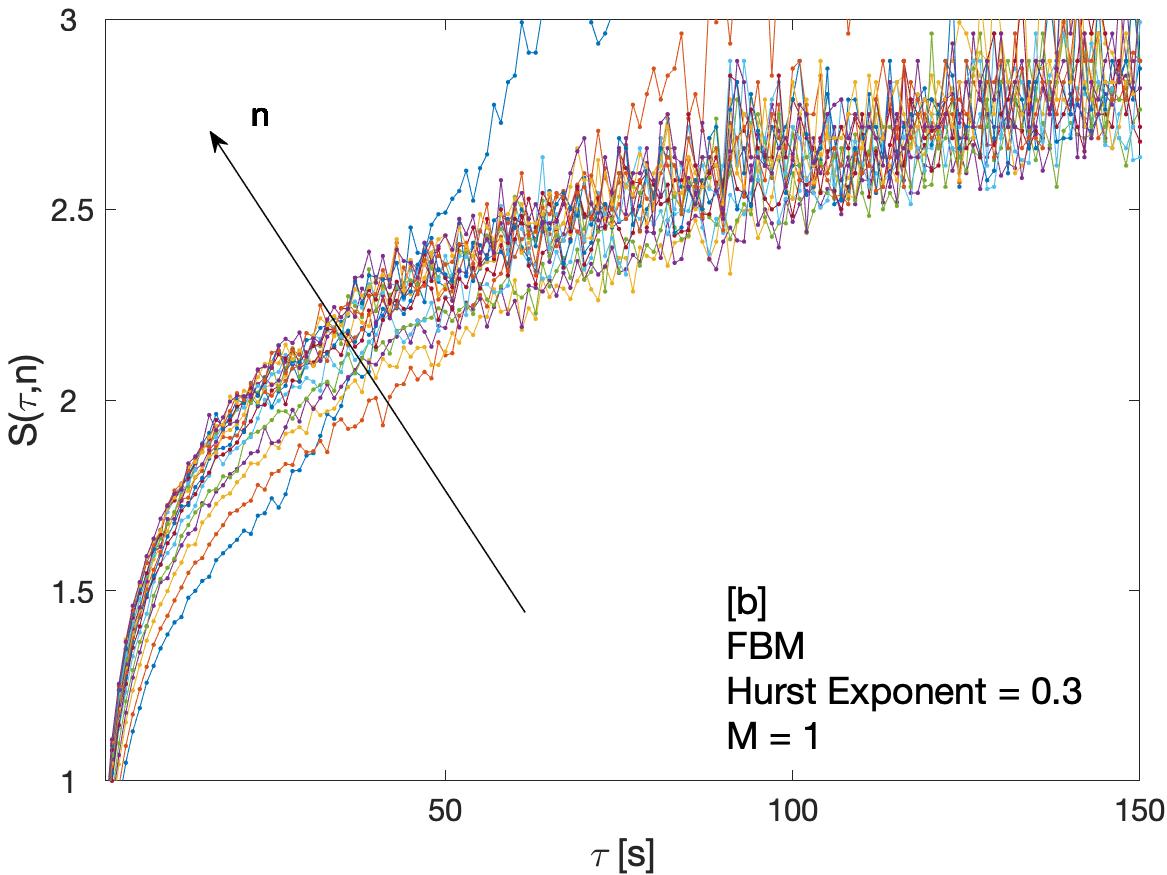}
    \includegraphics[width=4cm]{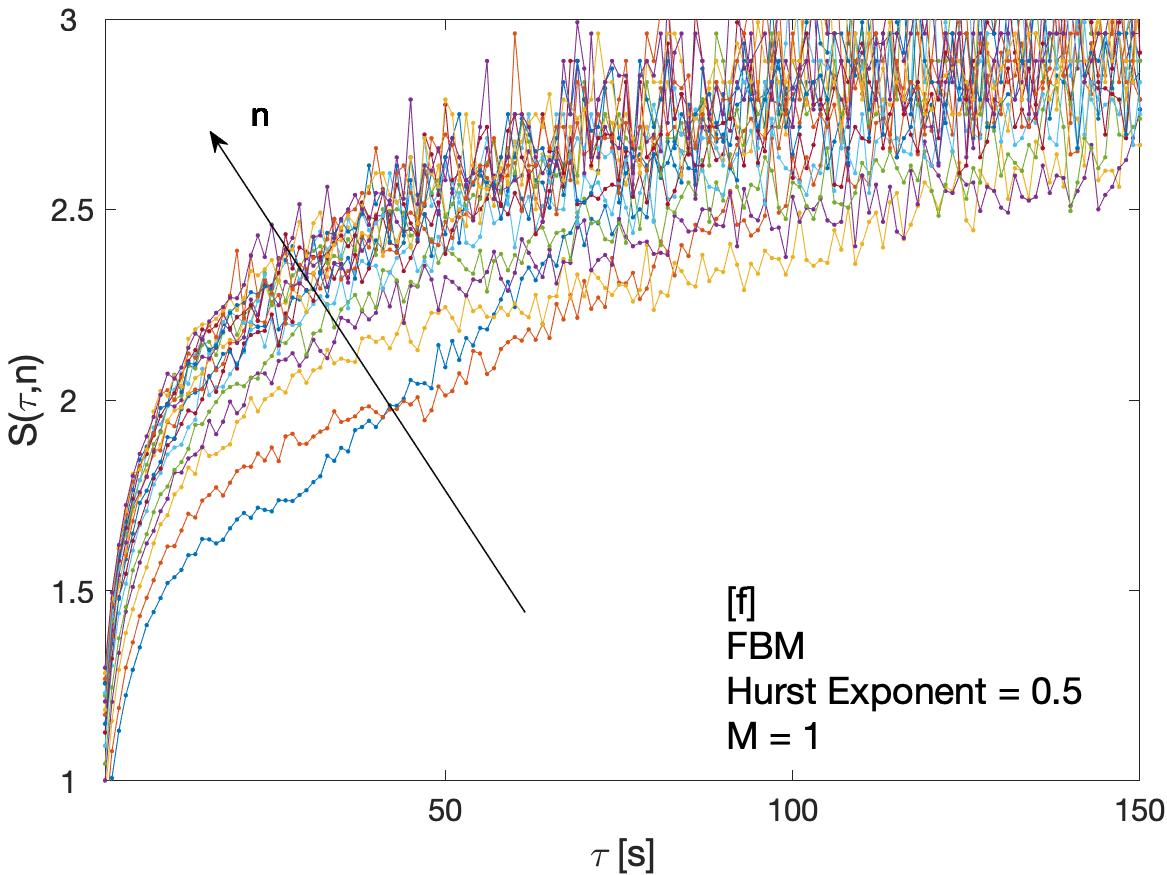}
    \includegraphics[width=4cm]{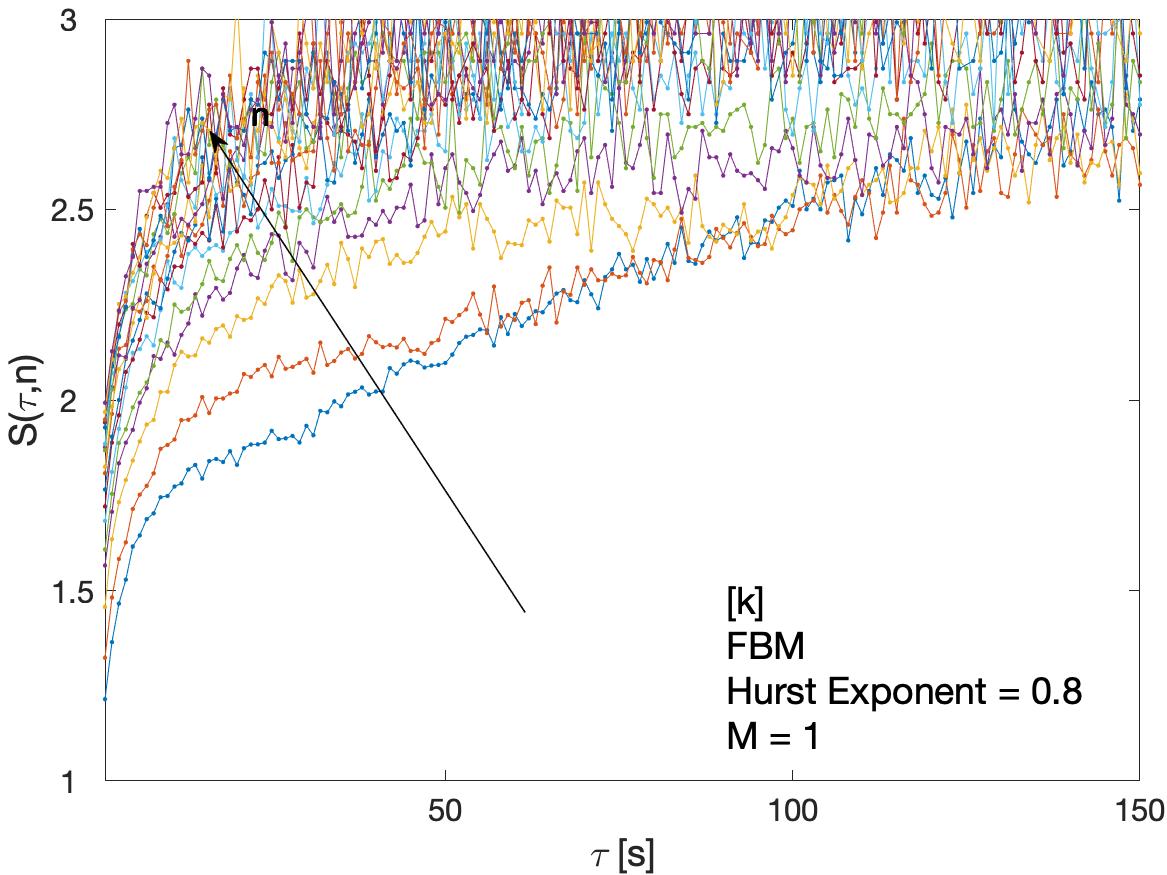}
    \\
    \includegraphics[width=4cm]{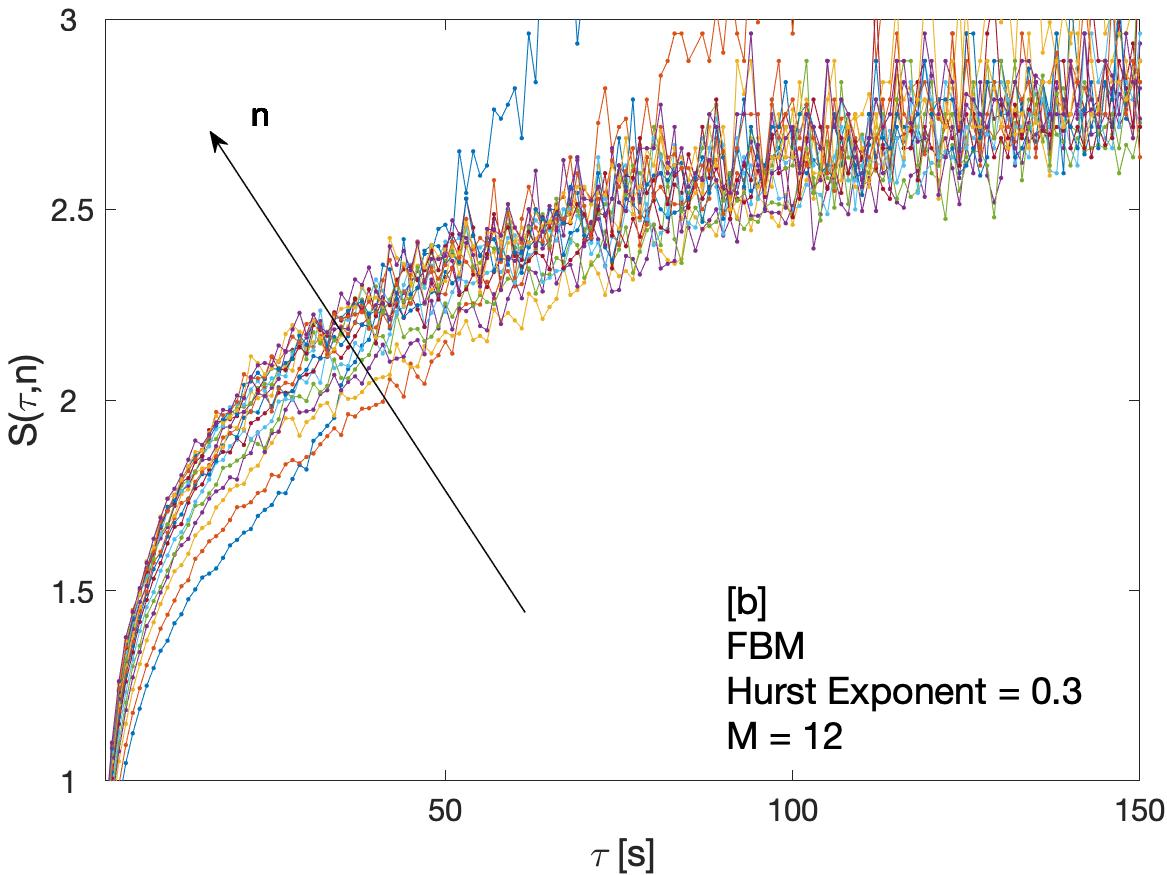}
    \includegraphics[width=4cm]{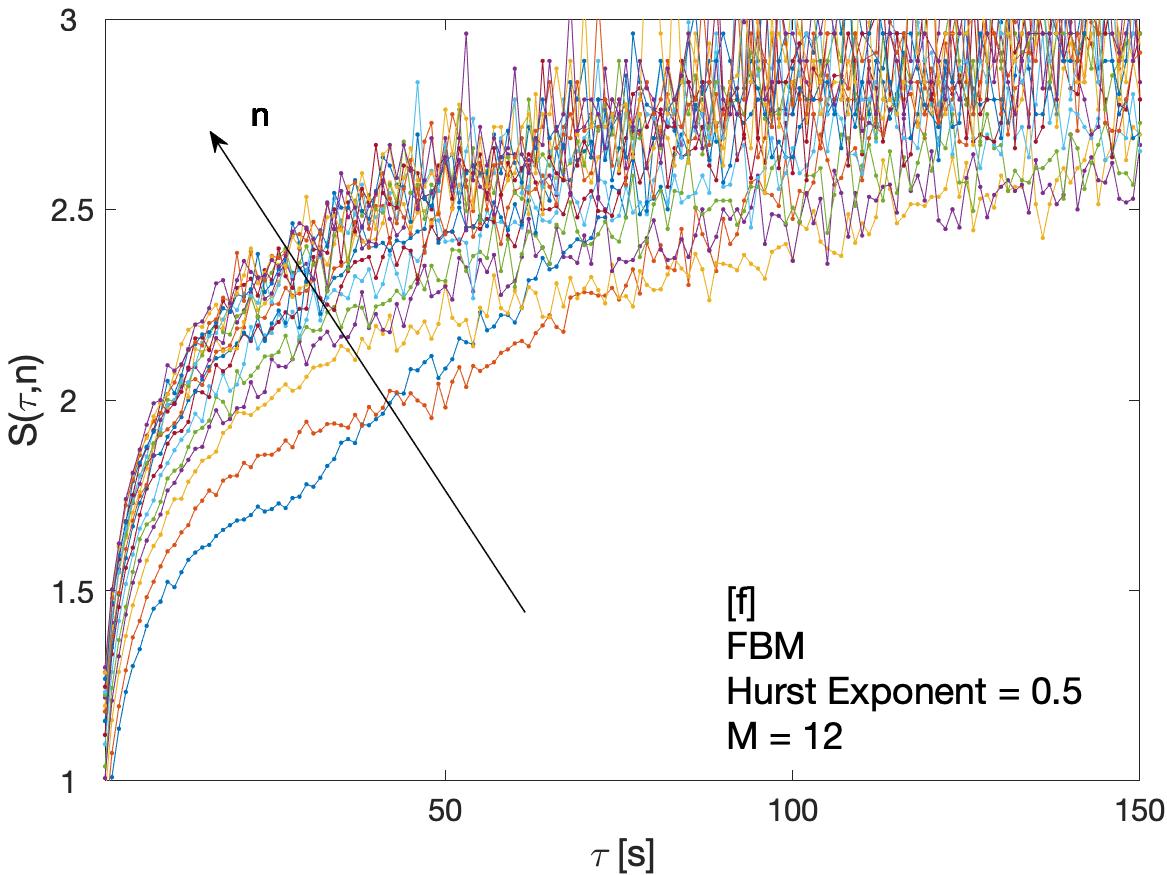}
    \includegraphics[width=4cm]{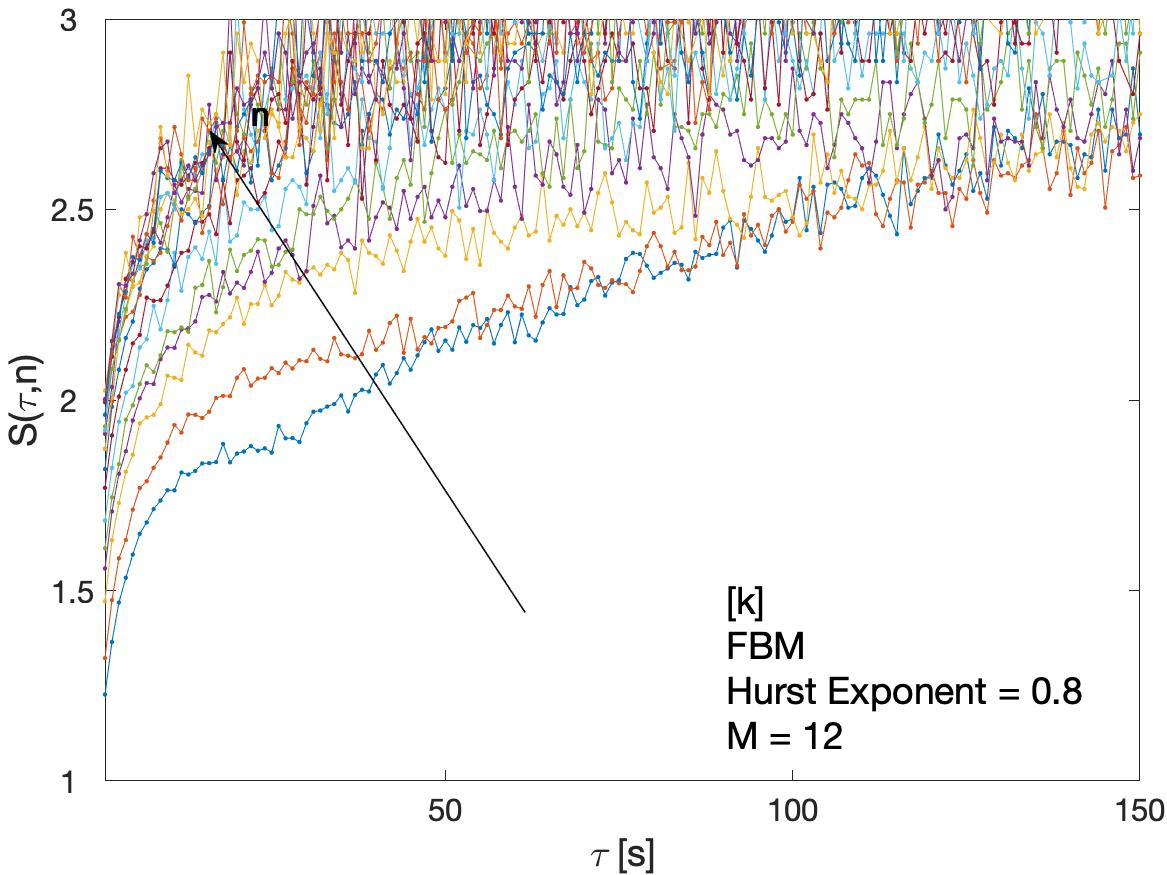}
    \\
    \caption{Cluster entropy results curves for  Fractional Brownian Motion (FBM) series with $H=0.3$, $H=0.5$, $H=0.8$. First row shows results for time horizon $M=1$ (approximately equivalent to the first month (January  2018) of raw data for NASDAQ, S\&P500, DIJA). The second row shows results for time horizon $M=12$ (approximately equivalent to twelve months of data in NASDAQ, S\&P500, DIJA, i.e the whole 2018 year).}
    \label{fig:FBM_CE}
\end{figure}

Results of the cluster entropy approach applied to fractional Brownian motion  are reported in Figures \ref{fig:FBM_CE}. The fractional Brownian motion series were generated by means of the FRACLAB  tool available at \cite{FRACLAB}. Several Fractional Brownian Motion series with Hurst exponent varying in the range $0.1 \leq H \leq 0.9$ are analysed. Figure \ref{fig:FBM_CE} shows the cluster entropy for time horizon $M=1$ and $M=12$, i.e. corresponding respectively to one period of data (one month) and twelve periods of data (one year) for FBM  series with $H = 0.3$, $H=0.5$ and $H=0.8$.
\par
In general, cluster entropy calculated at different time horizons $M$ presents a similar behavior. On account of Eq. (\ref{eq:carboneentropy}), one can expect power-law correlated clusters with a smooth logarithmic increase  of the entropy for $\tau < n$.  Conversely, for $\tau \geq n$, the exponentially correlated decay sets the entropy to increase linearly with the term $\tau/n$ dominating. However, a quite different behavior is observed for different $H$. For $H=0.3$ (anti-correlated FBM series) the  cluster entropy curves exhibit a very limited dependence on the moving average window $n$ over the range of investigated $\tau$.  For $H=0.5$ the cluster entropy curves vary more significantly as the moving average window $n$ changes. For $H = 0.8$ the cluster entropy curves vary even more remarkably and take increasing values for increasing $n$.
\par
\begin{figure}
    \centering
    \includegraphics[width=4cm]{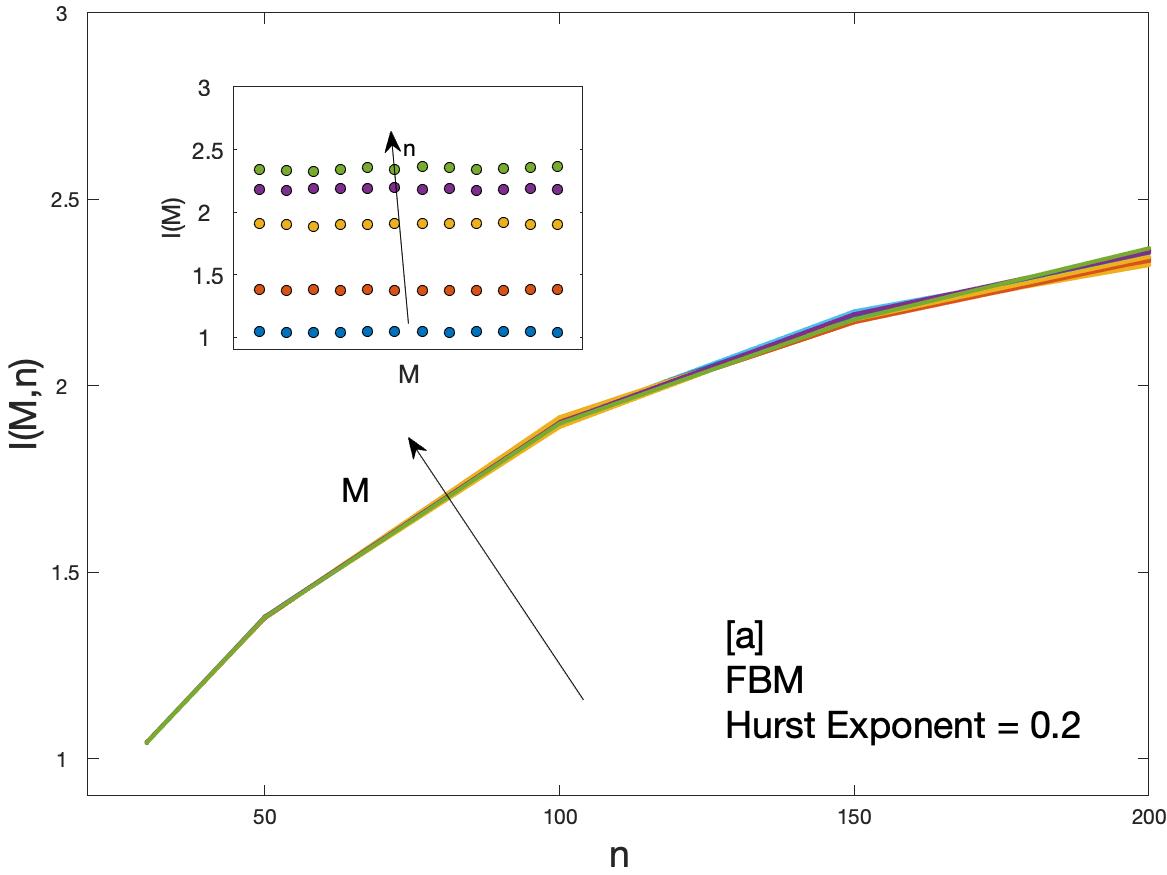}
    \includegraphics[width=4cm]{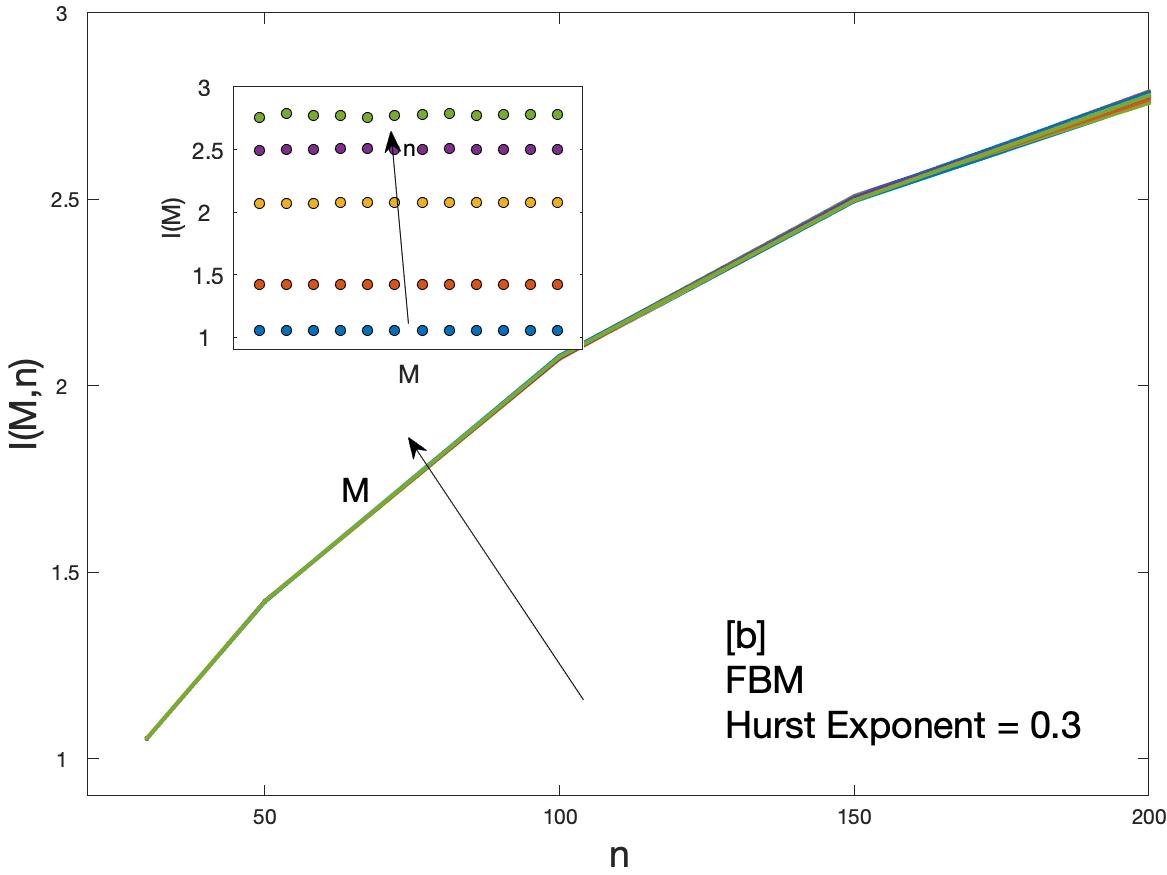}
    \includegraphics[width=4cm]{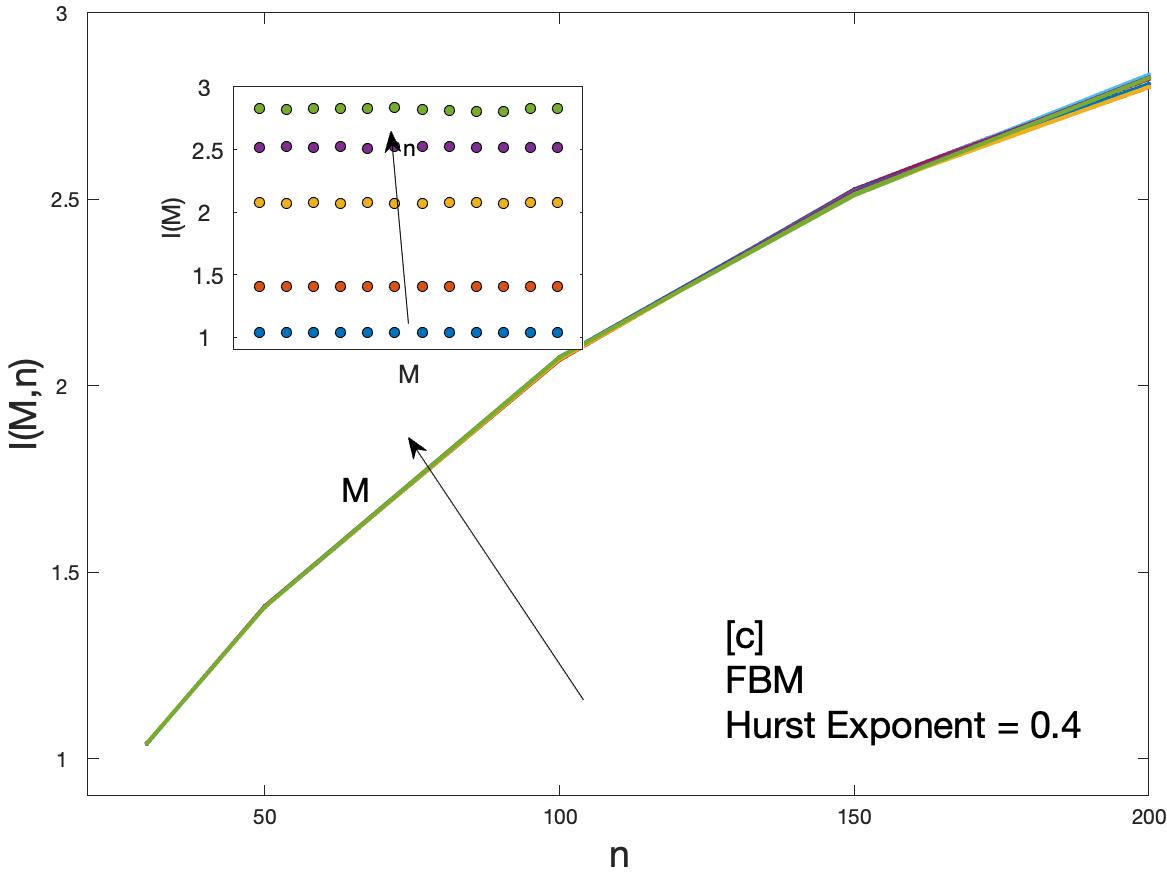}
    \\
    \includegraphics[width=4cm]{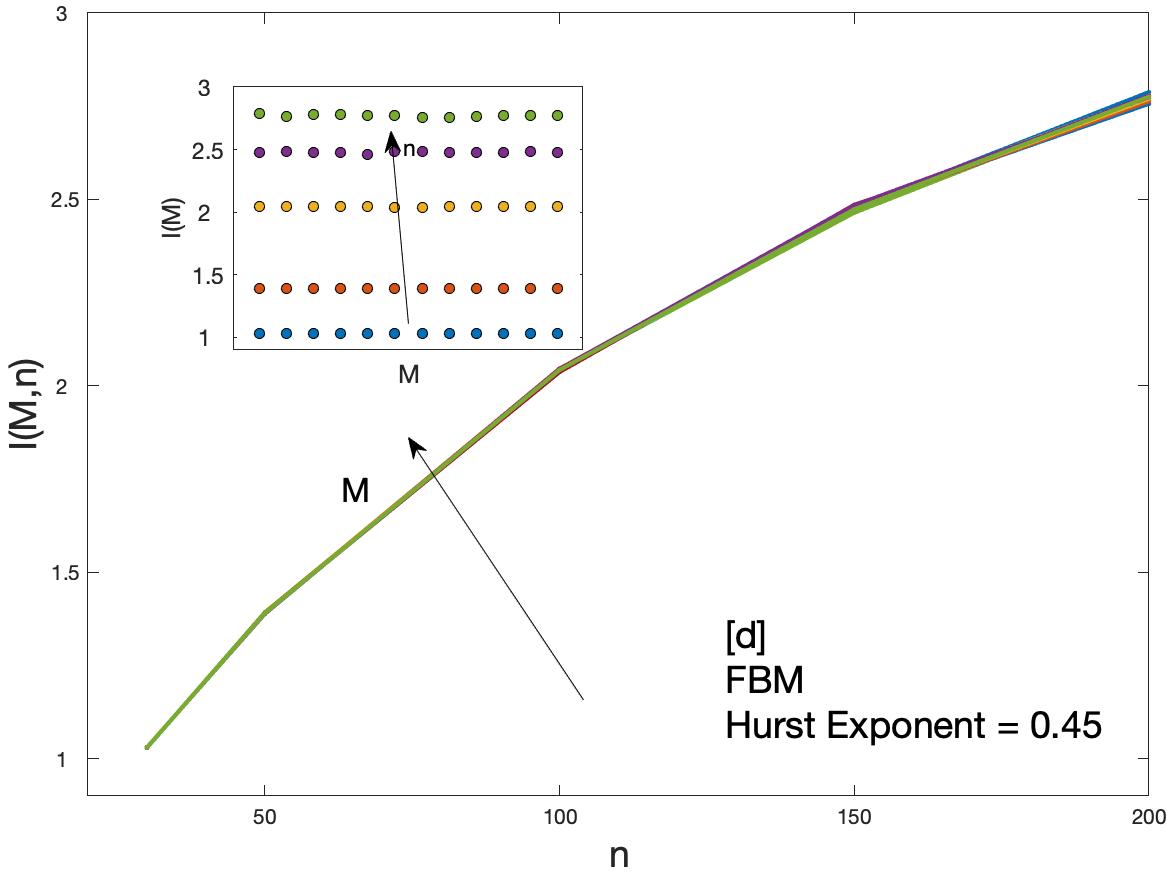}
    \includegraphics[width=4cm]{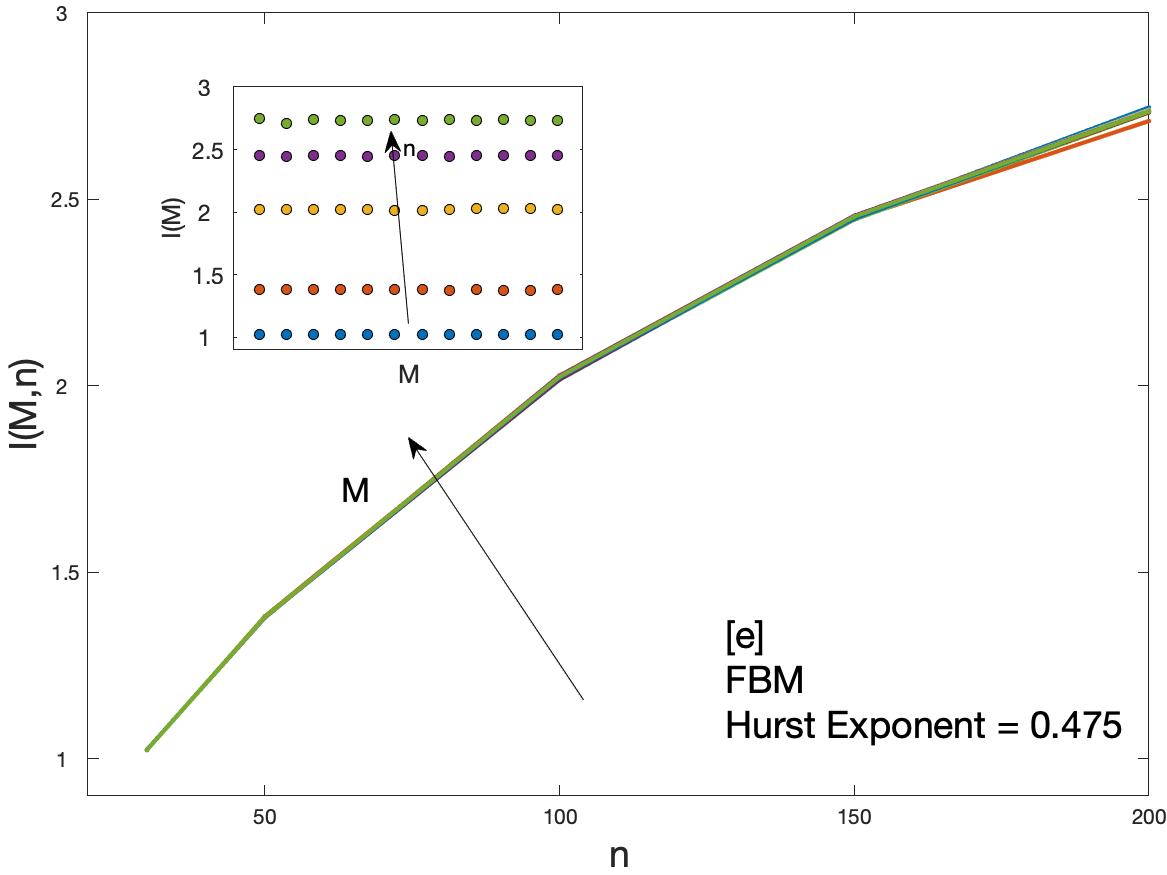}
    \includegraphics[width=4cm]{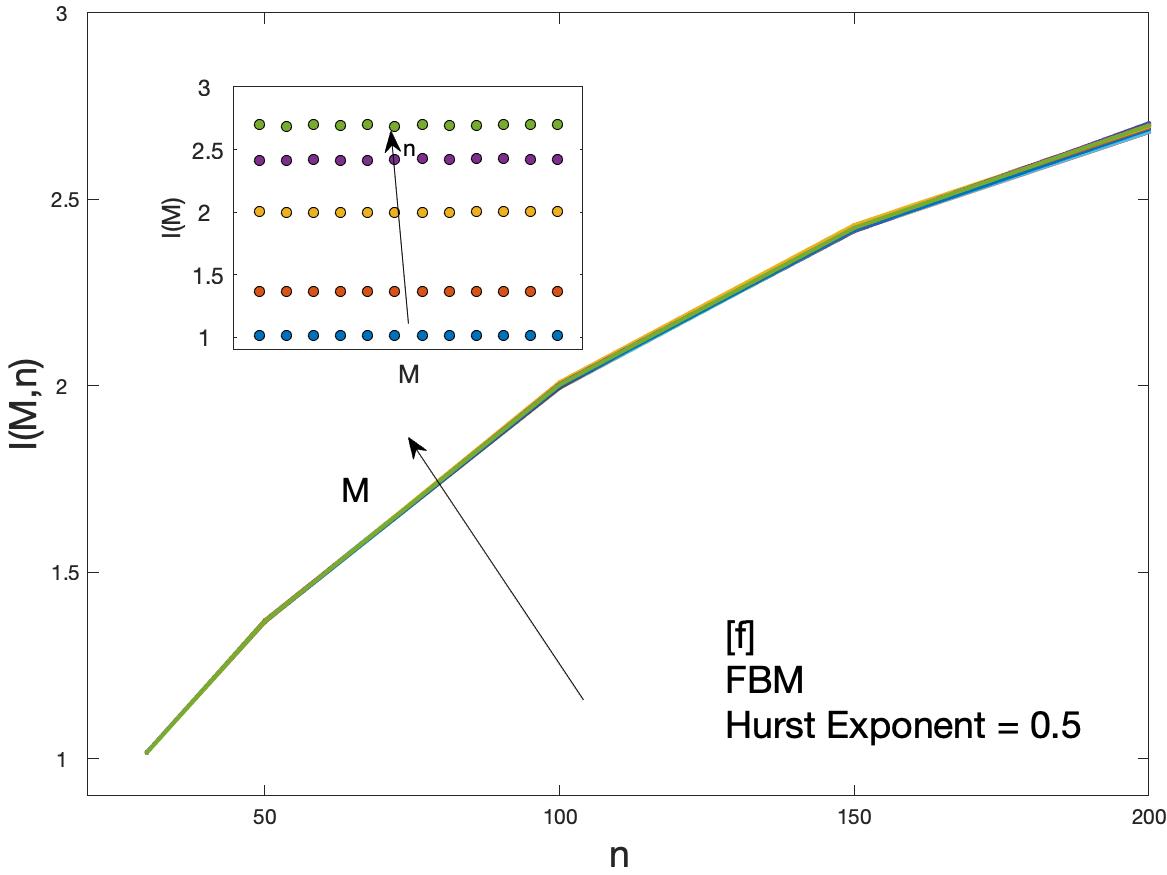}
    \\
    \includegraphics[width=4cm]{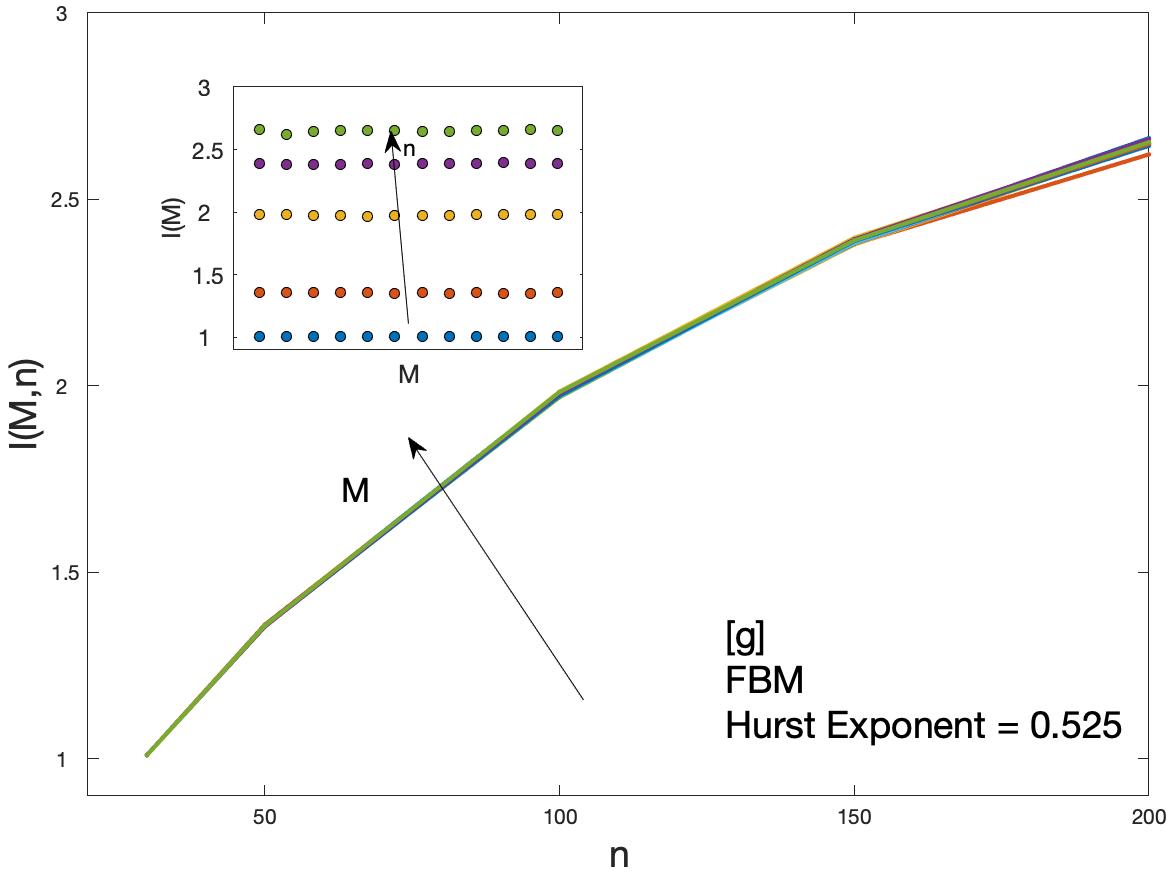}
    \includegraphics[width=4cm]{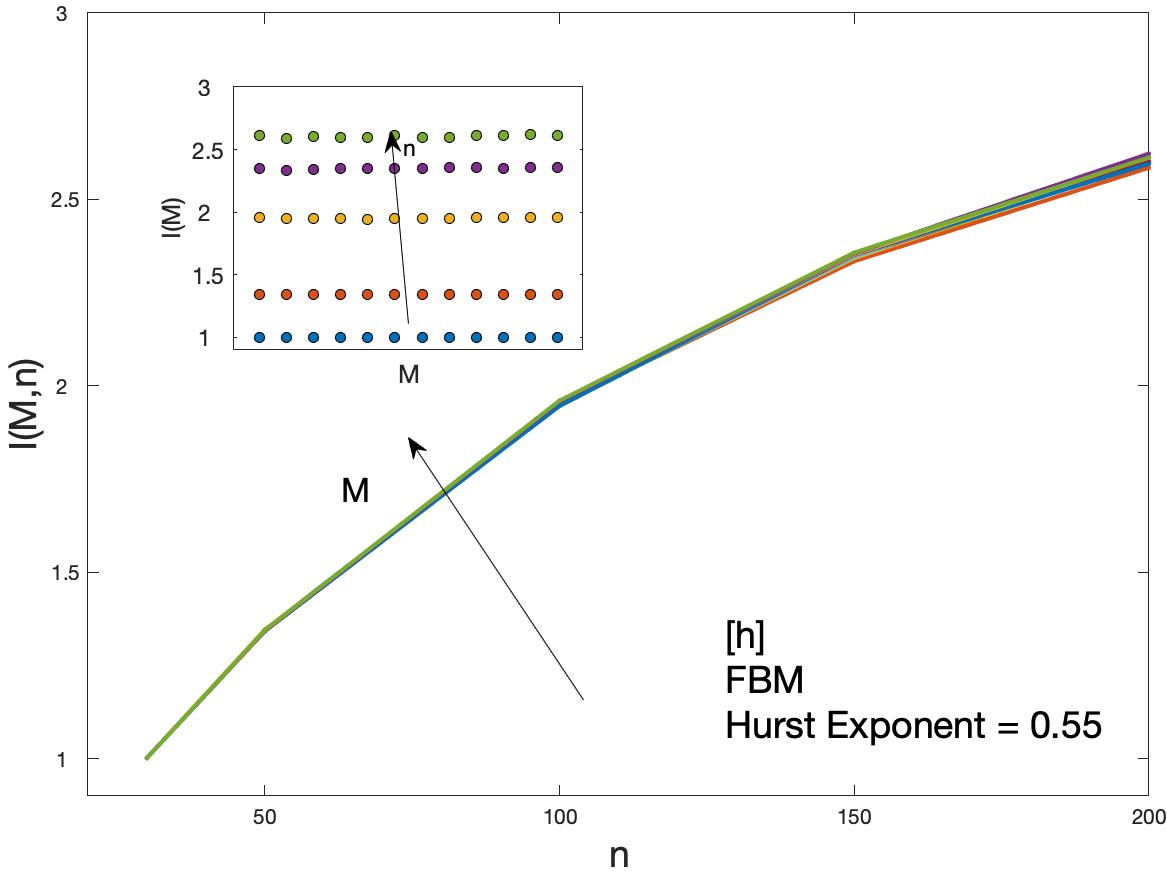}
    \includegraphics[width=4cm]{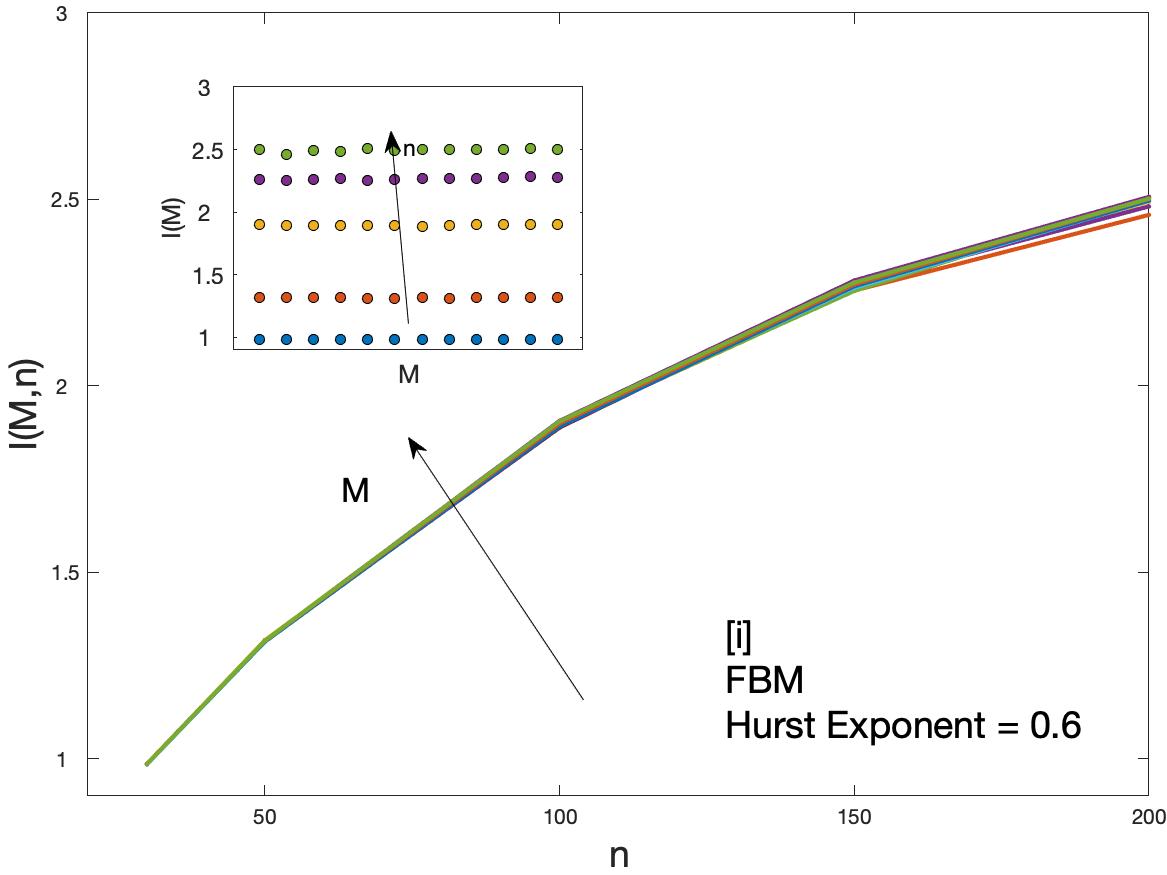}
    \\
    \includegraphics[width=4cm]{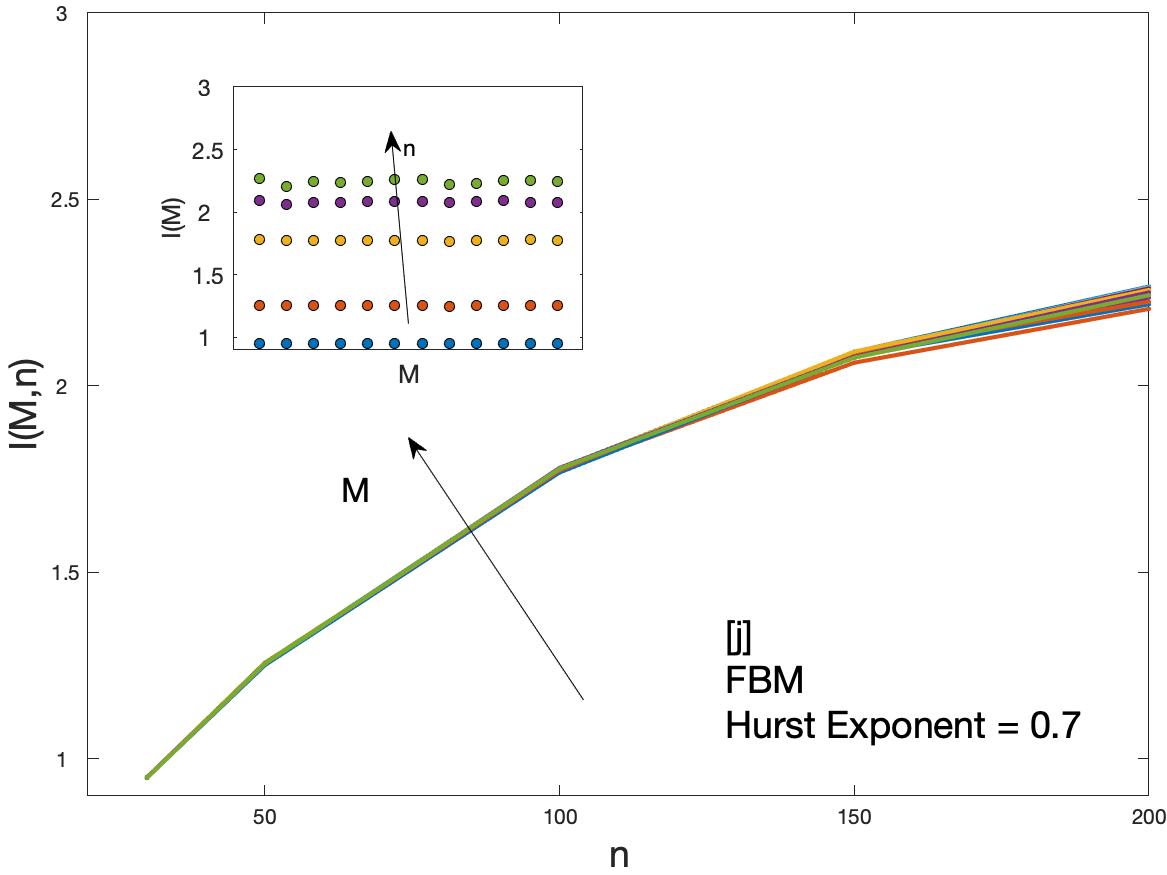}
    \includegraphics[width=4cm]{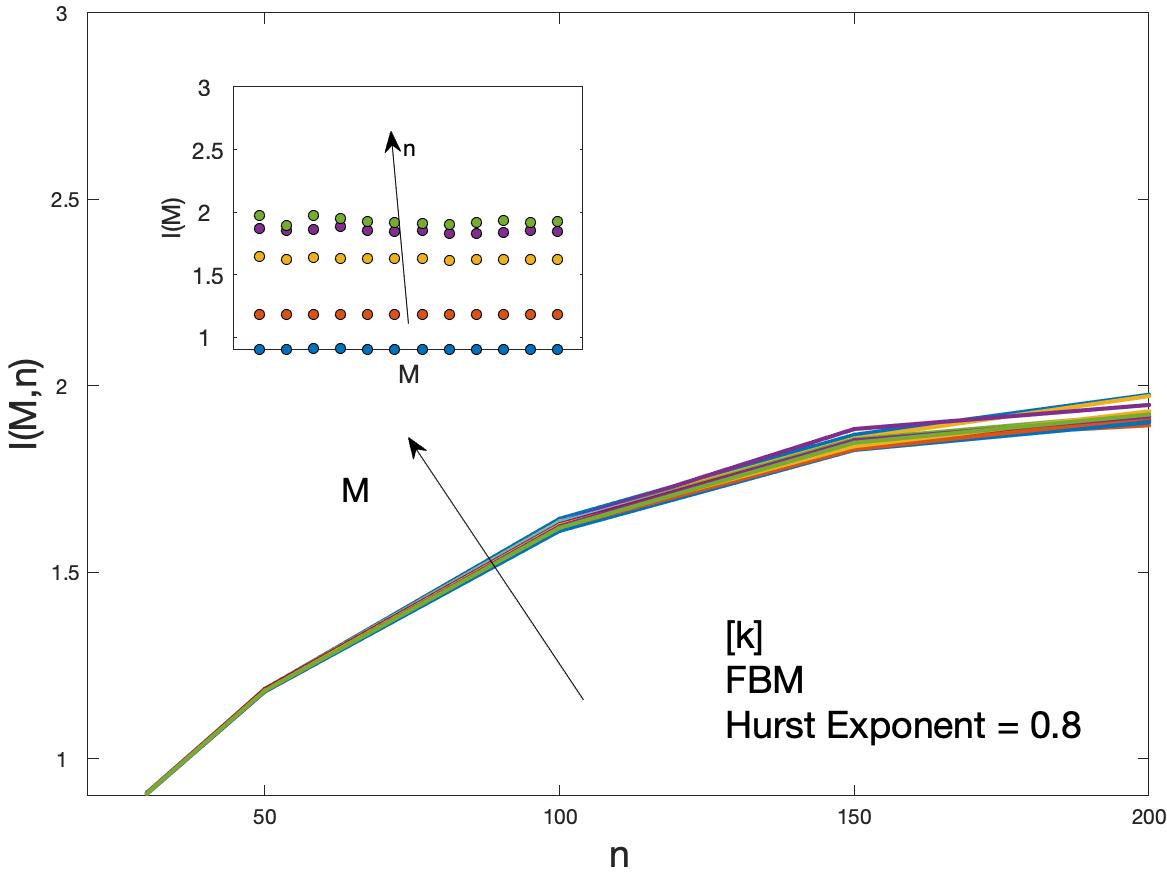}
    \includegraphics[width=4cm]{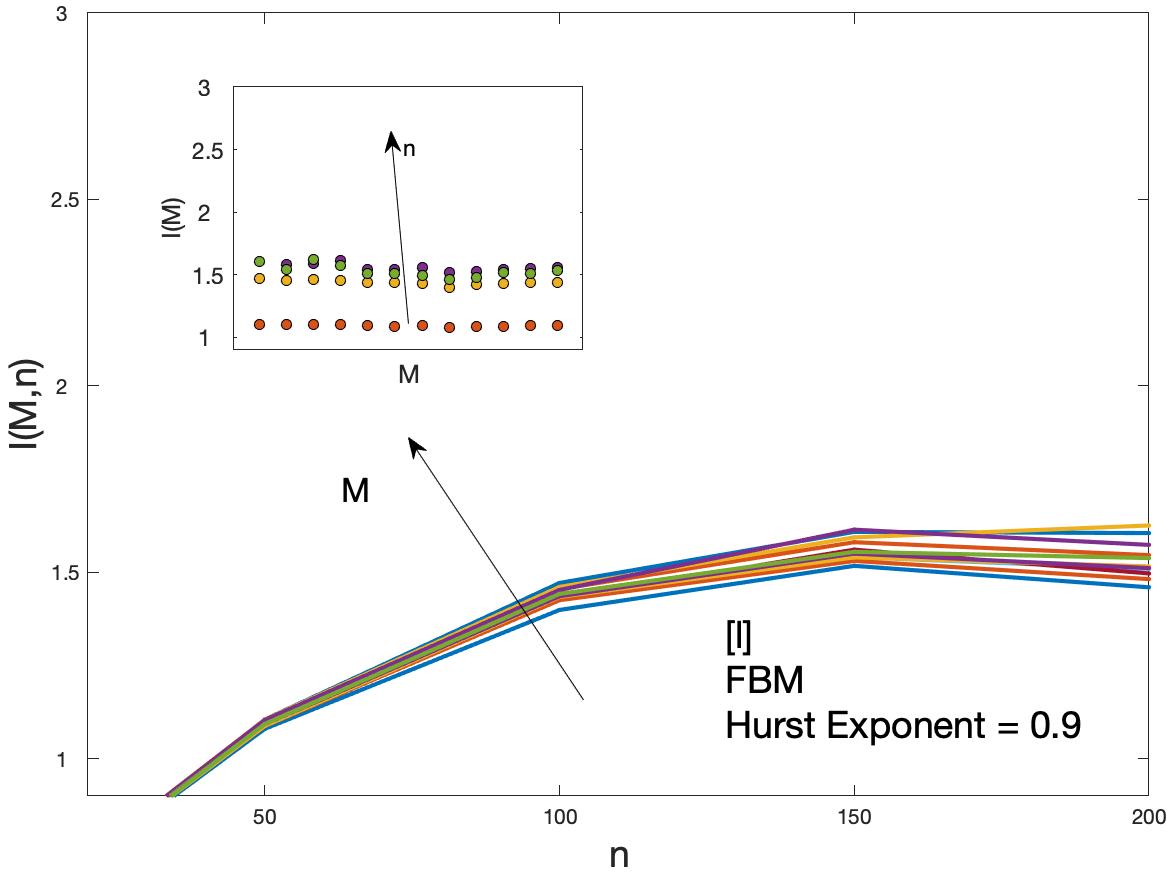}
    \caption{Market Dynamic Index $I(M,n)$ for Fractional Brownian Motion series with Hurst exponent  ranging from $H=0.2$ to $H=0.9$ respectively from $(a)$ to $(l)$.}
    \label{fig:FBM_MDI}
\end{figure}

The dependence of the cluster entropy on the Hurst exponent $H$ and the temporal horizon $M$ is reflected in the results of the Market Dynamic Index $I(M,n)$ plotted in Figure \ref{fig:FBM_MDI}.  The Market Dynamic Index $I(M,n)$ is estimated over several FBM series with different Hurst exponent. For anticorrelated series $0\leq H \leq0.5$ $I(M,n)$ curves overlap for all the moving average windows $n$ and time horizons $M$. For positively correlated series $0.5 \leq H \leq 0.9$, $I(M,n)$ exhibits slightly different values as a function of time horizons $M$. One can also note that the magnitude of the marginal increments in $I(M,n)$ at large $n$ increases as  $H$ increases for  $0\leq H \leq0.5$ , reaches a maximum for $H=0.5$  and then decreases  again for $0.5 \leq H \leq 0.9$. This effect is evident in the insets of  Figure \ref{fig:FBM_MDI}.

\par
The cluster entropy analysis  is implemented on \textit{Autoregressive Fractionally Integrated Moving Average} (ARFIMA) series obtained by means of simulations  for several combination of parameters \cite{Fatichi2020ARFIMA}. The extent of investigated parameters are marked  by alphabet labels and are reported in Table \ref{tab:arfima_models} for ARFIMA (1,d,1) and in Table  \ref{tab:arfima_models5} for ARFIMA (3,d,2) and ARFIMA(1,d,3).
\par
\begin{figure}
    \centering
    \includegraphics[width=4cm]{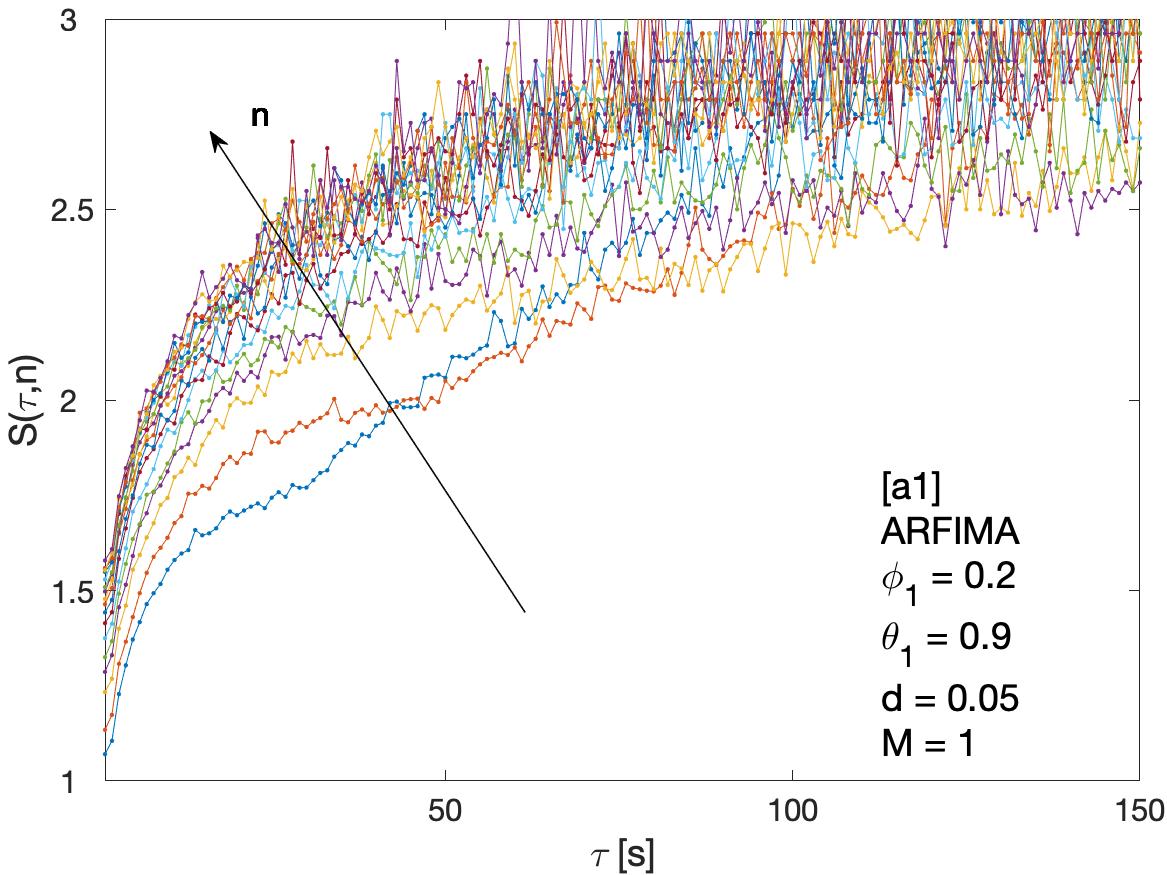}
    \includegraphics[width=4cm]{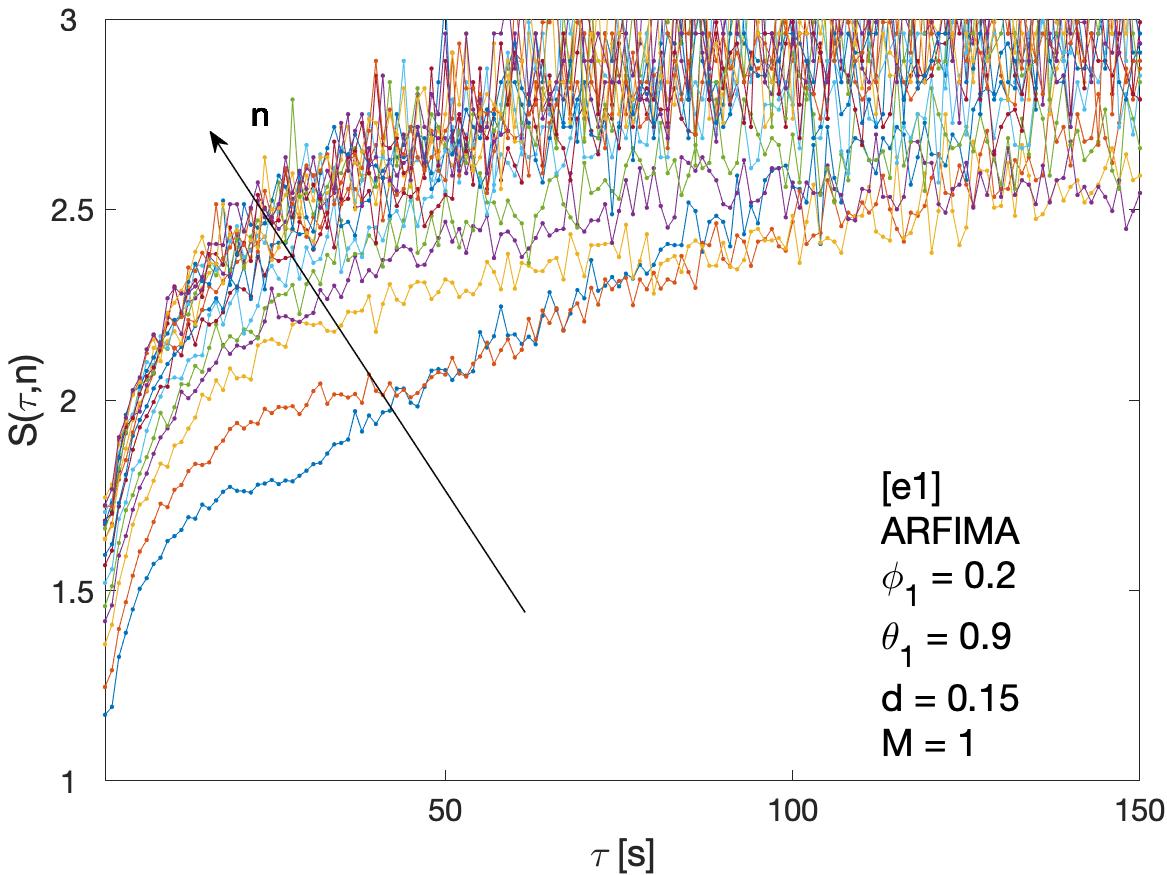}
    \includegraphics[width=4cm]{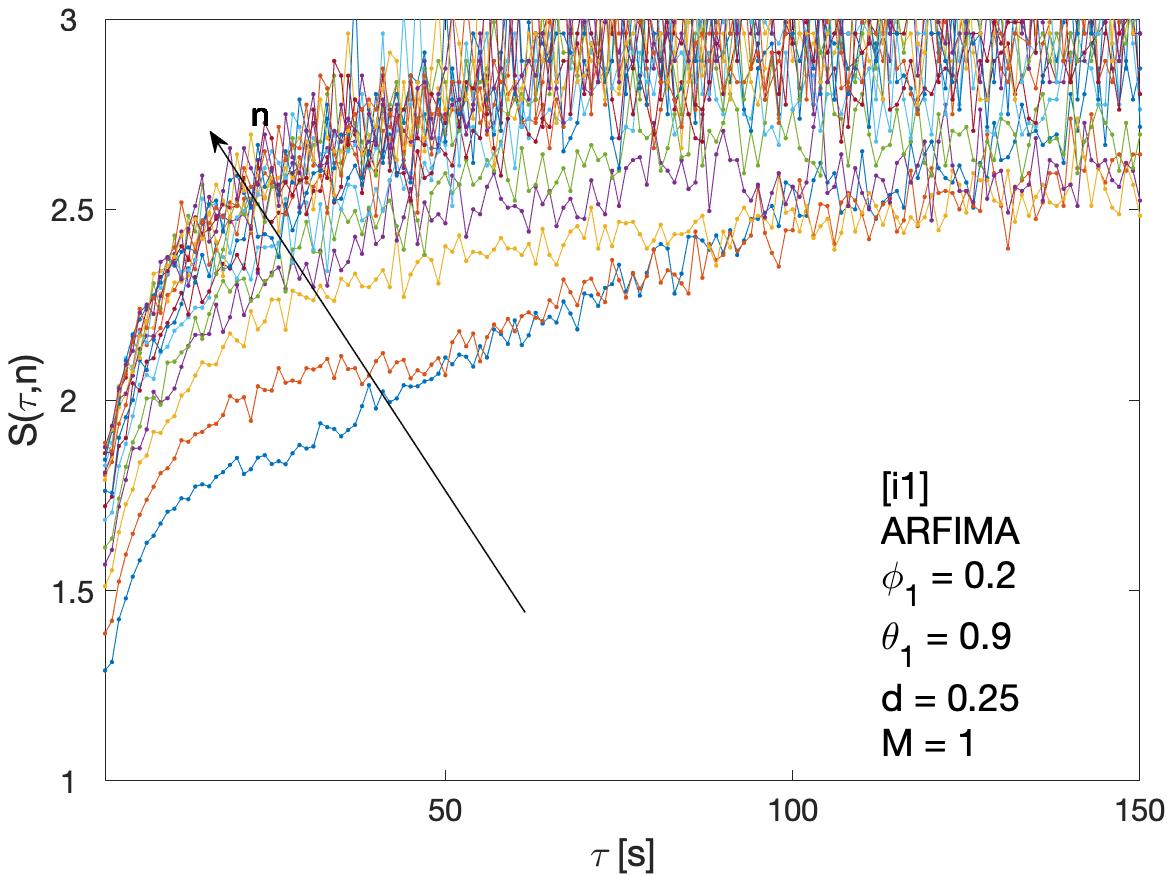}
    \\
    \includegraphics[width=4cm]{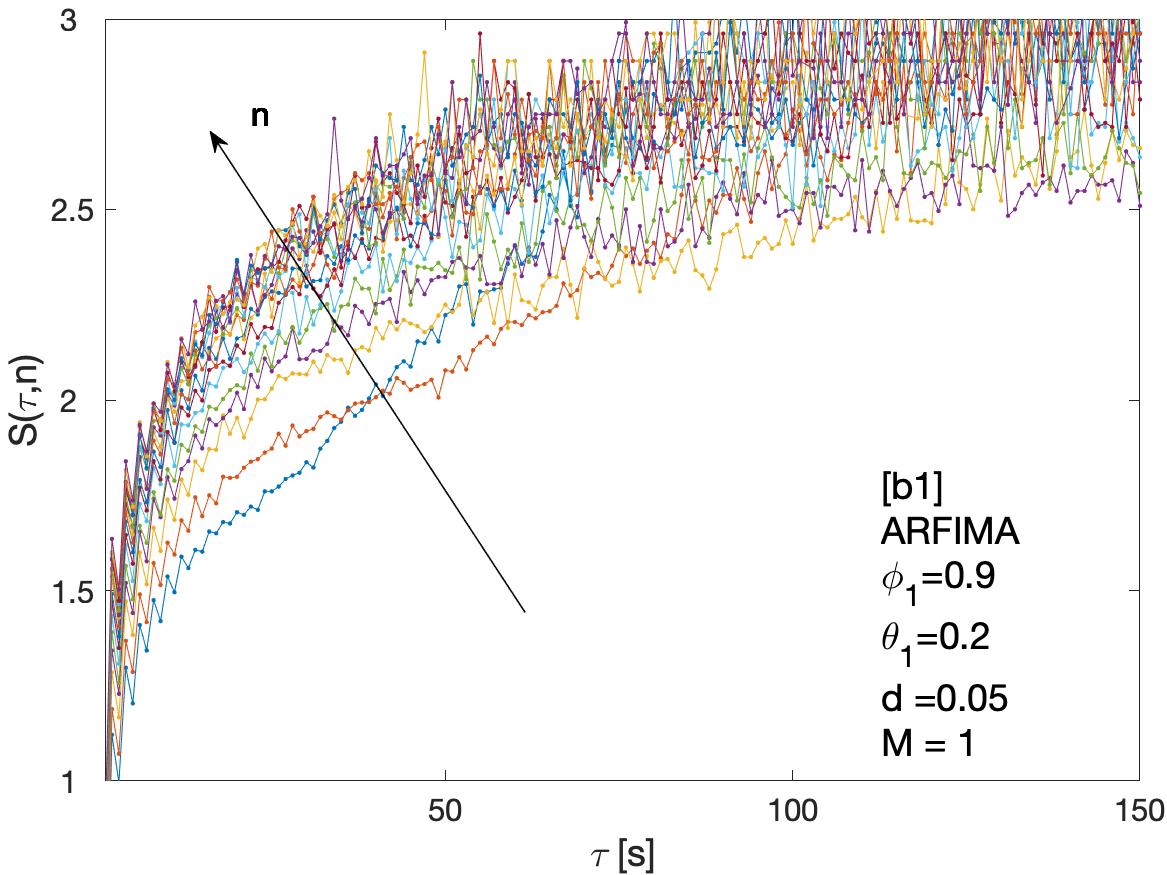}
    \includegraphics[width=4cm]{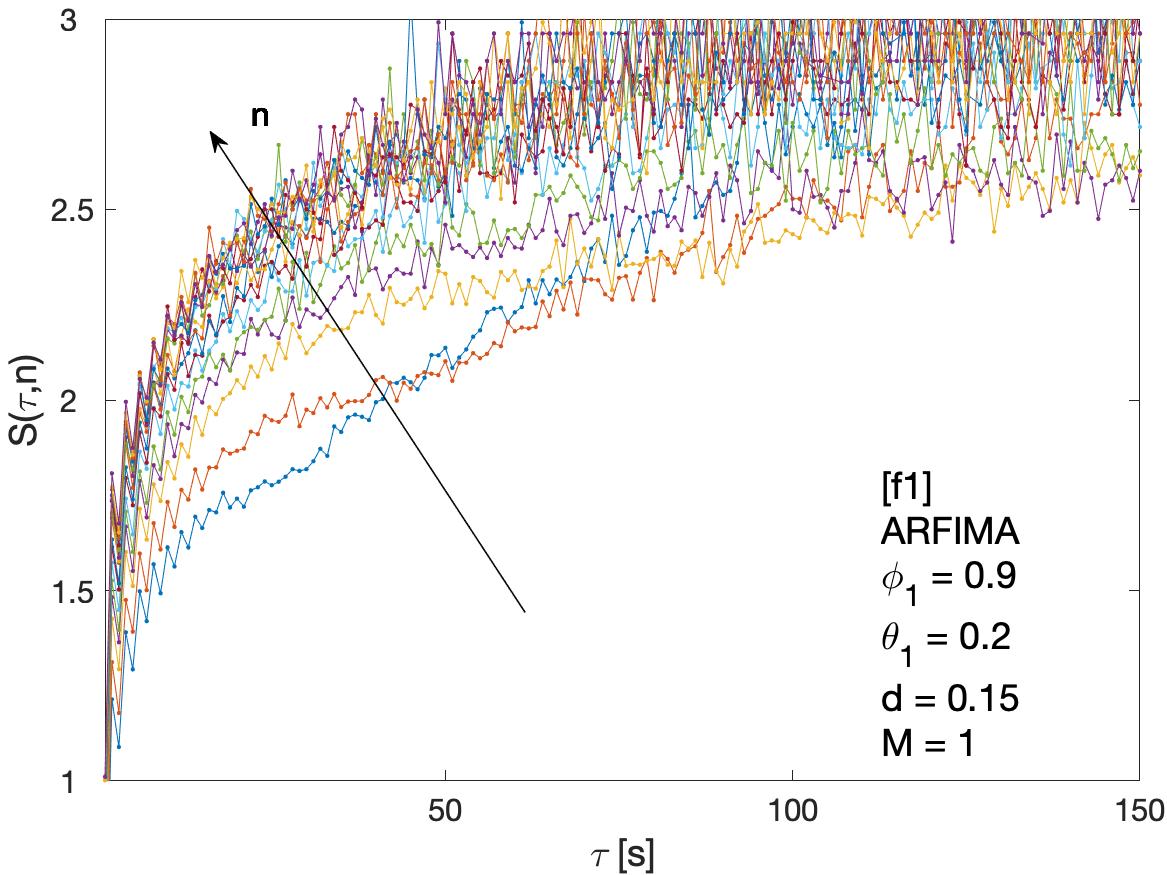}
    \includegraphics[width=4cm]{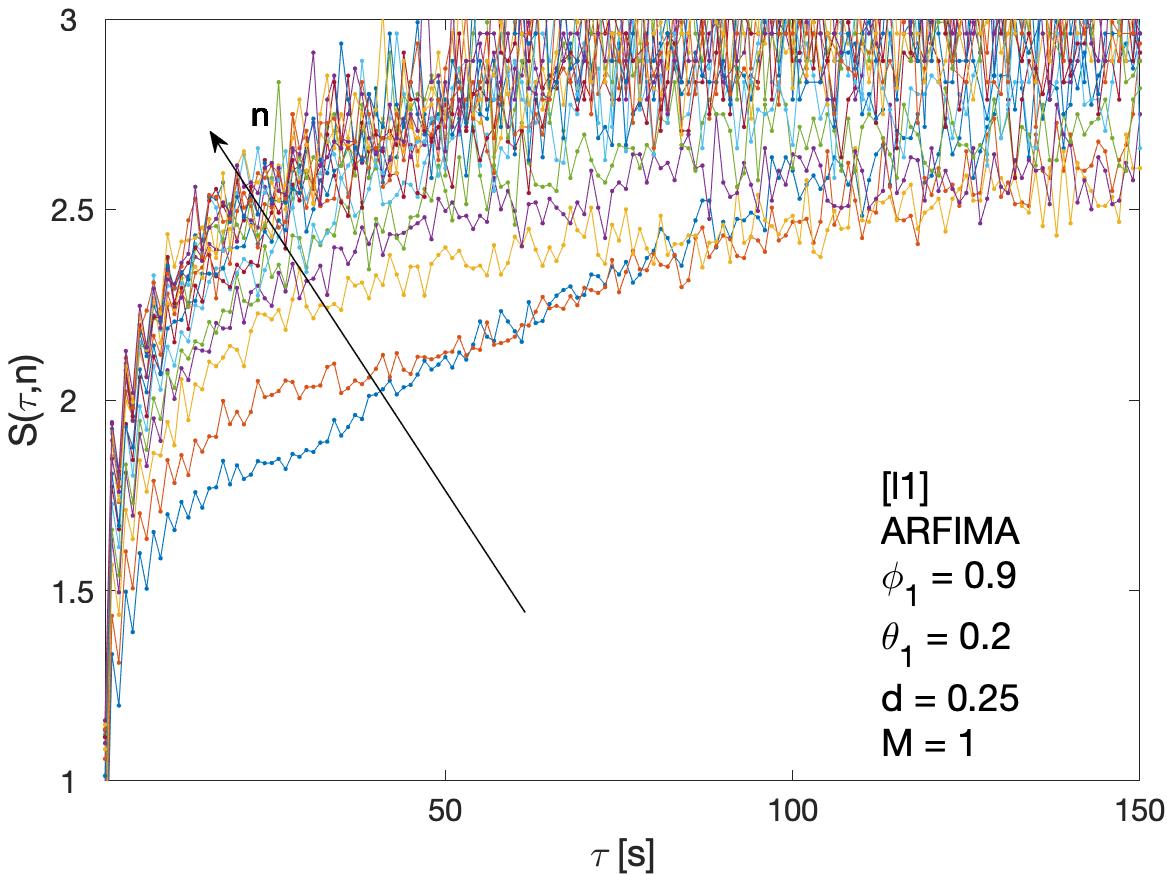}
    \caption{Cluster entropy results for horizon $M=1$ for ARFIMA series with different combinations of the differencing parameter $d$, autoregressive parameter $\phi$ and moving average parameter $\theta$. The differencing parameter takes values $d=0.05$, $d=0.15$, $d=0.25$ with a  different combinations of autoregressive and moving average parameter. The full set of analysed values of $d$, $\phi$ and $\theta$  is reported in Table \ref{tab:arfima_models}.}
    \label{fig:ARFIMA_CE_smpl_M1}
\end{figure}
\begin{figure}
    \centering
    \includegraphics[width=4cm]{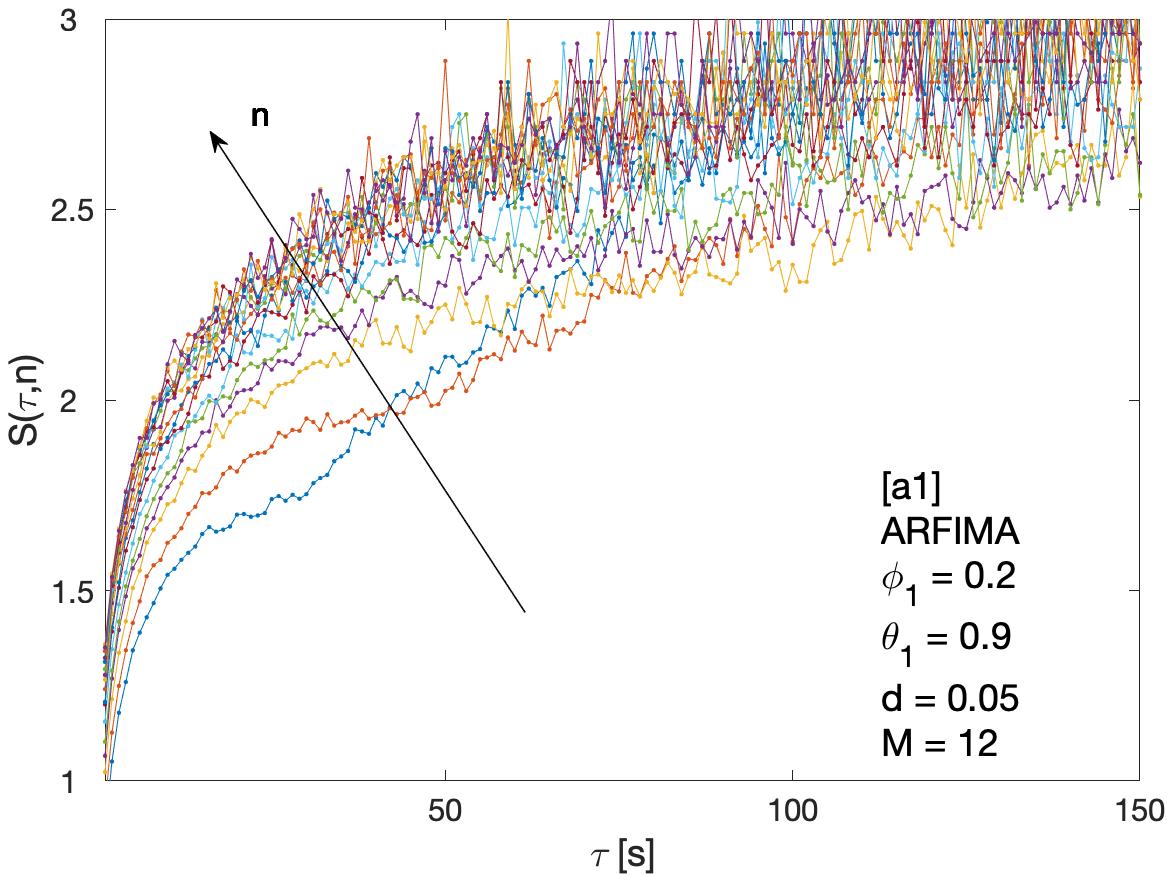}
    \includegraphics[width=4cm]{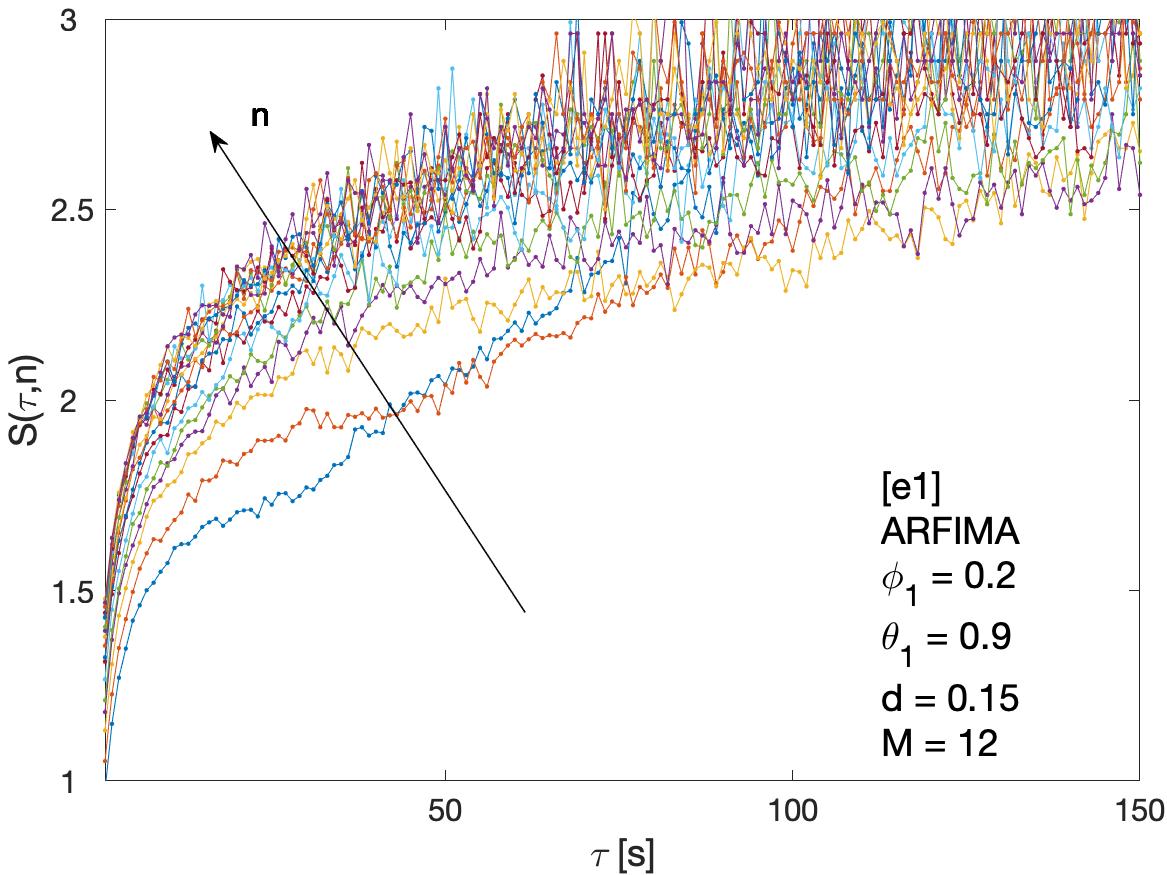}
    \includegraphics[width=4cm]{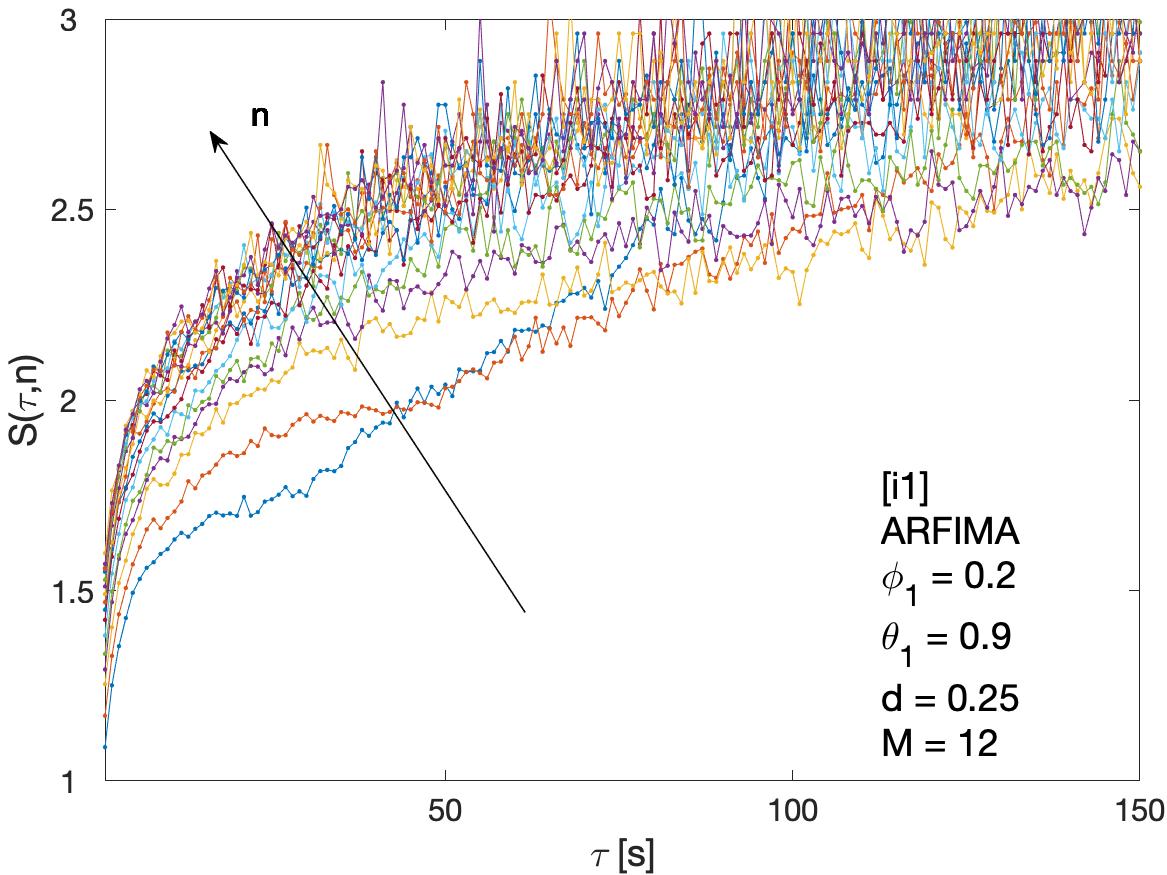}
    \\
    \includegraphics[width=4cm]{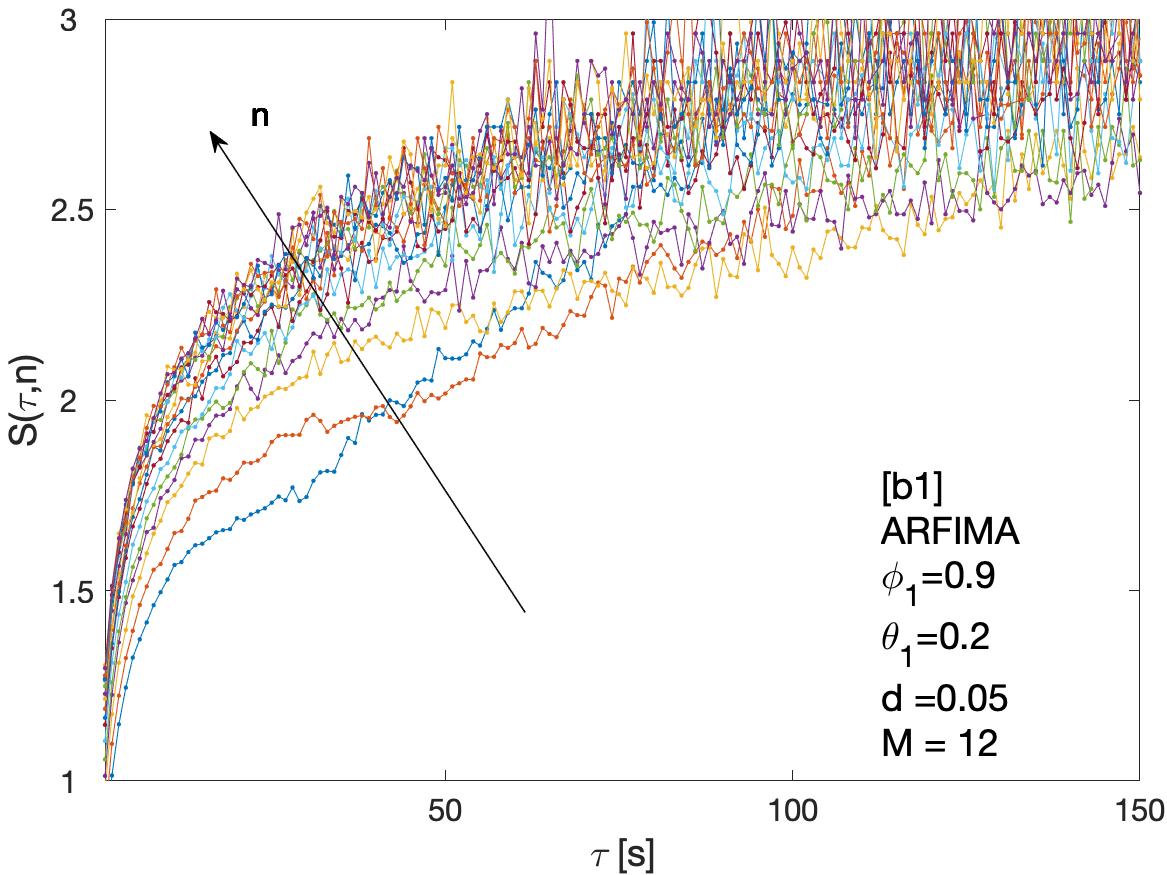}
    \includegraphics[width=4cm]{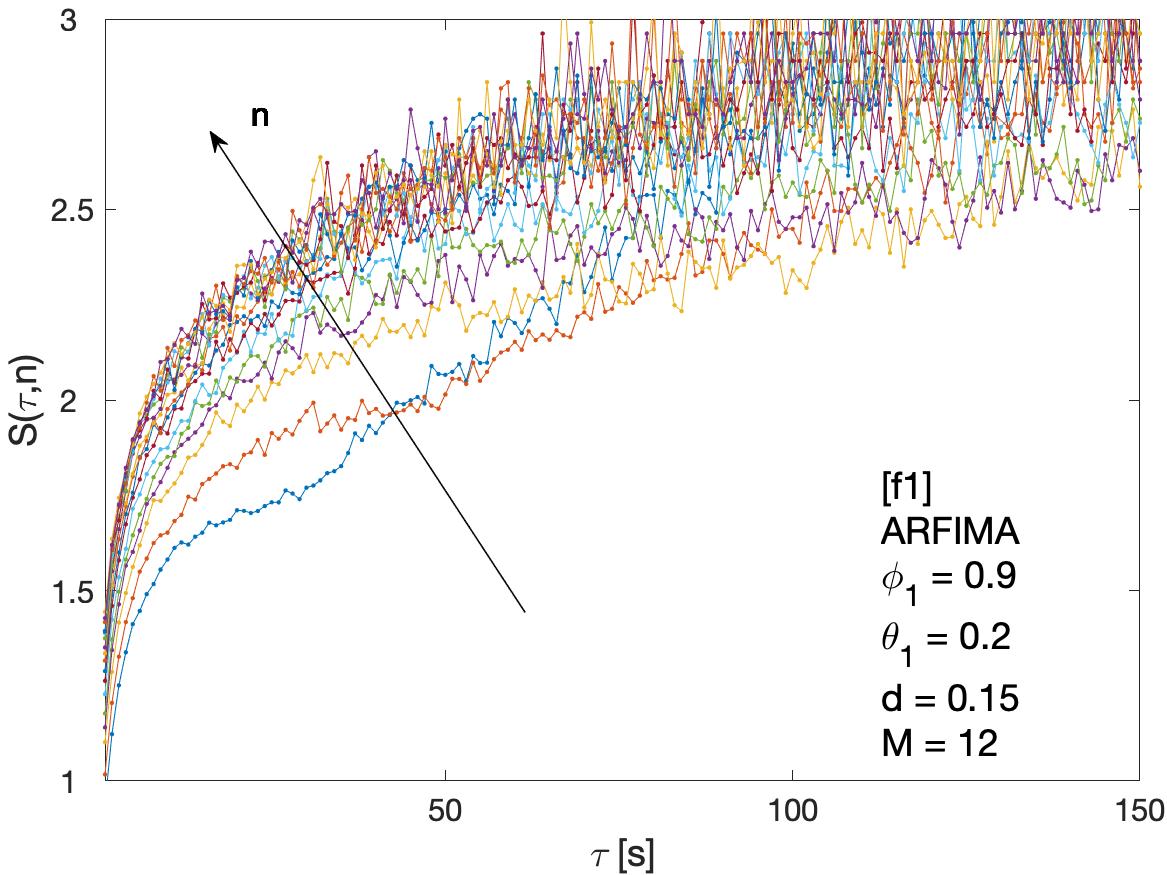}
    \includegraphics[width=4cm]{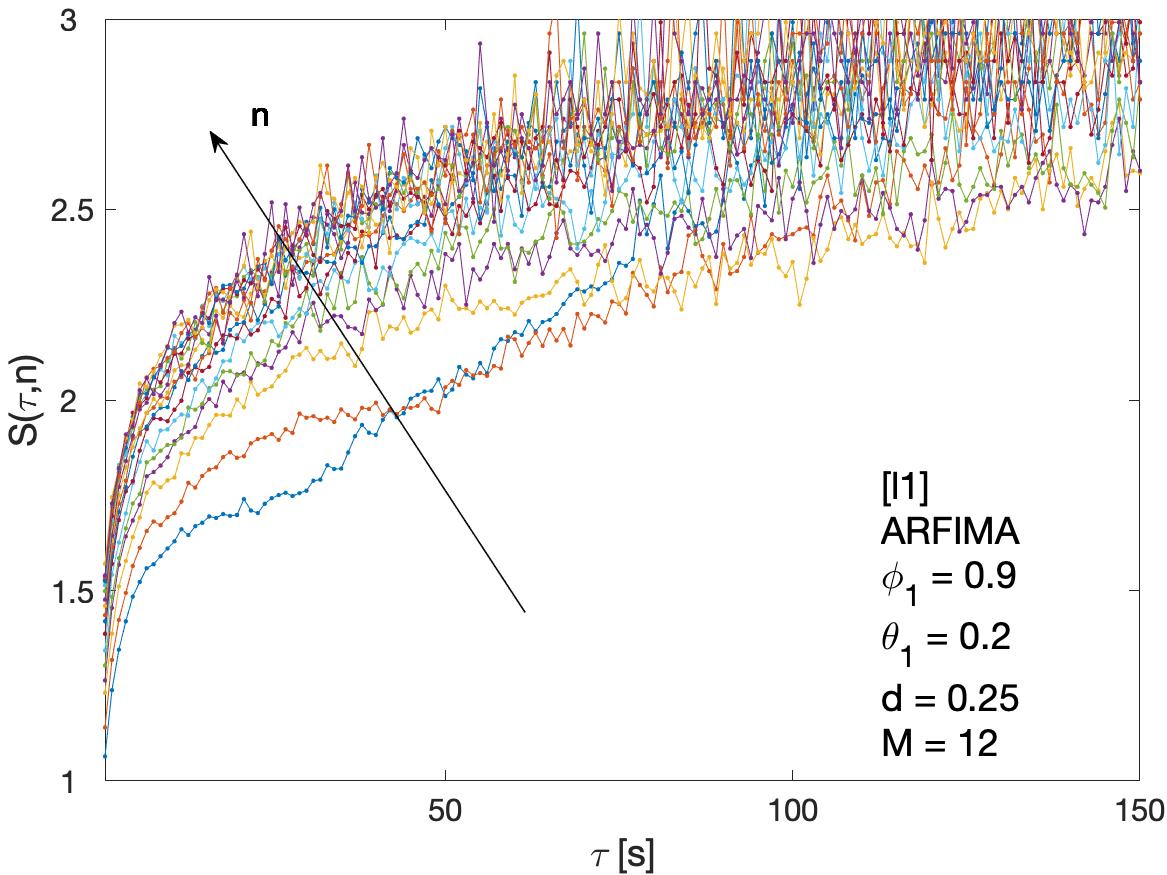}
    \caption{Cluster entropy results for horizon $M=12$ on ARFIMA series with different combinations of the differencing parameter $d$, autoregressive parameter $\phi$ and moving average parameter $\theta$. The differencing parameter takes values $d=0.05$, $d=0.15$ and $d=0.25$ with  a different combination of autoregressive and moving average parameters. The full set of analysed values of $d$, $\phi$ and $\theta$  is reported in Table \ref{tab:arfima_models}.}
    \label{fig:ARFIMA_CE_smpl_M12}
\end{figure}
\begin{figure}
    \centering
    \includegraphics[width=4cm]{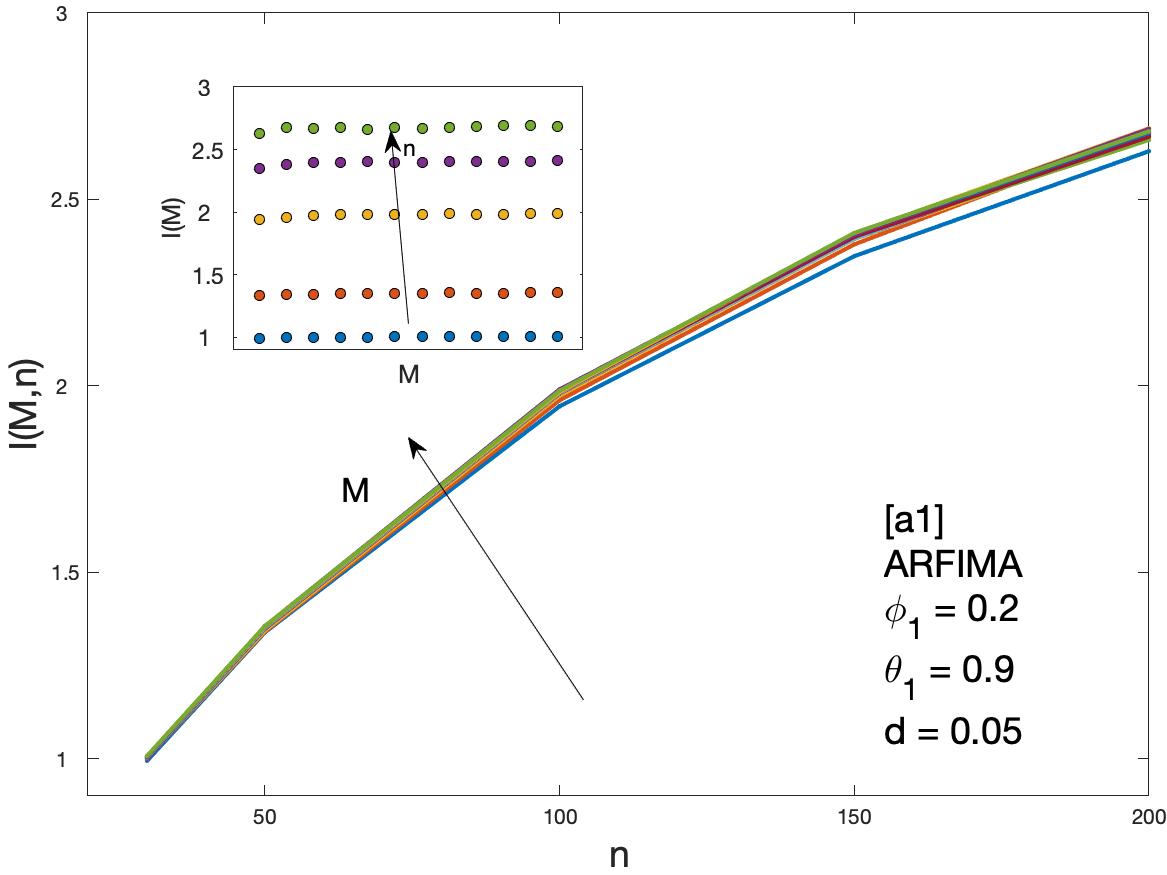}
    \includegraphics[width=4cm]{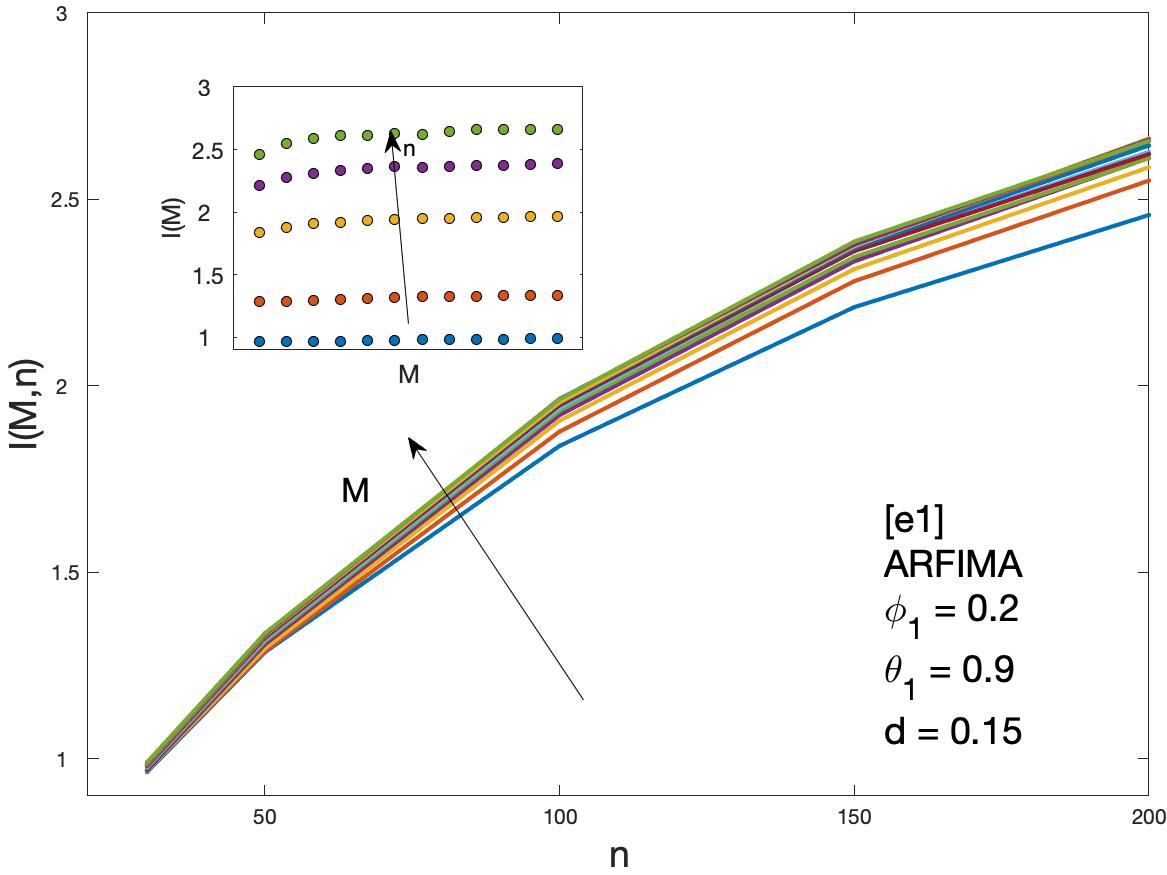}
    \includegraphics[width=4cm]{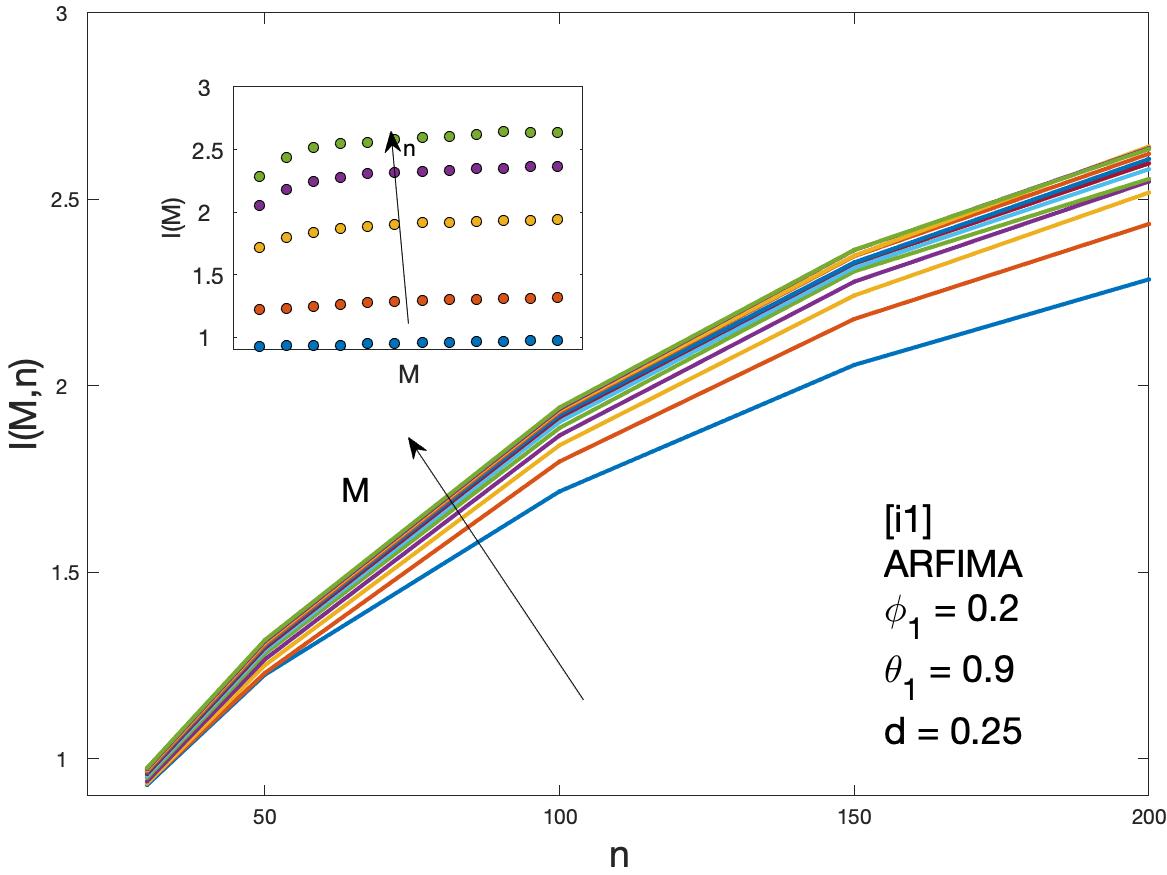}
    \\
    \includegraphics[width=4cm]{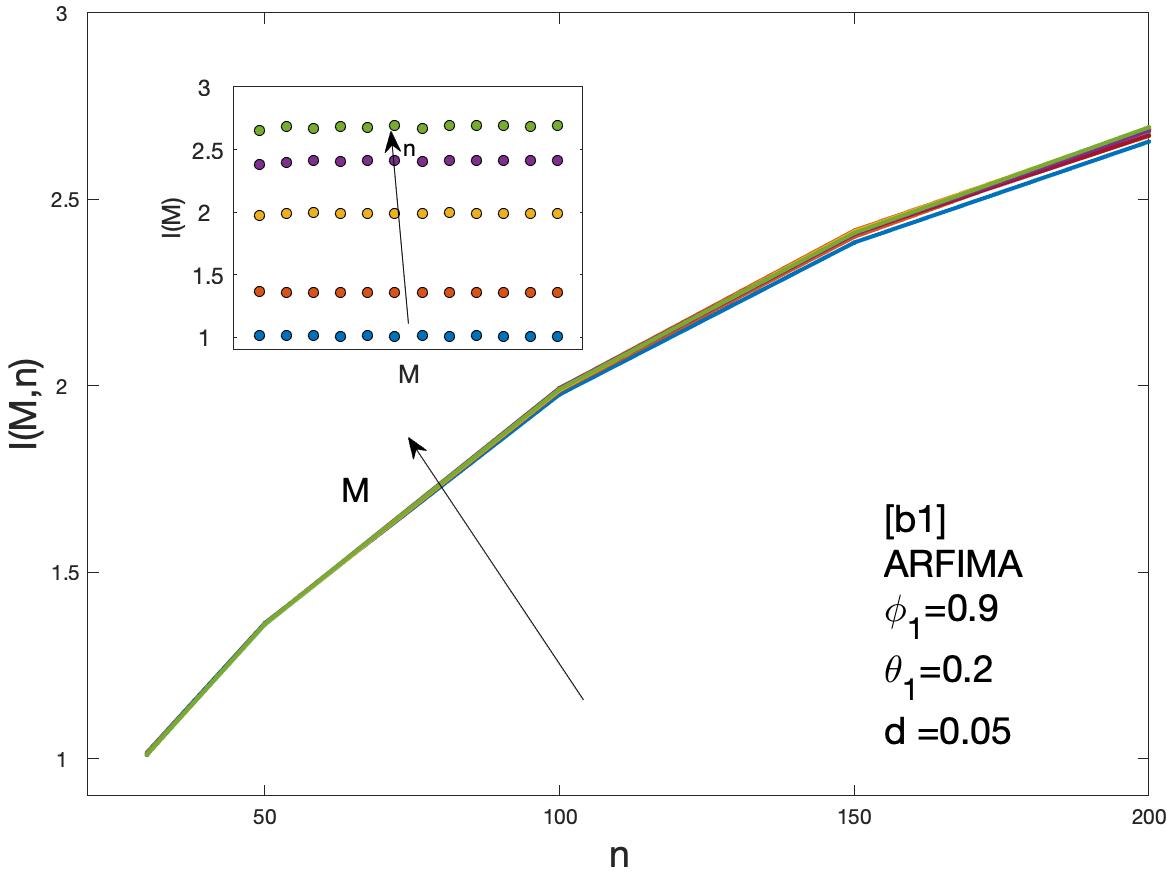}
    \includegraphics[width=4cm]{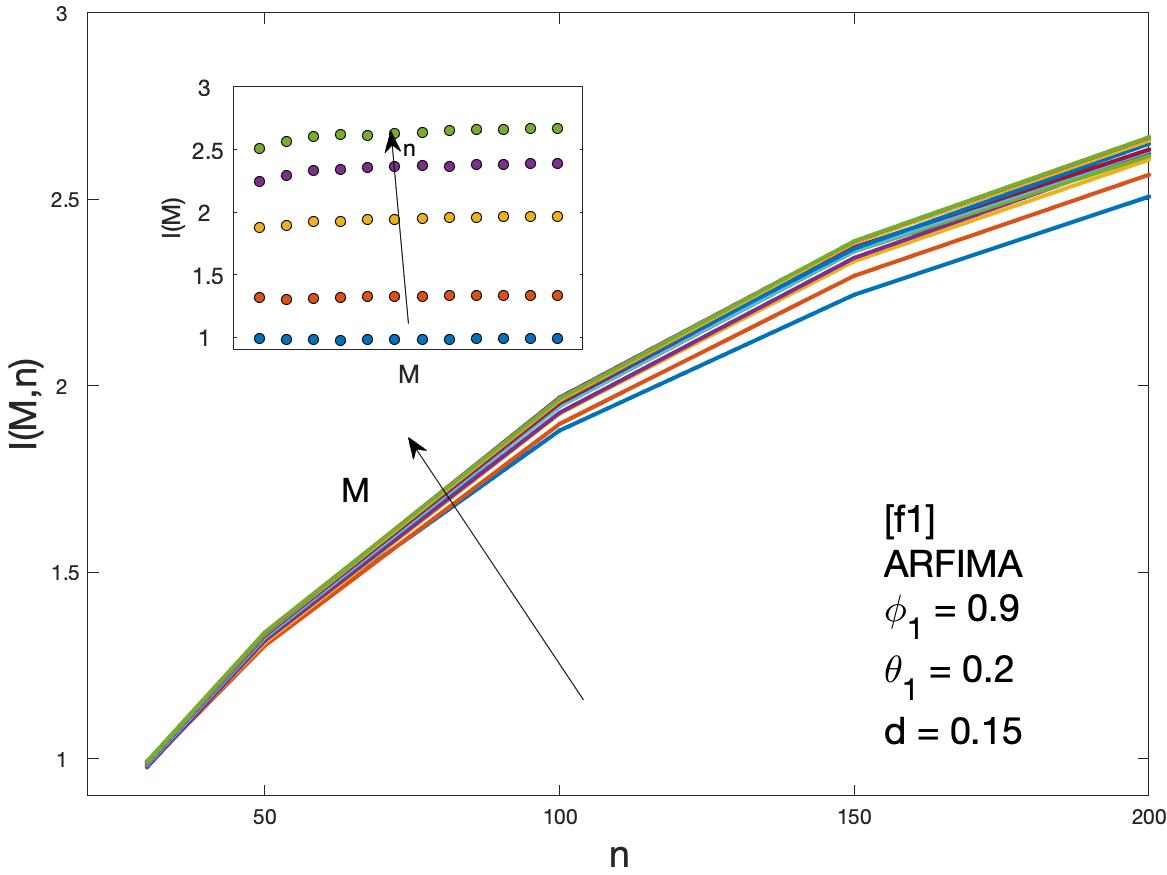}
    \includegraphics[width=4cm]{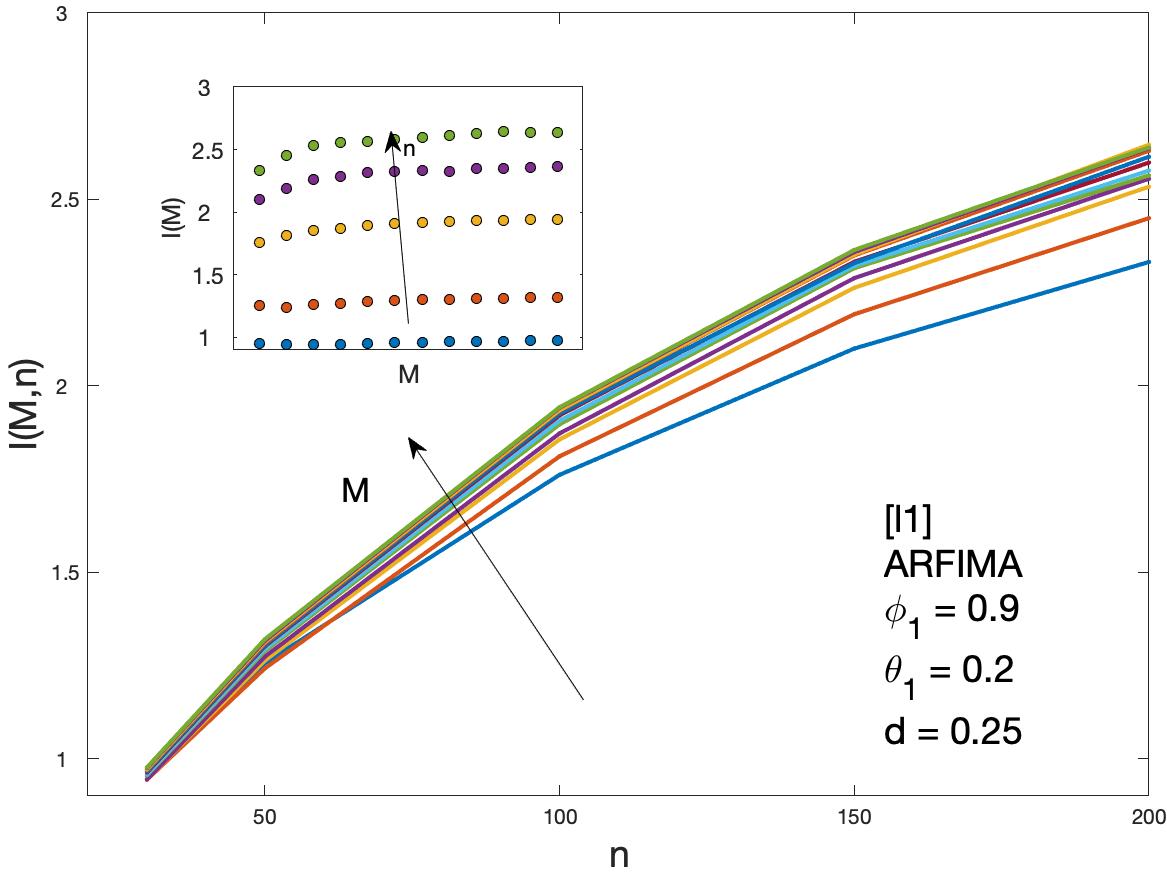}
    \caption{Market Dynamic Index $I(M,n)$ for ARFIMA series with different combinations of the differencing parameter $d$, autoregressive parameter $\phi$, and moving average parameter $\theta$. The differencing parameter  takes values $d=0.05$, $d=0.15$, $d=0.25$, with a different combination of autoregressive and moving average parameters. The full set of analysed values of $d$, $\phi$ and $\theta$  is reported in Table \ref{tab:arfima_models}}.
    \label{fig:ARFIMA_MDI}
\end{figure}
\begin{figure}
    \centering
    \includegraphics[width=4cm]{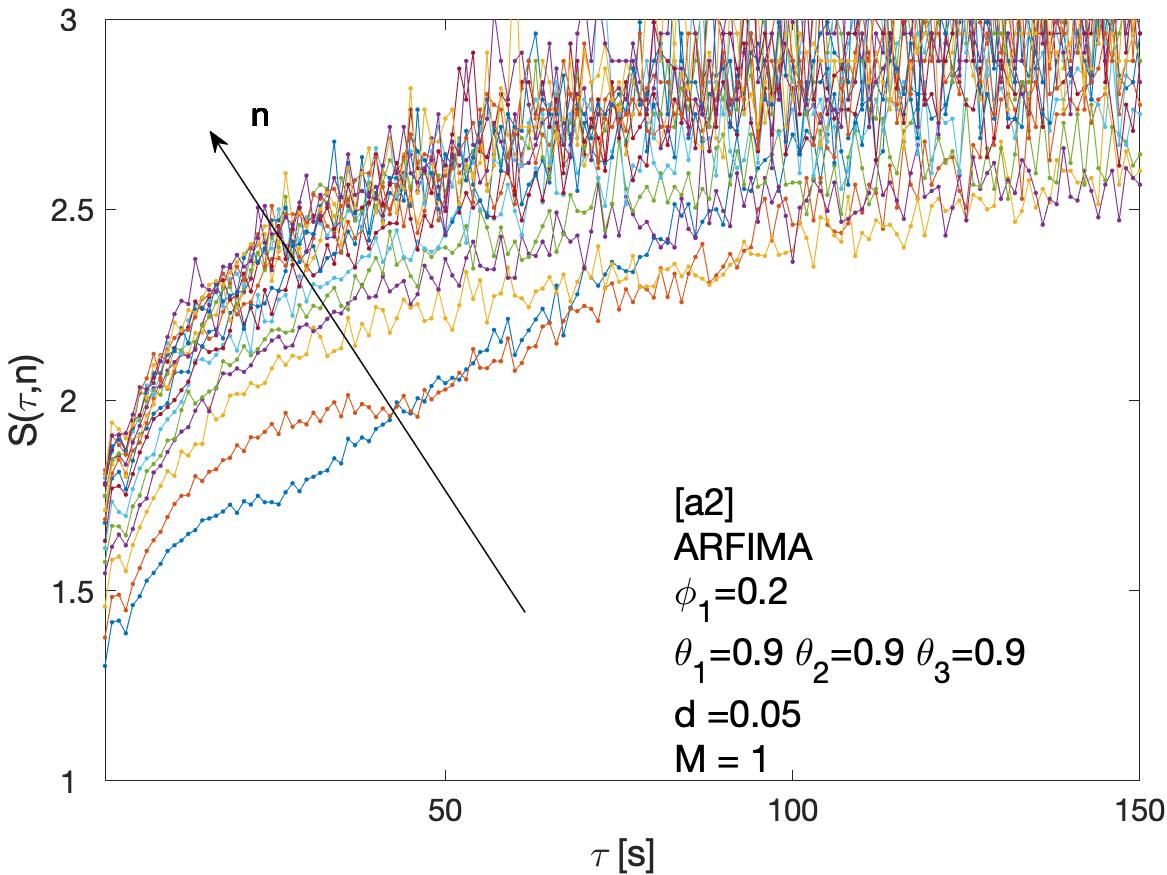}
    \includegraphics[width=4cm]{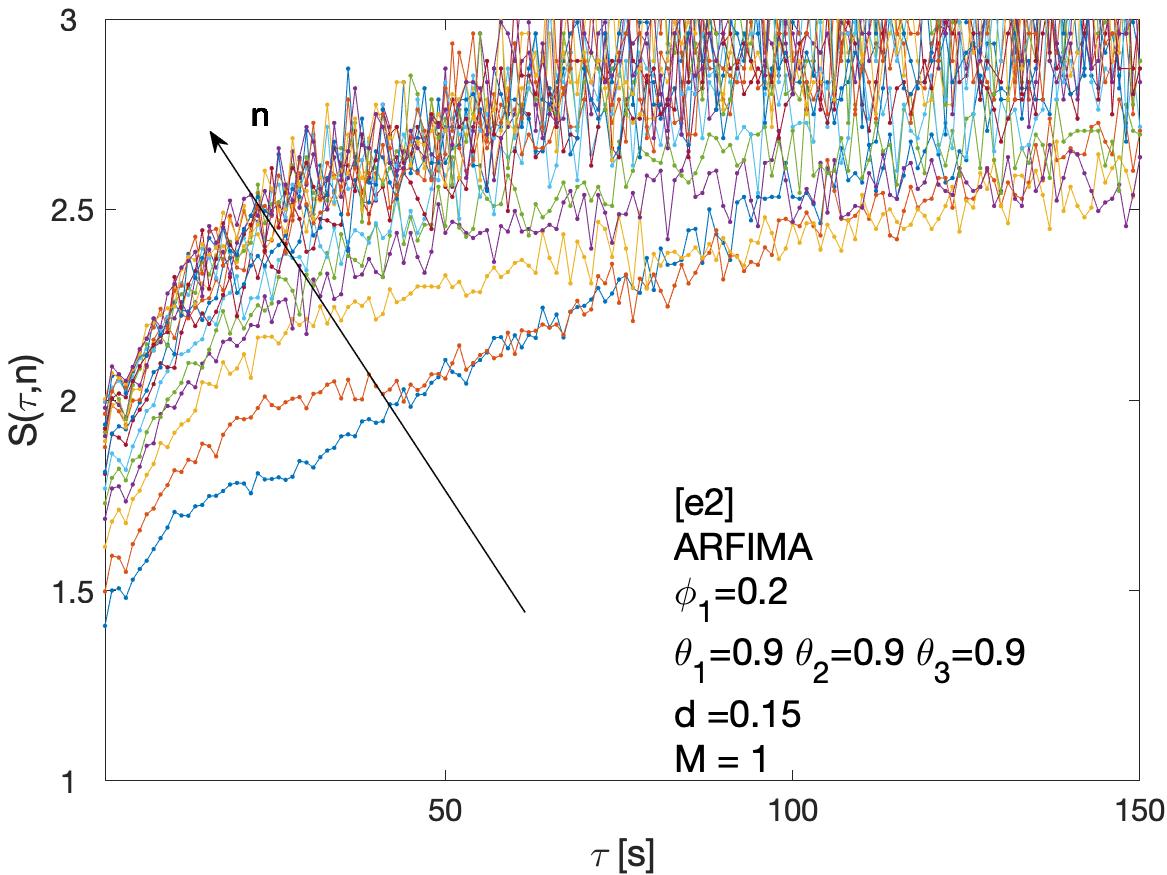}
    \includegraphics[width=4cm]{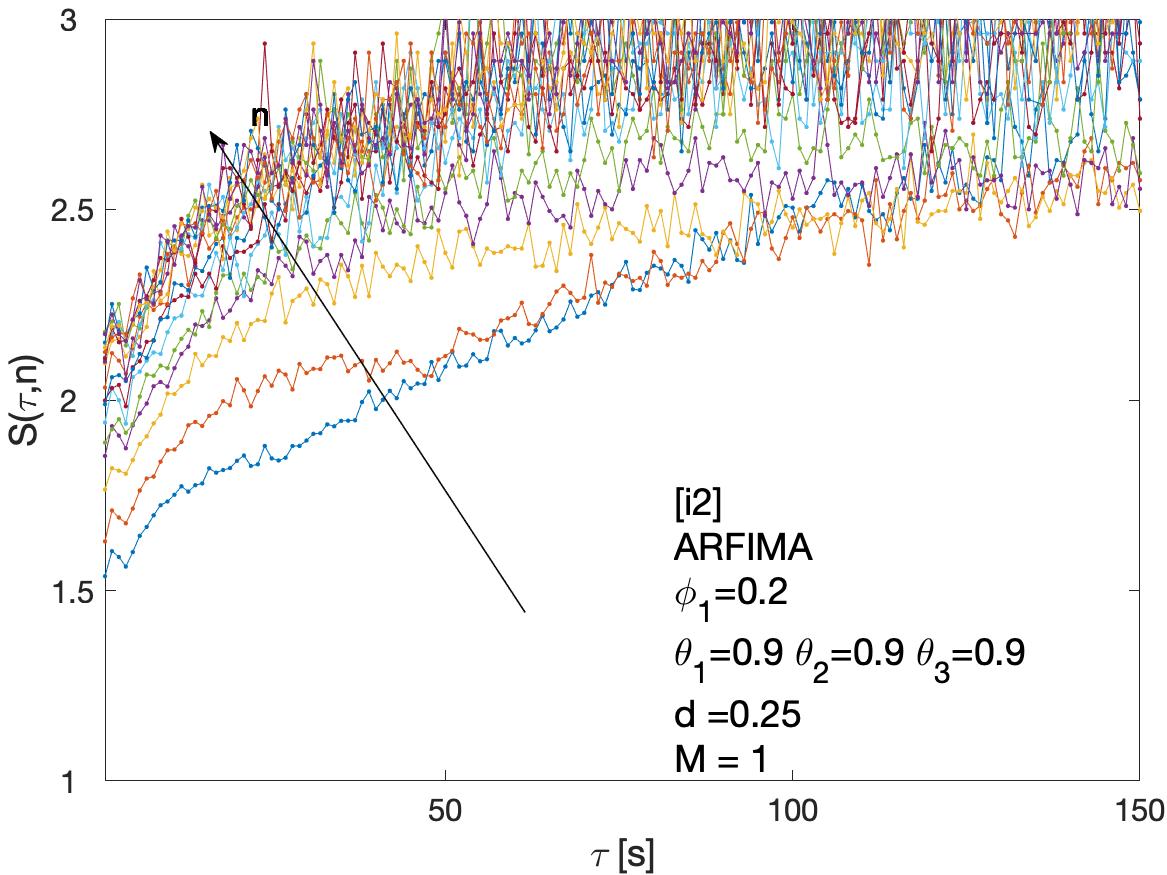}
    \\
    \includegraphics[width=4cm]{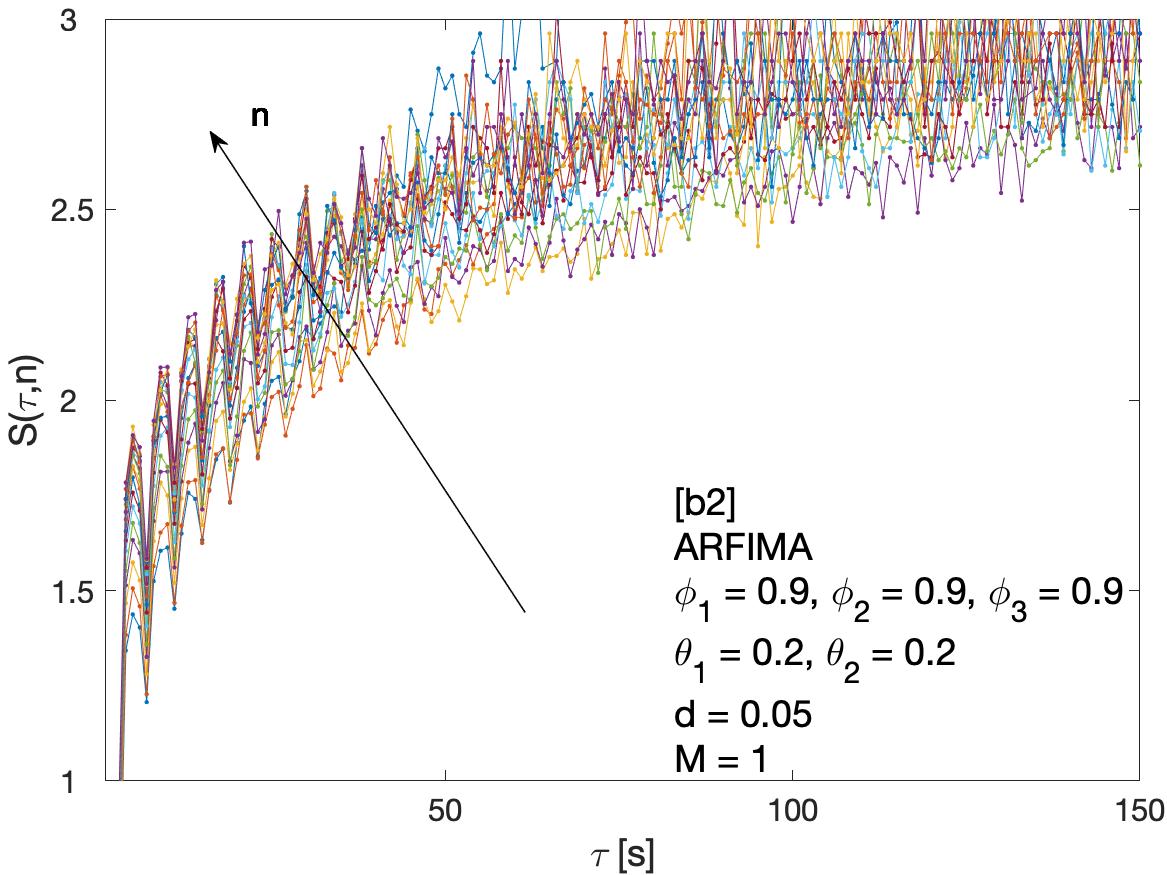}
    \includegraphics[width=4cm]{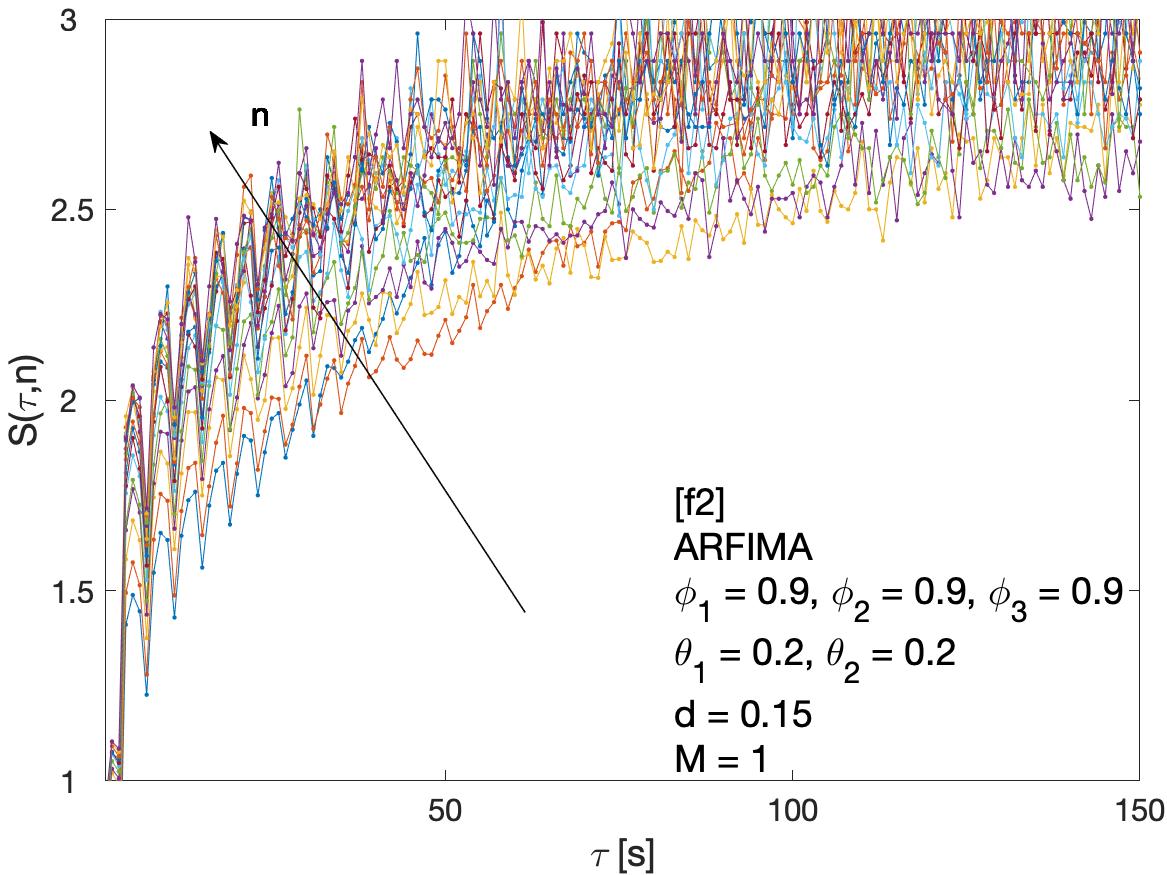}
    \includegraphics[width=4cm]{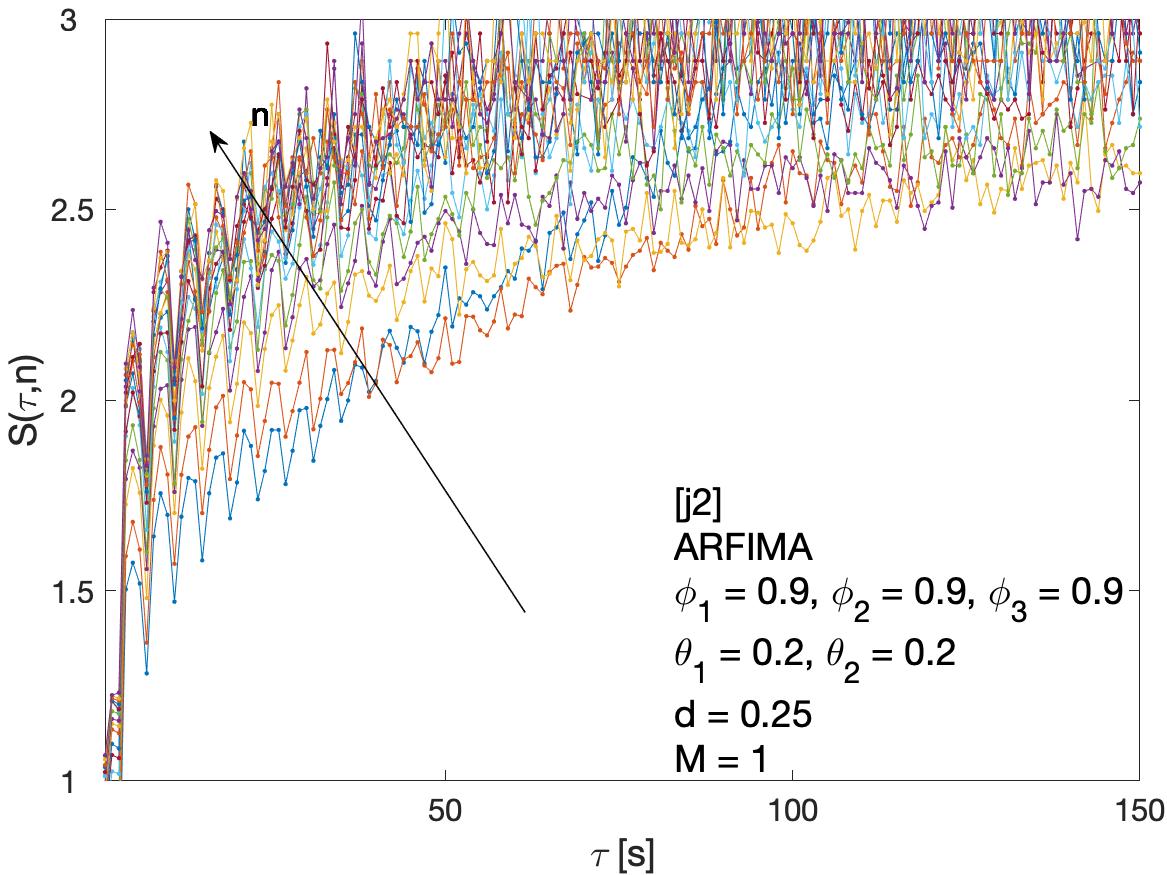}
    \caption{Cluster entropy results for horizon $M=1$ on ARFIMA series with different combinations of the differencing parameter $d$, autoregressive parameter $\phi_1$, $\phi_2$, and  $\phi_3$  and moving average parameter $\theta_1$, $\theta_2$ and $\theta_3$. The differencing parameter  takes values $d=0.05$, $d=0.15$, $d=0.25$, with a different combination of autoregressive and moving average parameters. The full set of analysed values of $d$, $\phi$ and $\theta$  is reported in Table \ref{tab:arfima_models5}.}
    \label{fig:ARFIMA_CE_cplx_M1}
\end{figure}
\begin{figure}
    \centering
    \includegraphics[width=4cm]{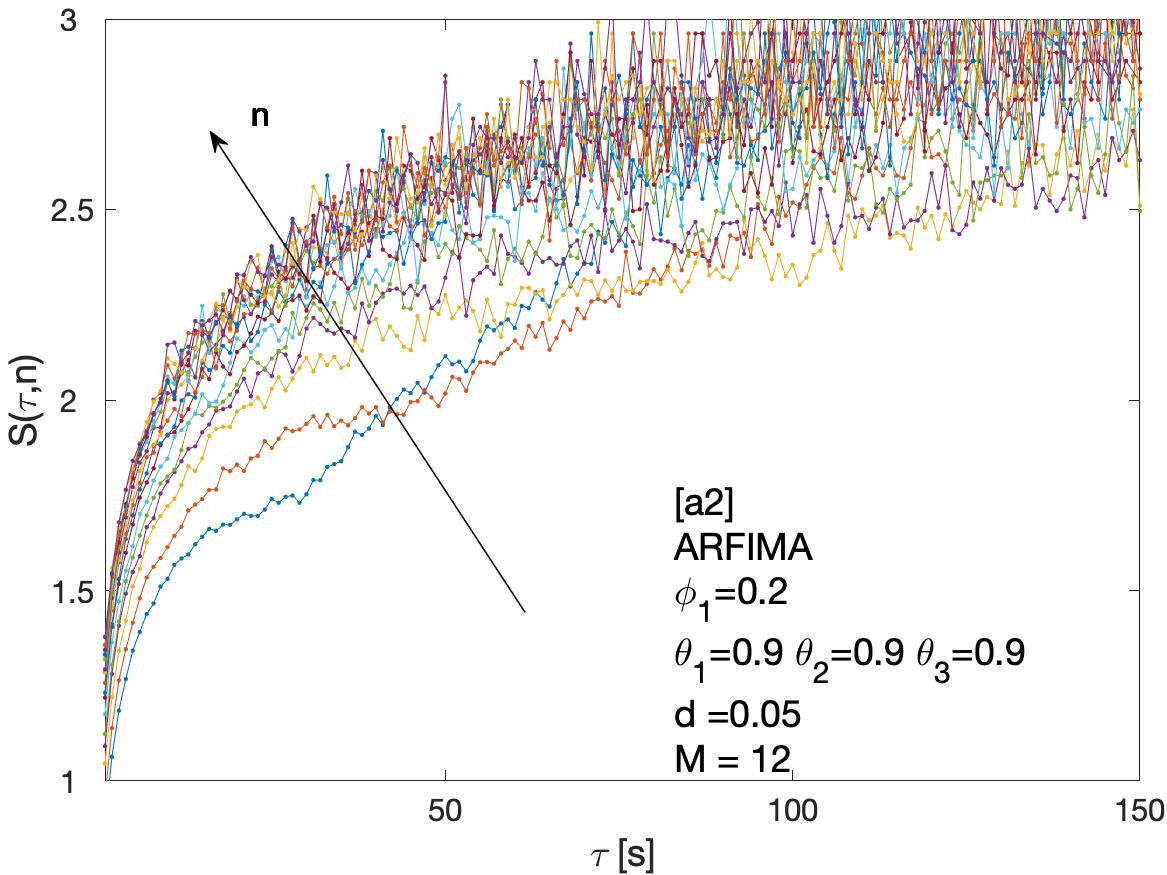}
    \includegraphics[width=4cm]{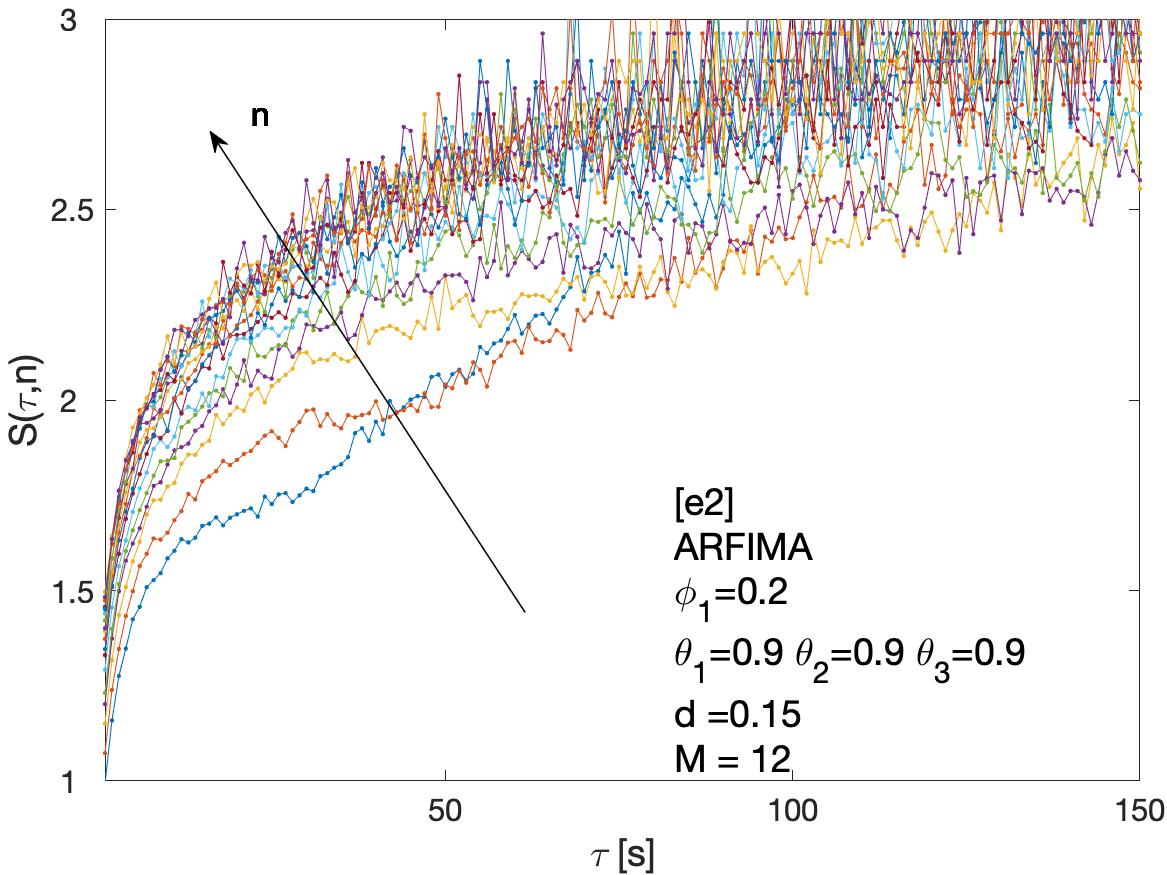}
    \includegraphics[width=4cm]{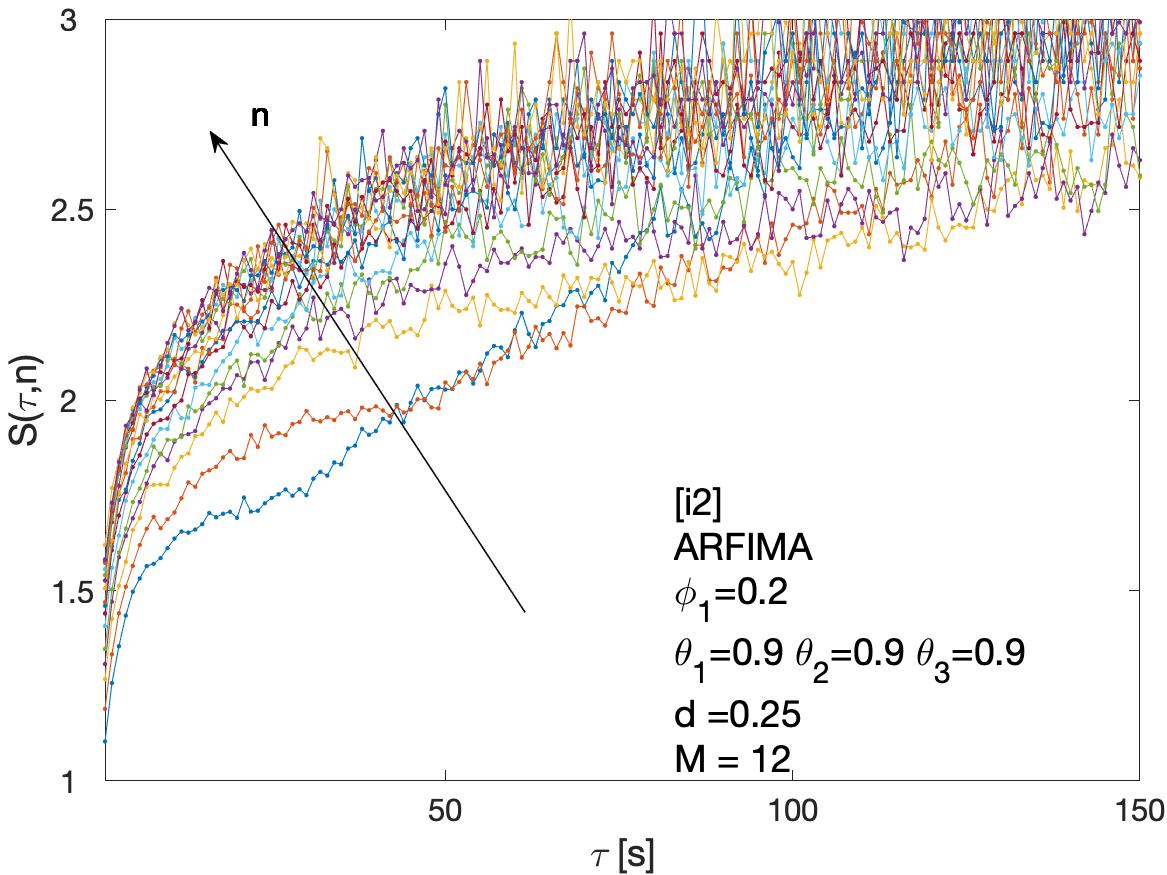}
    \\
    \includegraphics[width=4cm]{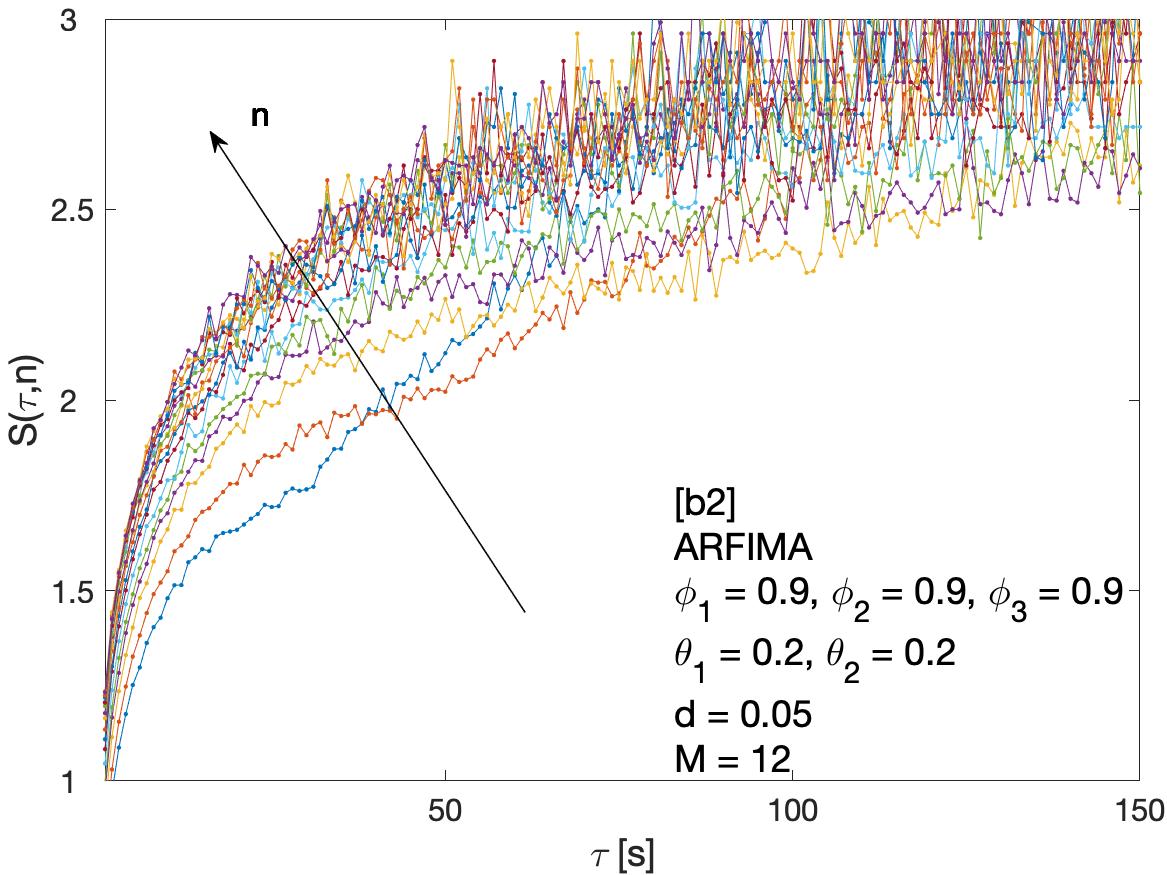}
    \includegraphics[width=4cm]{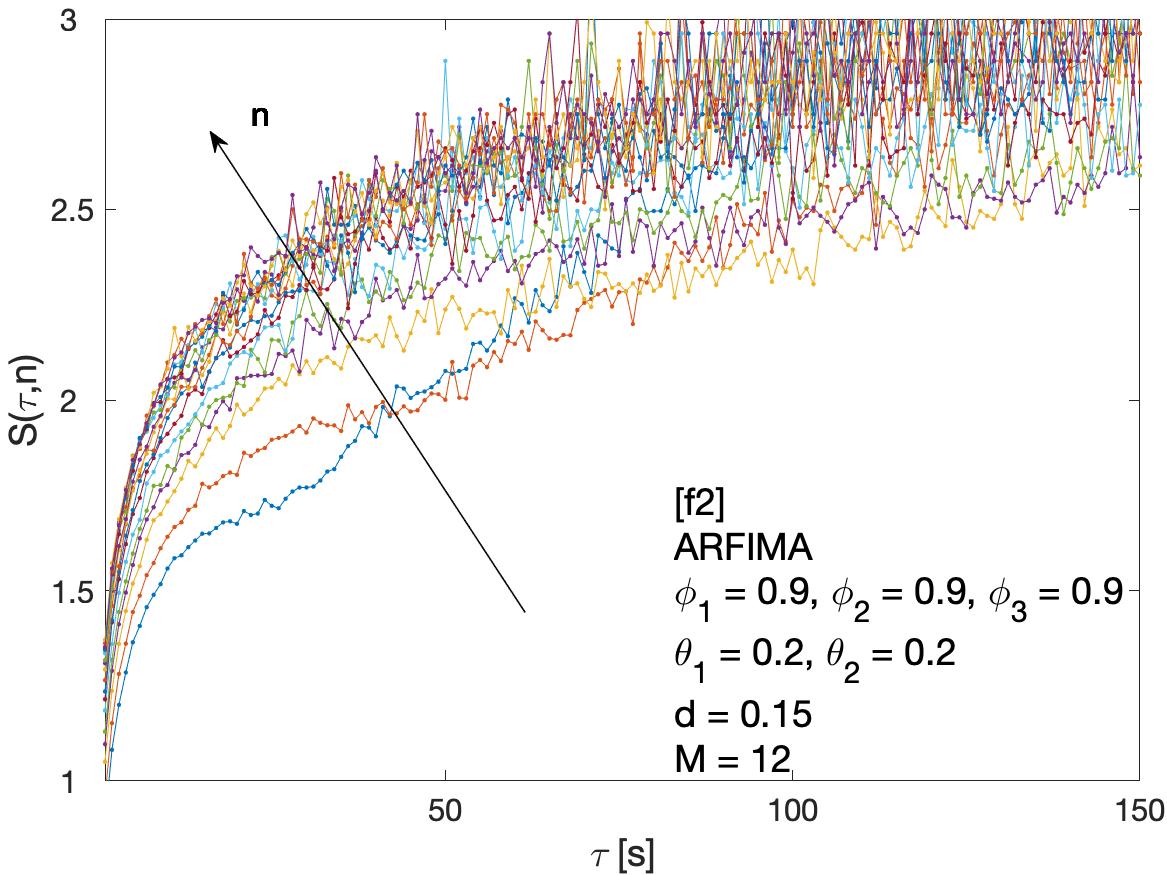}
    \includegraphics[width=4cm]{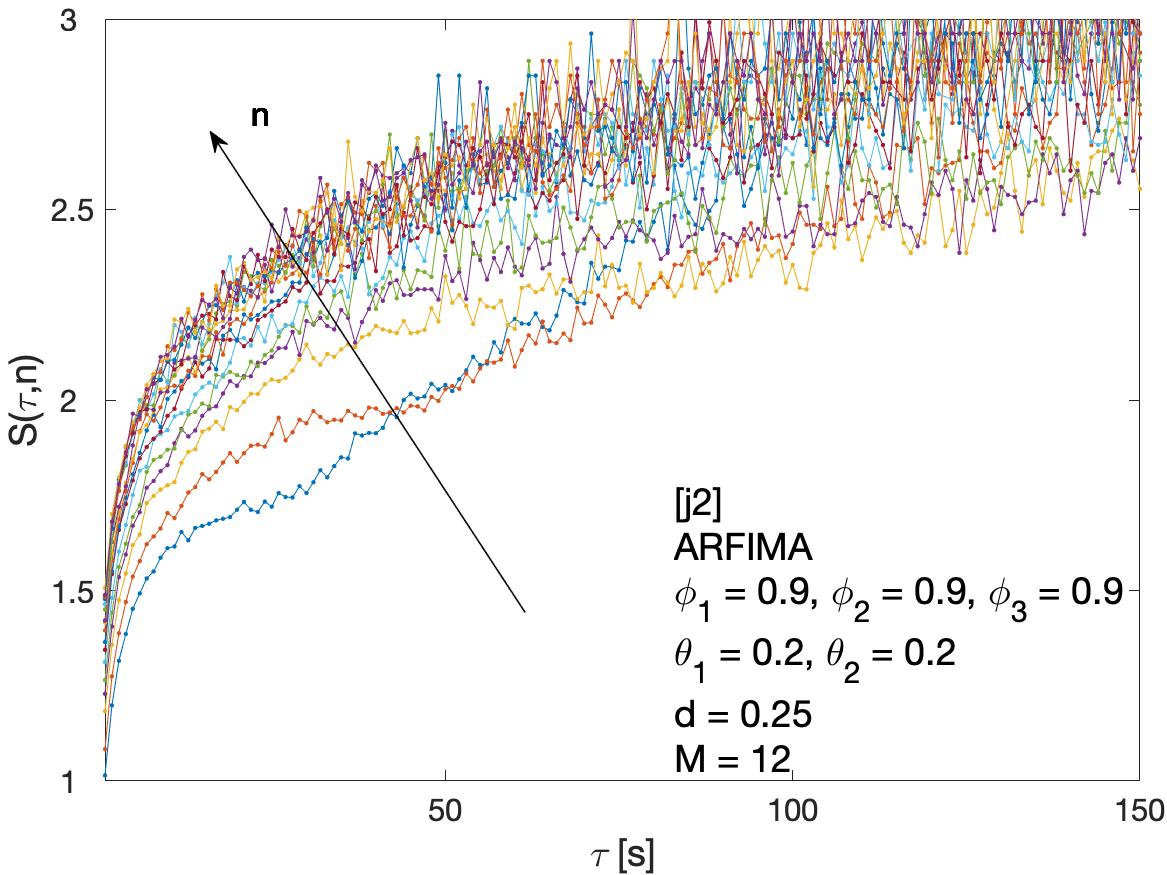}
    \caption{Cluster entropy results for horizon $M=12$ on ARFIMA series  with different combinations of the differencing parameter $d$, autoregressive parameter $\phi_1$, $\phi_2$, and  $\phi_3$  and moving average parameter $\theta_1$, $\theta_2$ and $\theta_3$. The differencing parameter takes values $d=0.05$, $d=0.15$ and $d=0.25$, with a different combination of autoregressive and moving average parameters. The full set of analysed values of $d$, $\phi$ and $\theta$  is reported in Table \ref{tab:arfima_models5}.}
    \label{fig:ARFIMA_CE_cplx_M12}
\end{figure}
\begin{figure}
    \centering
    \includegraphics[width=4cm]{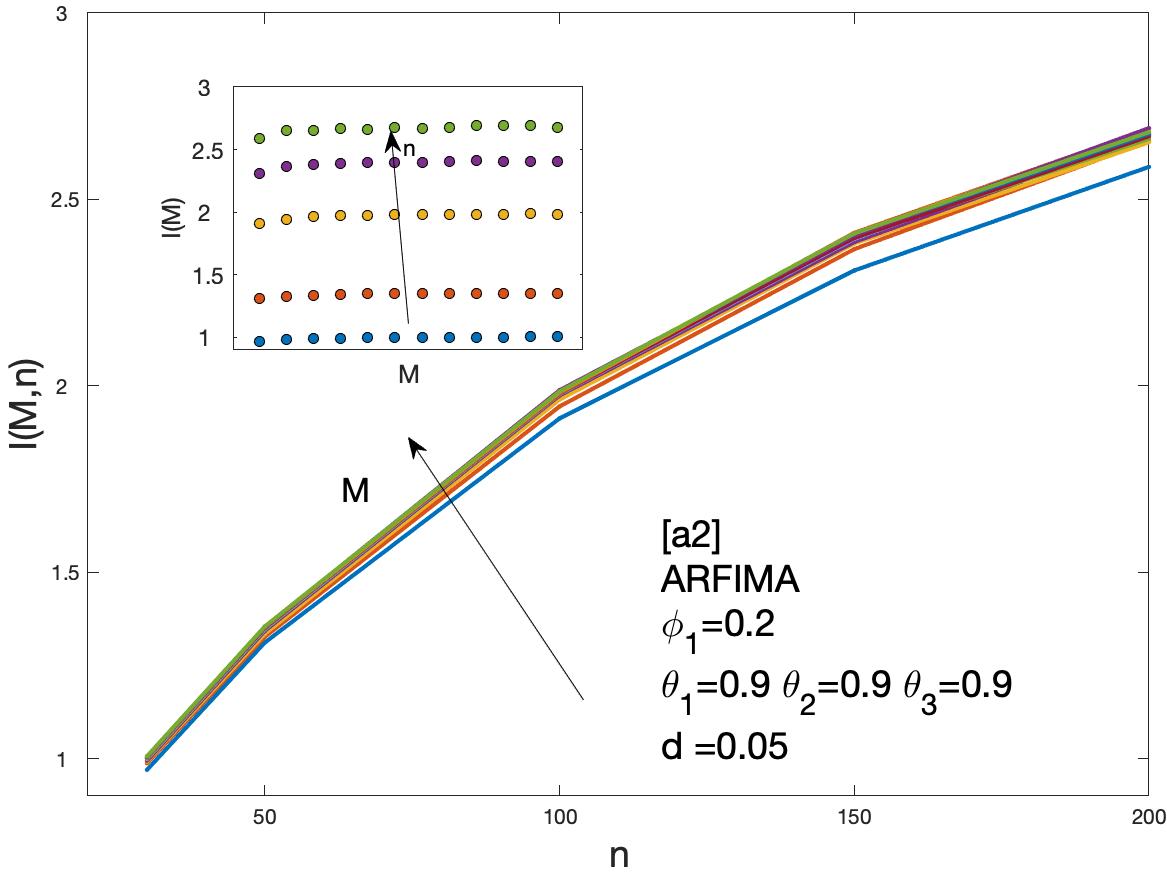}
    \includegraphics[width=4cm]{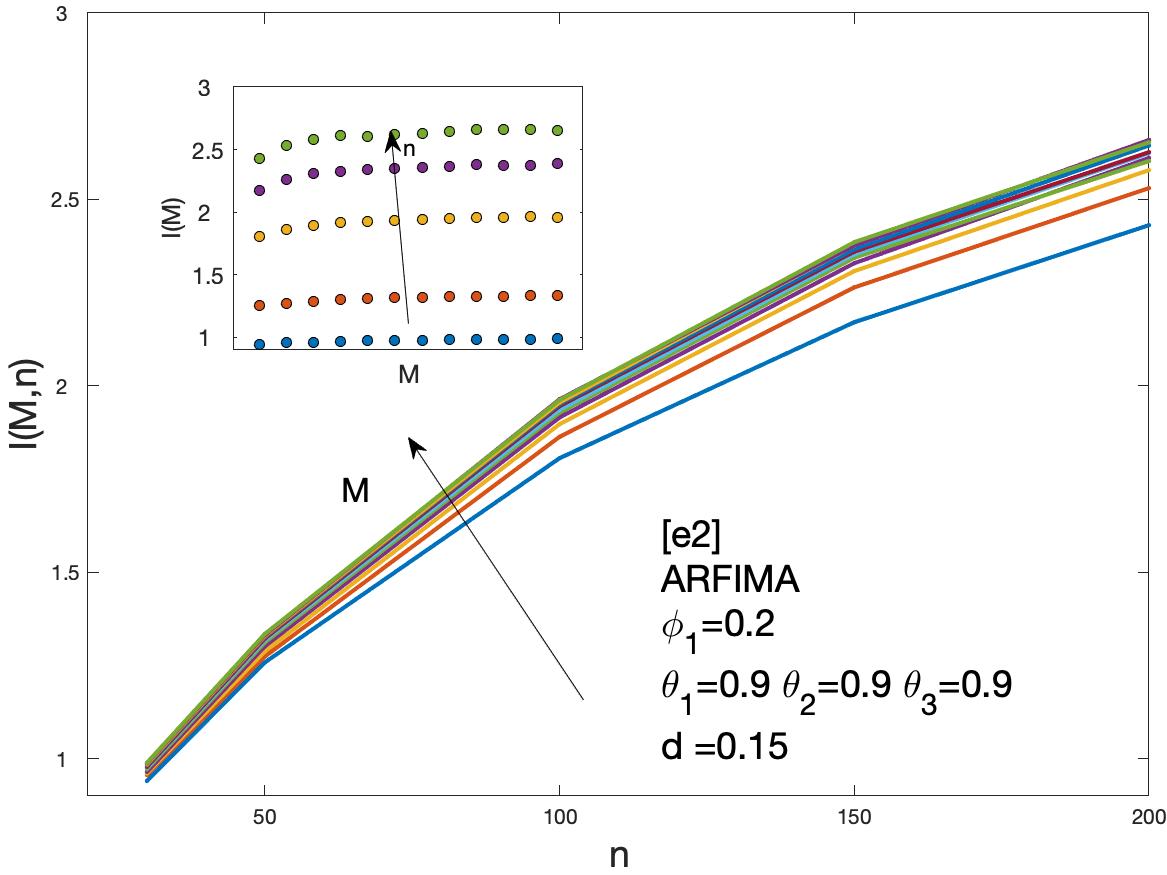}
    \includegraphics[width=4cm]{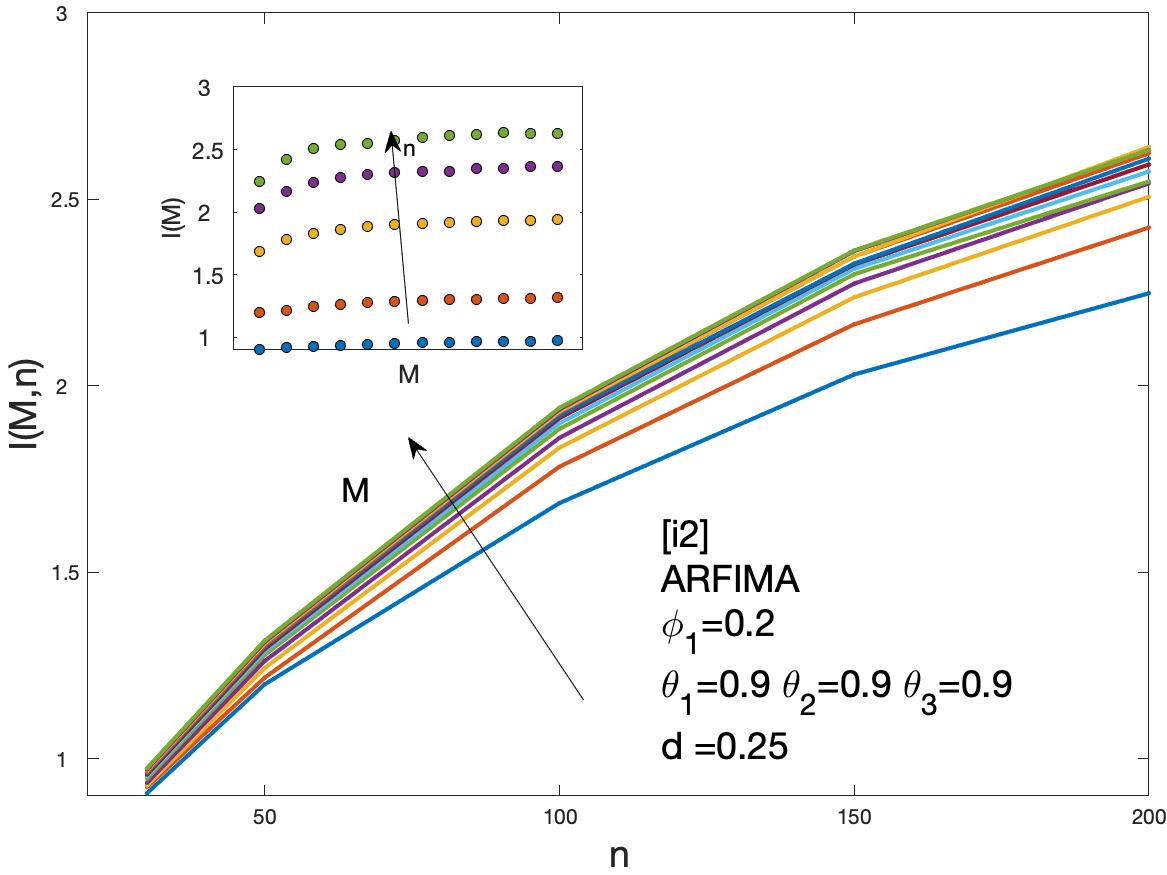}
    \\
    \includegraphics[width=4cm]{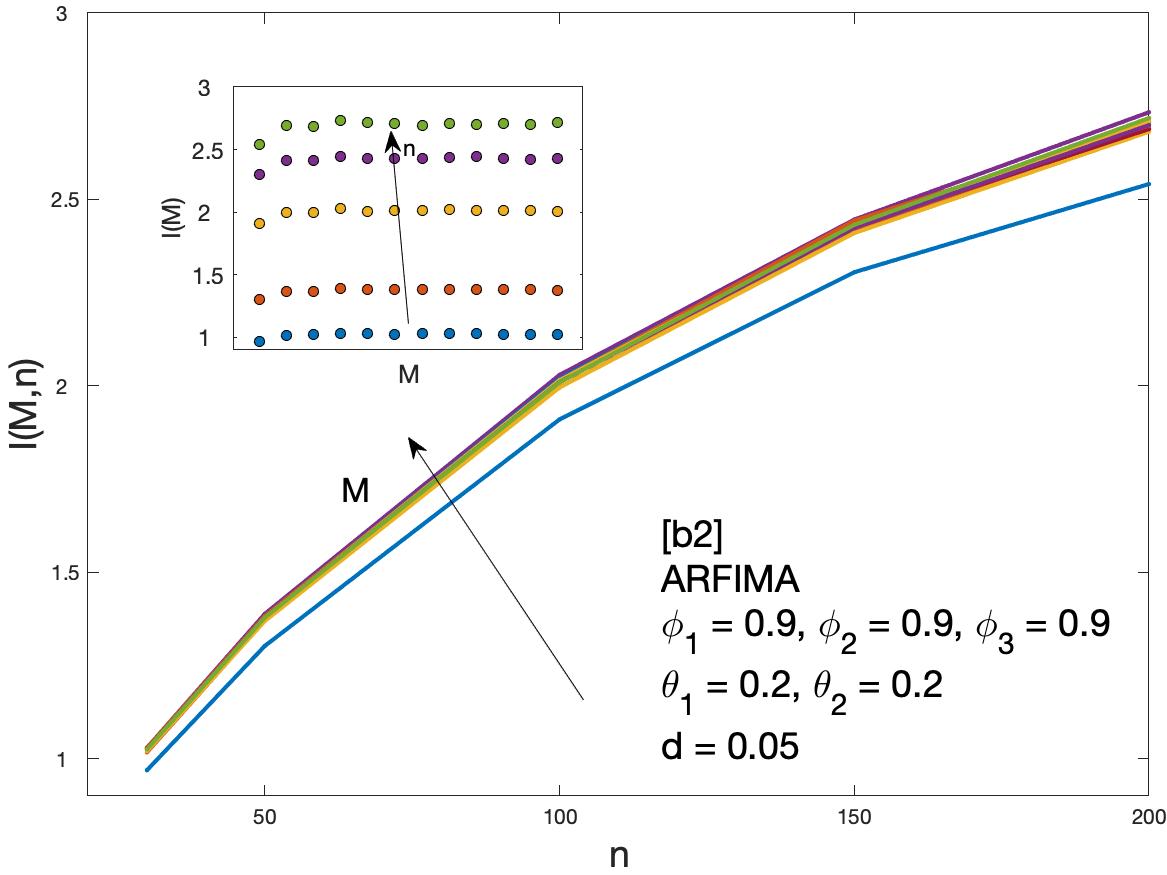}
    \includegraphics[width=4cm]{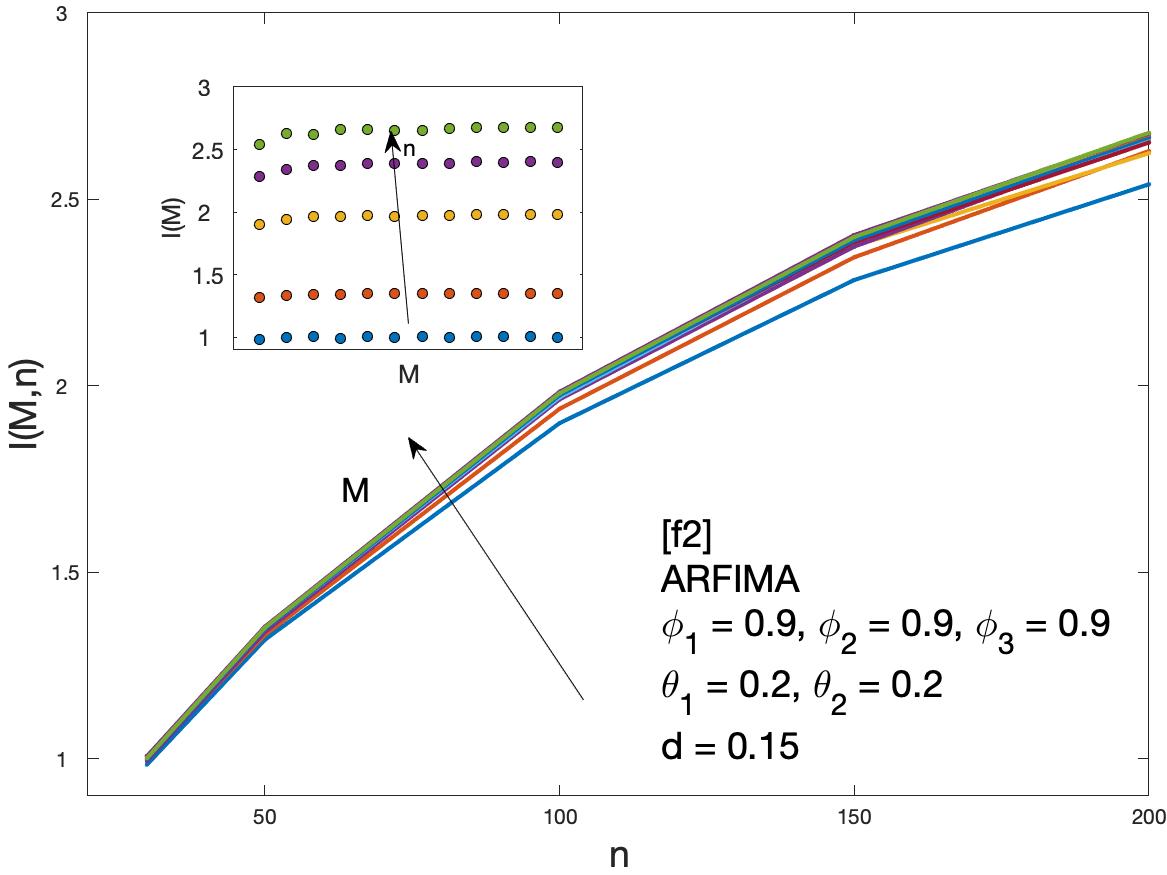}
    \includegraphics[width=4cm]{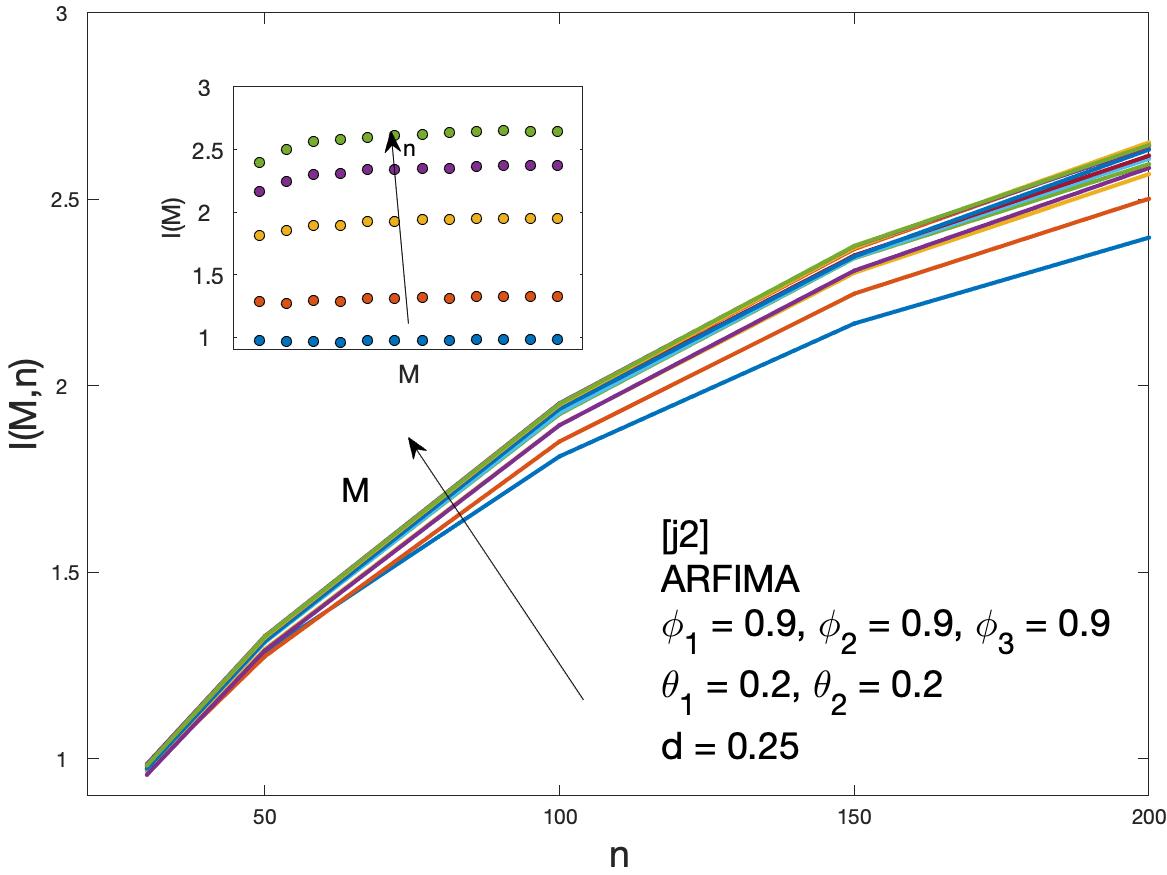}
    \caption{Market Dynamic Index $I(M,n)$ for ARFIMA series  with different combinations of the differencing parameter $d$, autoregressive parameter $\phi_1$, $\phi_2$, and  $\phi_3$  and moving average parameter $\theta_1$, $\theta_2$ and $\theta_3$. The differencing parameter takes values $d=0.05$, $d=0.15$ and $d=0.25$, with a different combination of autoregressive and moving average parameters as reported in Table \ref{tab:arfima_models5}.}
    \label{fig:ARFIMA_MDI_5}
\end{figure}

Cluster entropy results on ARFIMA (1,d,1),  are plotted in Figures \ref{fig:ARFIMA_CE_smpl_M1}, \ref{fig:ARFIMA_CE_smpl_M12}.
The corresponding market dynamic indexes $I(M,n)$  calculated by using the data of the cluster entropy results on ARFIMA (1,d,1) are shown in
\ref{fig:ARFIMA_MDI}.  Cluster entropy results  on  ARFIMA (3,d,2)  and  ARFIMA(1,d,3), corresponding to parameters marked  by alphabet labels in Table \ref{tab:arfima_models5}, are reported in Figures
\ref{fig:ARFIMA_CE_cplx_M1}, \ref{fig:ARFIMA_CE_cplx_M12}, \ref{fig:ARFIMA_MDI_5}.
Market Dynamic Index for series generated by ARFIMA processes are reported in Figure \ref{fig:ARFIMA_MDI}. With differencing parameter $0 < d < 0.2$, Market Dynamic Index curves are $n$-invariant for small values of $n$, but horizon dependence emerges at larger $n$. When $0.2 < d <0.5$ Market Dynamic Index curves show a significant horizon dependence even at small $n$. Therefore, according to the choice of the differencing parameter $d$, series generated by ARFIMA processes can reproduce the effect shown by the cluster entropy in real-world financial markets.

\section{Discussion and Conclusions}
\label{Sec:Discussion and Conclusions}
The cluster entropy behavior described by Equation (\ref{eq:carboneentropy}) has been replicated by simulations performed on artificially generated series, with results  reported in Section \ref{Sec:Results}. Figures show cluster entropy results for the following processes: Fractional Brownian Motion (Figure \ref{fig:FBM_CE}); Autoregressive Fractionally Integrated Processes (Figures \ref{fig:ARFIMA_CE_smpl_M1}, \ref{fig:ARFIMA_CE_smpl_M12}, \ref{fig:ARFIMA_CE_cplx_M1}, \ref{fig:ARFIMA_CE_cplx_M12}). The behavior of cluster entropy curves is well represented by Equation (\ref{eq:carboneentropy}), however deviations occur at extreme cases. In general, one can observe that power-law correlated clusters, characterized by length $\tau < n$, determine the logarithmic behavior of the entropy, regardless of the moving average window value $n$. On the other hand, exponentially correlated clusters, i.e. clusters with length $\tau \geq n$, are related to the linear behavior prescribed by the excess entropy term $\tau/n$, which depends on the moving average window $n$ and with slope decreasing as $1/n$.
 The Market Dynamic Index $I(M,n)$ is deduced from the  cluster entropy results by means of Equation (\ref{eq:realmdi}). Cumulative measures are useful to summarize key information in a single numerical index. $I(M,n)$ gathers the information present in the sequences at different time horizons $M$ and moving average windows $n$.
\par
The Market Dynamic Index $I(M,n)$ for series generated by means of Fractional Brownian Motion processes with Hurst exponent $0 < H < 0.5$ (anticorrelated FBMs) do not present any horizon dependence. Conversely, Fractional Brownian Motion series with $0.5 < H < 1$ (positively correlated FBMs) do show some horizon dependence. However, as it will be clarified below, Fractional Brownian Motion series fail to fully reproduce the financial markets behavior.
\par
The Market Dynamic Index estimated in long-range positively correlated sequences replicate the characteristic behaviour  observed in financial markets\cite{ponta2019quantifying}.
\par
In the case of ARFIMA processes, a significant horizon dependence emerges, as one can note by observing the Market Dynamic indexes plotted in Figures \ref{fig:ARFIMA_MDI} and \ref{fig:ARFIMA_MDI_5}.
Thus, cluster entropy for series generated by ARFIMA process exhibit horizon dependence as observed in real world  financial markets. The extent of long range dependence and its microscopic origin have been scrutinized in several studies
\cite{Bhardwaj2006An,BAILLIE2012Prediction} since the introduction of the ARFIMA process.
\par
To further validate our findings a statistical significance test has been performed by using  the paired t-test to check the null hypothesis $h=0$ that the cluster entropy values obtained by ARFIMA simulations come from distributions with equal mean and same variance with a probability $p$ with simple Brownian Motion assumed as benchmark $H=0.5$ results are reported in Table \ref{tab:p-test}.
\par
We report the results of the \emph{T-paired test}  performed on  NASDAQ, DJIA and S$\&$P500 markets in Table \ref{tab:p-testm} \cite{ponta2019quantifying}.
A qualitative comparison between Table \ref{tab:p-test} and Table \ref{tab:p-testm} suggests an overall similarity of the ARFIMA and real world markets.  In particular, one can note that the $p$ values in column [f1] are quite close to those of the S\&P500   suggesting  a correlation degree  with Hurst exponent $H\simeq 0.65$ and differencing parameter $d\simeq 0.15$  for S\&P500. Probability values in column [e2] are close to S\&P500, confirming the value $H\simeq 0.65$ and $d \simeq 0.15$.
The probability values for DJIA  are better approximated by the set of ARFIMA parameters in column [b1] and column [a2] suggesting lower values of the correlation exponents: $H\simeq 0.55$ and $ d\simeq 0.05$.
The lower values of the probability $p$ indicate a more complex behavior of the NASDAQ with stronger deviation from the fully uncorrelated Brownian motion. By looking at the Table \ref{tab:p-test}, one can relate the NASDAQ behaviour to higher values of the long-range parameters. In particular, the NASDAQ probability values become closer to  parameter sets corresponding to higher correlation degrees [i2] and [n2]. The values of the correlation exponents are expected to increase and reach values $H\simeq 0.75$ and $ d\simeq 0.25$.
\par
The cluster entropy behavior appears deeply linked to the positive persistence and long-range correlation. In real-world financial series horizon dependence deviates from the case of absolutely random series, such as those generated by means of stochastic differential equations.  The Market Dynamic Index, obtained via an integration performed on cluster entropy, provides this result in a cumulative and, thus, particularly robust form. Moreover, the different horizon dependence of NASDAQ and DJIA, where the former is a diversified stock market with a high degree of heterogeneity and the latter is an index representative of a chosen set of industrial stocks, is consistent with the ability of the cluster entropy index to quantify market heterogeneity.
Therefore, contrary to the traditional financial market theories, the hypothesis of efficient markets and rational investor behavior do not hold.
\clearpage
\newpage
\section*{Funding}
Pietro Murialdo acknowledges financial support from  \href{http://futurict2.eu/}{FuturICT 2.0} a FLAG-ERA Initiative within the Joint Transnational Calls 2016.


\section*{Conflicts of interest} 
The authors declare no conflict of interest. The funders had no role in the design of the study; in the collection, analyses, or interpretation of data; in the writing of the manuscript, or in the decision to publish the results.
\bibliographystyle{elsarticle-num}
\bibliography{Information2017}



\clearpage
\newpage
\begin{table}[]
    \centering
    \begin{tabular}{ccccc}
\toprule
\textit{M}&$N$&$N_M$&$t_S$&$t_S^*$\\
\midrule
1&586866&586866&1.0000&1\\
2&1117840&586866&1.9048&1\\
3&1704706&586866&2.9048&2\\
4&2291572&586866&3.9048&3\\
5&2906384&586866&4.9524&4\\
6&3493250&586866&5.9524&5\\
7&4069315&586866&6.9340&6\\
8&4712062&586866&8.0292&8\\
9&5243029&586866&8.9339&8\\
10&5885781&586866&10.0292&10\\
11&6461845&586866&11.0108&11\\
12&6982017&586866&11.8971&11\\
\bottomrule
    \end{tabular}
    \caption{Example of cumulative series. Lengths for each time horizon $M$ are reported for NASDAQ data in 2018. Each row corresponds to the number of transactions that took place in month $M$ in 2018 plus the number of transactions that occurred in past months of 2018. These lengths are used as a reference to generate artificial series and allow a correct comparison between results obtained on real and artificial data.}
    \label{tab:sampling}
\end{table}

\begin{table}[h!]
    \centering
    \begin{tabular}{cccc|c}
    \toprule
$H$&$d$&$\phi$&$\theta$\\[0.5ex]
    \midrule
\multirow{2}{*}{0.55}&\multirow{2}{*}{0.05}&0.20&0.90& a1\\
&&0.90&0.20& b1\\
\hline
\multirow{2}{*}{0.60}&\multirow{2}{*}{0.10}&0.20&0.90& c1\\
&&0.90&0.20& d1\\
\hline
\multirow{2}{*}{0.65}&\multirow{2}{*}{0.15}&0.20&0.90& e1\\
&&0.90&0.20& f1\\
 \hline
\multirow{2}{*}{0.70}&\multirow{2}{*}{0.20}&0.20&0.90& g1\\
&&0.90&0.20& h1\\
\hline
\multirow{6}{*}{0.75}&\multirow{6}{*}{0.25}&0.20&0.90& i1\\
\cline{3-4}
&&\multirow{2}{*}{0.30}&0.40& j1\\
&&&0.85& k1\\
 \cline{3-4}
&&\multirow{3}{*}{0.90}&0.20& l1\\
&&&0.40& m1\\
&&&0.85& n1\\
\hline
\multirow{2}{*}{0.80}&\multirow{2}{*}{0.30}&0.20&0.90& o1\\
&&0.90&0.20& p1\\
\hline
\multirow{4}{*}{0.98}&\multirow{4}{*}{0.48}&\multirow{2}{*}{0.30}&0.40& q1\\
&&&0.85& r1\\
 \cline{3-4}
&&\multirow{2}{*}{0.90}&0.40& s1 \\
&&&0.85& t1\\
\bottomrule
    \end{tabular}
    \caption{Full set of parameter range for the ARFIMA (1,d,1) processes simulated in this work. Specifically $H$ is the Hurst exponent and  $d$ is the differencing parameter ($1^{st}$ and 2$^{nd}$ columns) which are related by Equation (20), $\phi$ is the autoregressive parameter (3$^{rd}$ column), and $\theta$ is the  moving average parameter (4$^{th}$ column).}
    \label{tab:arfima_models}
\end{table}

\begin{table}[h!]
\centering
\begin{tabular}{cccccccc|c}
\toprule
$H$ & $d$ & $\phi_1$ & $\phi_2$ & $\phi_3$ & $\theta_1$ & $\theta_2$ & $\theta_3$\\[0.5ex]
\midrule
\multirow{2}{*}{0.55}&\multirow{2}{*}{0.05}&0.20&-&-&0.90&0.90&0.90& a2\\
&&0.90&0.90&0.90&0.20&0.20&-& b2\\
 \hline
\multirow{2}{*}{0.60}&\multirow{2}{*}{0.10}&0.20&-&-&0.90&0.90&0.90& c2\\
&&0.90&0.90&0.90&0.20&0.20&-& d2\\
\hline
\multirow{2}{*}{0.65}&\multirow{2}{*}{0.15}&0.20&-&-&0.90&0.90&0.90& e2\\
&&0.90&0.90&0.90&0.20&0.20&-& f2\\
 \hline
\multirow{2}{*}{0.70}&\multirow{2}{*}{0.20}&0.20&-&-&0.90&0.90&0.90& g2\\
&&0.90&0.90&0.90&0.20&0.20&-& h2\\
\hline
\multirow{2}{*}{0.75}&\multirow{2}{*}{0.25}&0.20&-&-&0.90&0.90&0.90& i2\\
&&0.90&0.90&0.90&0.20&0.20&-&j2\\
 \hline
\multirow{3}{*}{0.80}&\multirow{3}{*}{0.30}&0.20&-&-&0.90&0.90&0.90& k2\\
&&0.40&0.16&-&0.90&0.81&0.73& l2\\
&&0.90&0.90&0.90&0.20&0.20&-& m2\\
 \hline
0.85&0.35&0.20&-&-&0.90&0.90&0.90& n2\\
 \hline
0.98&0.48&0.40&0.16&-&0.90&0.81&0.73& o2\\
\bottomrule
\end{tabular}
\caption{Full set of parameter range for  ARFIMA (3,d,2) and ARFIMA(1,d,3) processes simulated in this work.  Specifically $H$ is the Hurst exponent and  $d$ is the differencing parameter which are related by Equation (20) ($1^{st}$ and 2$^{nd}$ columns); $\phi_1$, $\phi_2$ and  $\phi_3$  are the autoregressive parameters (3$^{rd}$, 4$^{th}$ and 5$^{th}$ columns); $\theta_1$, $\theta_2$ and $\theta_3$  are the  moving average parameters (6$^{th}$, 7$^{th}$ and 8$^{th}$ columns).}
\label{tab:arfima_models5}
\end{table}
\clearpage
\newpage

\begin{table}[h]
     \centering
    \begin{tabular}{ccccccccc}
    \toprule
   $M$ & \textit{[b1]} & \textit{[f1]} & \textit{[l1]} & \textit{[a2]} & \textit{[e2]} & \textit{[i2]} & \textit{[n2]}  & \textit{[o2]}\\
    \midrule
1&0.9597&0.7938&0.6013&0.8519&0.6779&0.4956&0.3542&0.2314\\  2&0.9863&0.8429&0.6985&0.9293&0.7883&0.6566&0.5414&0.4304\\  3&0.982&0.8789&0.7743&0.938&0.8346&0.7362&0.6468&0.5576\\  4&0.9848&0.8922&0.8031&0.956&0.8689&0.7827&0.7147&0.638\\  5&0.9878&0.9062&0.8325&0.9608&0.8809&0.8102&0.7528&0.6911\\  6&0.994&0.9197&0.8517&0.9724&0.9043&0.8417&0.784&0.7322\\  7&0.9785&0.9186&0.8633&0.9617&0.9038&0.8521&0.8036&0.7614\\  8&0.993&0.9321&0.8775&0.9762&0.9229&0.871&0.8333&0.7931\\  9&0.9867&0.937&0.889&0.9737&0.9273&0.8809&0.8438&0.8100\\  10&0.9813&0.9333&0.8952&0.971&0.9261&0.8880&0.8533&0.8195\\  11&0.9816&0.9436&0.9011&0.9749&0.9326&0.8965&0.8643&0.8342\\  12&0.9853&0.9451&0.9072&0.9741&0.9353&0.9019&0.8764&0.8508\\
    \bottomrule
    \end{tabular}
     \caption{Probability $p$ to reject the null hypothesis that the cluster entropy values for the ARFIMA processes  at varying horizons $M$, have same mean and variance of the Fractional Brownian Motion with $H=0.5$. The probability $p$ has been estimated by standard \emph{T-paired test}. First column reports the temporal horizon $M$.  The other columns refers to parameter sets  \textit{[b1]}, \textit{[f1]}, \textit{[l1]}, \textit{[a2]}, \textit{[e2]}, \textit{[i2]}, \textit{[n2]}, \textit{[o2]} of Table \ref{tab:arfima_models}  and Table \ref{tab:arfima_models5}.}
    \label{tab:p-test}
\end{table}

\begin{table}[h]
\centering
\begin{tabular}{cccc}
\toprule
\textbf{$M$}&\textbf{NASDAQ}	&\textbf{S\&P500}&\textbf{DJIA}\\
\midrule
 1&0.5154&0.7399&0.8892\\
 2&0.6026&0.8335&0.9257\\
3       &0.647      &0.8588         &0.9332\\
 4&0.6631&0.8814&0.9283 \\  5&0.6823&0.9018&0.9417 \\  6&0.7124&0.9246&0.9534 \\  7&0.7162&0.9224&0.9461 \\ 8&0.7288&0.9309&0.9618 \\  9&0.7370&0.9479&0.9645\\
\bottomrule
\end{tabular}
\caption{Probability $p$ to reject the null hypothesis that the cluster entropy values for the NASDAQ, DJIA and S$\&$P500 at varying horizons $M$ have same mean and variance of the Fractional Brownian Motion with $H=0.5$. First column reports the temporal horizon $M$. The probability $p$ has been estimated by standard \emph{T-paired test}  \cite{ponta2019quantifying}}
\label{tab:p-testm}
\end{table}


\clearpage

\clearpage

\clearpage

\clearpage

\clearpage

\end{document}